\documentclass[a4paper,11pt]{article}
\pdfoutput=1
\usepackage{amsfonts, amsmath, amssymb}
\usepackage[utf8]{inputenc}
\usepackage[a4paper,top=3cm,bottom=2.5cm,left=2.5cm,right=2.5cm, bindingoffset=5mm]{geometry}
\usepackage{color}
\usepackage{booktabs}
\usepackage{subfigure}
\usepackage{graphicx}
\usepackage{float}
\usepackage{hyperref}
\usepackage{multirow}
\hypersetup{citecolor=blue}
\usepackage{array}
\usepackage{multirow}
\usepackage{colordvi}

\newcommand{\be}{\begin{equation}}
\newcommand{\ee}{\end{equation}}
\newcommand{\beq}{\begin{equation*}}
\newcommand{\eeq}{\end{equation*}}
\newcommand{\ba}{\begin{eqnarray}}
\newcommand{\ea}{\end{eqnarray}}
\newcommand{\mefff}{\mbox{$ \langle\!~m~\!\rangle $}}
\newcommand{\betabeta}{\mbox{$(\beta \beta)_{0 \nu}$}}
\newcommand{\meff}{\mbox{$\left|  \langle\!~m~\!\rangle\right| \ $}}

\newcommand{\Dmq}{\ensuremath{\Delta m^2}}

\def\ltap{\ \raisebox{-.4ex}{\rlap{$\sim$}} \raisebox{.4ex}{$<$}\ }
\def\gtap{\ \raisebox{-.4ex}{\rlap{$\sim$}} \raisebox{.4ex}{$>$}\ }

\newcommand{\un}{\underline}
\newcommand{\ov}{\overline}

\begin{document}

\hfill{{\small SISSA 37/2013/FISI}}
%
%
%
\vspace{20pt}
\begin{center}
{\bf{\large  Neutrinoless Double Beta Decay in the Presence of Light
Sterile Neutrinos}}

\vspace{0.4cm} I. Girardi$\mbox{}^{a,b)}$,
A. Meroni$\mbox{}^{a,b)}$ and S. T. Petcov$\mbox{}^{a,b,c)}$
\footnote{Also at: Institute of Nuclear Research and Nuclear Energy,
Bulgarian Academy of Sciences, 1784 Sofia, Bulgaria}

\vspace{0.1cm} $\mbox{}^{a)}${\em  SISSA, Via Bonomea 265, 34136
Trieste, Italy.\\}

\vspace{0.1cm} $\mbox{}^{b)}${\em  INFN, Sezione di Trieste, 34126
Trieste, Italy.\\}

\vspace{0.1cm} $\mbox{}^{c)}${\em Kavli IPMU (WPI), The University
of Tokyo, Kashiwa,
Japan.\\
}

\end{center}

\begin{abstract}
We investigate the predictions for
neutrinoless double beta (\betabeta-) decay
effective Majorana mass
$\meff$ in the $3+1$ and $3+2$
schemes with one and two additional sterile neutrinos
with masses at the eV scale. The two schemes are
suggested by the neutrino oscillation
interpretation of the reactor neutrino and Gallium
``anomalies'' and of the data of the LSND and
MiniBooNE experiments. We analyse
in detail the possibility
of a complete or partial cancellation between the
different terms in $\meff$, leading to a
strong suppression of $\meff$. We determine
the regions of the relevant
parameter spaces where such a suppression
can occure. This allows us to derive the conditions
under which the effective Majorana mass satisfies
$\meff > 0.01$ eV, which is the range
planned to be exploited by the next generation of
$\betabeta$-experiments.

\end{abstract}

\section{Introduction}

  All compelling neutrino oscillation data can be described
within the reference 3-flavour neutrino mixing scheme
with 3 light neutrinos $\nu_j$ having masses $m_j$ not
exceeding approximately 1 eV, $m_j \ltap 1$ eV, $j=1,2,3$
(see, e.g., \cite{PDG2012}). These data allowed to determine
the parameters which drive the observed solar,
atmospheric, reactor and accelerator flavour
neutrino oscillations - the three neutrino mixing angles
of the standard parametrisation of the Pontecorvo, Maki,
Nakagawa and Sakata (PMNS) neutrino mixing matrix,
$\theta_{12}$, $\theta_{23}$ and $\theta_{13}$, and the two
neutrino mass squared differences $\Delta m^2_{21}$ and
$\Delta m^2_{31}$ (or $\Delta m^2_{32}$) - with a relatively
high precision \cite{Fogli:2012ua,Gonzales-Garcia:2012}.
In Table 1 we give the values of the 3-flavour neutrino
oscillation parameters as determined in the global
analysis performed in \cite{Fogli:2012ua}.

    At the same time at present there are a number of hints
for existence of light sterile neutrinos with
masses at the eV scale.
They originate from the re-analyses of the
short baseline (SBL) reactor neutrino oscillation
data using newly calculated fluxes of reactor
$\bar{\nu}_e$, which show a possible ``disappearance''
of the reactor $\bar{\nu}_e$ (``reactor neutrino anomaly''),
from the results of the calibration experiments of
the Gallium solar neutrino detectors
GALLEX and SAGE (``Gallium anomaly'') and from
the results of the LSND and MiniBooNE experiments
(for a summary of the data and complete list of references see, e.g.,
\cite{SterWPaper11}). The evidences for sterile neutrinos from
the different data are typically at the level of up to
approximately $3\sigma$, 
except in the case of the results of the 
LSND experiment which give much higher C.L.

Significant constraints on the parameters characterising the 
oscillations involving sterile neutrinos follow from the negative 
results of the searches for  $\nu_{\mu} \rightarrow \nu_e$ and/or
$\bar{\nu}_{\mu} \rightarrow \bar{\nu}_e$ oscillations
in the KARMEN \cite{KARMEN2002}, NOMAD \cite{NOMAD2003} and 
ICARUS \cite{ICARUS13} experiments,  
and from the nonobservation of effects of oscillations 
into sterile neutrinos in the solar neutrino experiments and in the 
studies of $\nu_{\mu}$ and/or $\bar{\nu}_{\mu}$ disappearance 
in the CDHSW \cite{CDHSW84}, MINOS \cite{MINOSster} 
and SuperKamiokande \cite{SuperKster} experiments.

Constraints on the number and masses of sterile neutrinos
are provided by cosmological data. The recent  
Planck results, in particular, on
the effective number of relativistic degrees of
freedom at recombination epoch $N_{\text{eff}}$, 
can be converted into a constraint on the number of 
(fully thermalised) sterile neutrinos \cite{Ade:2013lta} 
(see also, e.g.,  \cite{Mirizzi:2013kva,Wyman:2013lza} 
and references quoted therein).
The result one obtains depends on the model complexity and the 
input data used in the analysis. Assuming the validity of the 
$\Lambda$ CDM (Cold Dark Matter) model and combining the 
i) Planck and WMAP CMB data,
ii) Planck,  WMAP and Baryon Acoustic Oscillation (BAO) data,
iii) Planck,  WMAP, BAO and high multipole CMB data, 
for the best fit value and 95\% C.L. interval of allowed values of 
$N_{\text{eff}}$ it was found \cite{Ade:2013lta}:
i) 3.08, (2.77 - 4.31), 
ii) 3.08, (2.83 - 3.99),
ii) 3.22, (2.79 -3.84). 
The prediction in the case of three light (active) 
neutrinos reads  $N_{\text{eff}}=3.046$. 
The quoted values are compatible at $2\sigma$ with the existence of extra
radiation corresponding to one (fully thermalised) 
sterile neutrino, while the possibility of 
existence of two (fully thermalised) 
sterile neutrinos seems to be disfavored 
by the available cosmological data.
In what concerns the combined cosmological limits
on the mass and number of sterile neutrinos, 
they depend again on data used as input in the analysis: 
in the case of one fully thermalised sterile neutrino, 
the upper limits at 95\% C.L. are typically 
of approximately 0.5 eV, 
but is relaxed to 1.4 eV if one includes in 
the relevant data set the results of 
measurements of the local galaxy cluster mass 
distribution \cite{Giunti1309}.
The existence of one sterile neutrino with a mass 
in the 1 eV range and couplings tuned to explain 
the anomalies described briefly above 
would be compatible with 
the cosmological constraints 
if the production of sterile neutrinos 
in the Early Universe is suppresses by some non-standard mechanism 
(as like a large lepton asymmetry, see, e.g., \cite{NSaviano1302}), 
so that  $N_{\text{eff}} < 3.8$ \cite{Giunti1309}.  

The bounds on $N_{\text{eff}}$ and on the sum of the light neutrino
masses will be improved by current or forthcoming observations. For
instance, the EUCLID survey \cite{Laureijs:2011mu}
is planned to determine the sum of neutrino
masses  with a $1\sigma$ uncertainty of $\sim 0.01$ eV, combining 
the EUCLID data with measurements of the CMB anisotropies from 
the Planck mission. This would lead to strong constrains 
on extra sterile neutrino states.

 Two possible ``minimal'' phenomenological
models (or schemes) with light sterile neutrinos
are widely used in order
to explain the reactor neutrino and Gallium anomalies, 
the LSND and MiniBooNE data as well as the results of 
the negative searches for active-sterile neutrino 
oscillations: the so-called ``$3 + 1$'' and ``$3 + 2$''
models, which contain respectively one and two sterile neutrinos
(right-handed sterile neutrino fields). The latter are assumed to
mix with the 3 active flavour neutrinos (left-handed flavour
neutrino fields) (see, e.g.,
\cite{GiuntiTalk:2013vaa,Kopp:2013vaa}). Thus, the ``$3 + 1$'' and
``$3 + 2$'' models have altogether 4 and 5 light massive neutrinos
$\nu_j$, which in the minimal versions of these models are Majorana
particles. The additional neutrinos $\nu_4$ and $\nu_4$, $\nu_5$,
should have masses $m_4$ and $m_4$, $m_5$ at the eV scale (see
further). It follows from the data that if $\nu_4$ or $\nu_4$,
$\nu_5$ exist, they couple to the electron and muon in the weak
charged lepton current with couplings $U_{e k}$ and $U_{\mu k}$,
$k=4;~4,5$, which are approximately  $|U_{e k}|\sim 0.1$ and
$|U_{\mu k}|\sim 0.1$. The hypothesis of existence of light sterile
neutrinos with eV scale masses and the indicated charged current
couplings to the electron and muon will be tested in a number of
experiments with reactor and accelerator neutrinos, and neutrinos
from artificial sources, some of which are under preparation and
planned to start taking data already this year (see, e.g.,
\cite{SterWPaper11} for a detailed list and discussion of the
planned experiments).
%
\begin{table}[h]
\centering
\renewcommand{\arraystretch}{1.1}
\begin{tabular}{lcc}
\toprule
 Parameter  &  best-fit ($\pm 1\sigma$) & 3$\sigma$ \\ \midrule
 $\Delta m^{2}_{21} \; [10^{-5}\text{ eV}^2]$  & 7.54$^{+0.26}_{-0.22}$ &
               6.99 - 8.18 \\
$ |\Delta m^{2}_{31}|~(NO)  \; [10^{-3}\text{ eV}^2]$ &
   2.47$^{+0.06}_{-0.10}$  &
         2.19 - 2.62\\
$ |\Delta m^{2}_{32}|~(IO)  \; [10^{-3}\text{ eV}^2]$   & 
2.46$^{+0.07}_{-0.11}$ 
&  2.17 - 2.61\\
 $\sin^2\theta_{12}$ (NH or IH)  & 0.307$^{+0.018}_{-0.016}$
            & 0.259 - 0.359\\
$\sin^2\theta_{23}$  (NH) & 0.386$^{+0.024}_{-0.021}$ &  0.331 - 0.637 \\
\phantom{aaaaaa}  (IH)  &  0.392$^{+0.039}_{-0.022}$  & 0.335 - 0.663\\
$\sin^2\theta_{13}$ (NH)  &  0.0241$^{+0.0025}_{-0.0025}$  & 0.0169 - 0.0313\\
\phantom{aaaaaa}  (IH)  & 0.0244$^{+0.0023}_{-0.0025}$ & 0.0171 - 0.0315 \\
\bottomrule
\end{tabular}
\caption{The best-fit values and $3\sigma$
allowed ranges of the 3-flavour neutrino oscillation parameters derived from
a global fit of the current neutrino oscillation data
(from \cite{Fogli:2012ua}). The $3\sigma$ allowed range for
$\theta_{23}$  takes into account the statistical octant degeneracy
resulting from the analysis.
}
 \label{tabNudata}
\end{table}

  It was noticed in \cite{BPP2,SGWRode05} and more recently in
\cite{Barry:2011,Li:2012,Goswami:2013}
that the contribution of the additional light
Majorana neutrinos $\nu_4$ or $\nu_{4,5}$ to the
neutrinoless double beta ($\betabeta$-) decay amplitude,
and thus to the $\betabeta$-decay effective
Majorana mass $\meff$ (see, e.g.,
\cite{BiPet87,WRodej11}),
can change drastically the predictions for
$\meff$ obtained in the reference
3-flavour neutrino mixing scheme, $|\mefff^{(3\nu)}|$.
We recall that the predictions for  $|\mefff^{(3\nu)}|$
depend on the type of the neutrino mass spectrum
\cite{BPP1,PPSNO2bb}. As is well known,
depending on the sign of $\Delta m^2_{31(2)}$,
which cannot be determined from
the presently available
neutrino oscillation data,
two types of neutrino mass spectrum are possible:\\
{\it i) spectrum with normal ordering (NO)}:
$m_1 < m_2 < m_3$,
$\Delta m^2_{31} >0$,
$\Delta m^2_{21} > 0$,
$m_{2(3)} = (m_1^2 + \Delta m^2_{21(31)})^{1\over{2}}$; \\~~
{\it ii) spectrum with inverted ordering (IO)}:
$m_3 < m_1 < m_2$,
$\Delta m^2_{32}< 0$,
$\Delta m^2_{21} > 0$,
$m_{2} = (m_3^2 + \Delta m^2_{23})^{1\over{2}}$,
$m_{1} = (m_3^2 + \Delta m^2_{23} - \Delta m^2_{21})^{1\over{2}}$.\\
Depending on the value of the lightest neutrino mass,
${\rm min}(m_j)$, the neutrino mass spectrum can be:\\
%
{\it a) Normal Hierarchical (NH)}:
$m_1 \ll m_2 < m_3$, $m_2 \cong (\Delta m^2_{21})^{1\over{2}}
\cong 8.68 \times 10^{-3}$ eV,
$m_3 \cong (\Delta m^2_{31})^{1\over{2}}
\cong 4.97\times 10^{-2}~{\rm eV}$; or  \\
%
{\it b) Inverted Hierarchical (IH)}: $m_3 \ll m_1 < m_2$,
with $m_{1,2} \cong |\Delta m^2_{32}|^{1\over{2}}\cong 4.97\times 10^{-2}$ eV; or  \\
%
{\it c) Quasi-Degenerate (QD)}: $m_1 \cong m_2 \cong m_3 \cong m_0$,
$m_j^2 \gg |\Delta m^2_{31(32)}|$, $m_0 \gtap 0.10$ eV, $j=1,2,3$.

The precision of the current data do not allow to determine 
the type of the neutrino mass spectrum and thus we have 
$\Delta m^2_{31}(NO) \cong - \Delta m^2_{32}(IO)$.

 Using the values of the neutrino oscillation
parameters and their $3\sigma$ allowed ranges
one finds that \cite{Petcov:2013poa} (see also,
e.g., \cite{PPSNO2bb}) $|\mefff^{(3\nu)}| \ltap 0.005$ eV
in the case of 3-neutrino mass spectrum of NH type, while
if the spectrum is of the IH type
one has $0.014~{\rm eV} \ltap |\mefff^{(3\nu)}| \ltap 0.050$ eV.
These predictions are significantly modified, e.g.,
in the 3+1 scheme due to the contribution of $\nu_4$
to $\meff$ \cite{Barry:2011}. Now $\meff$ in the NH case
satisfies  $\meff \geq  0.01$ eV and
can lie in the interval $(0.01 - 0.05)$ eV,
and we can have $\meff \ltap 0.005$ eV
if the 3-neutrino mass spectrum is of the IH type.

 In the present article we investigate
the predictions for $\meff$ in the $3+1$ and $3+2$ schemes. More
specifically, we analyze in detail the possibility of a complete or
partial cancellation between the different terms in $\meff$, leading
to a strong suppression of $\meff$.
Whenever possible (e.g., in the cases of the $3+1$ scheme and
for the CP conserving values of the CP violation (CPV) phases in the
$3+2$ scheme), we determine analytically the
region of the relevant parameter spaces where such a suppression can
occur. In both the $3+1$ and $3+2$ schemes we perform also
a numerical analysis to derive the values of the
CPV phases for which a complete cancellation can take place. This allows
us to derive the conditions under which the effective Majorana mass
satisfies $\meff
> 0.01$ eV, which is the range planned to be exploited by the next
generation of $\betabeta$-experiments. Our study is a natural
continuation of the earlier studies \cite{BPP2,SGWRode05} and
\cite{Barry:2011,Li:2012,Goswami:2013} on the subject.

%
\section{One Sterile Neutrino: the 3+1 Model}
%

In this Section we study the case of existence of one
extra sterile neutrino. We will use the parametrisation of the
PMNS matrix adopted in \cite{Kopp:2013vaa}:
\be
U =  O_{24} O_{23} O_{14} V_{13} V_{12}\,\,
diag (1,e^{i\alpha/2},e^{i\beta/2},e^{i\gamma/2}),
\ee
%
where $O_{ij}$ and $V_{kl}$ describe real and complex rotations in
$i-j$ and $k-l$ planes, respectively, and $\alpha$, $\beta$ and
$\gamma$ are three CP violation (CPV) Majorana phases \cite{BHP80}.
Each of the matrices  $V_{12}$ and  $V_{13}$ contains one CPV phase,
$\delta_{12}$ and $\delta_{13}$, respectively, in their only two
nonzero nondiagonal elements. The phases $\delta_{12}$ and
$\delta_{13}$  enter into the expression for $\meff$ in the
combinations $\alpha/2 - \delta_{12}$ and $\beta/2 - \delta_{13}$.
Therefore for the purposes of the present study we can set
$\delta_{12}=0$ and $\delta_{13}=0$ without loss of generality. With
this set-up for the CPV phases, the elements of the first row of the
PMNS matrix, which are relevant for our further discussion, are
given by
\be
\begin{split}
U_{e1} & = c_{12} c_{13} c_{14}, \\
U_{e2} & = e^{i \alpha/2} c_{13} c_{14} s_{12}, \\
U_{e3} & = e^{i\beta/2} c_{14} s_{13}, \\
U_{e4} & = e^{i\gamma/2} s_{14}\,,\\
\label{Uek31}
\end{split}
\ee
%
where we have used the standard notation
$c_{ij} \equiv \cos\theta_{ij}$ and
$s_{ij} \equiv \sin\theta_{ij}$.
The element $U_{e4}$, and thus the angle
$\theta_{14}$, describes the coupling of 4th
neutrino $\nu_4$ to the electron in the
weak charged lepton current.

  The masses of all neutrinos of interest
for the present study
satisfy $m_j \ll 1$ MeV, $j=1,2,3,4$.
Therefore, the expression for the
effective Majorana mass
in the $3 + 1$ model has the form
(see, e.g., \cite{BiPet87,WRodej11}):
\be
\meff =\left|  m_1 |U_{e1}|^2 +  m_2 |U_{e2}|^2 e^{i \alpha} +
m_3 |U_{e3}|^2e^{i \beta} + m_4 |U_{e4}|^2 e^{i \gamma}
\right|.
\ee
%

 In this study we will use two reference sets of  values of
the two sterile neutrino oscillation parameters $\sin^2\theta_{14}$
and $\Dmq_{41}$, which enter into the expression for $\meff$ and
which are obtained in the analyses performed in
\cite{GiuntiTalk:2013vaa,Kopp:2013vaa}. Some of the results obtained
in \cite{Kopp:2013vaa} using different data sets are given in Table
\ref{tab:1nu}. We will use the best fit values\footnote{We will
use throughout all the text the notation $\Dmq_{41}$ in the case of
NO spectrum and $\Dmq_{43}$ for the IO spectrum.}
\be
\sin^2\theta_{14}= 0.0225\,,~~~\Dmq_{41} = 0.93~{\rm eV^2}\,~~{\rm (A)}\,,
\label{A4}
\ee
%
found in \cite{Kopp:2013vaa} in the global analysis of all
the data (positive evidences and negative results)
relevant for the tests of the sterile neutrino hypothesis, and
\be
\sin^2\theta_{14}= 0.023\,,~~~\Dmq_{41} = 1.78~{\rm eV^2}\,~~{\rm (B)}\,,
\label{B4}
\ee
%
obtained in \cite{Kopp:2013vaa} from the fit of all the $\nu_e$ and
$\bar{\nu}_e$ disappearance data (reactor neutrino and Gallium
anomalies, etc.) and quoted in Table \ref{tab:1nu}. Global analysis
of the sterile neutrino related data was performed recently, as we
have already noticed, also in \cite{GiuntiTalk:2013vaa} (for earlier
analyses see, e.g., \cite{Archi12}). The authors of
\cite{GiuntiTalk:2013vaa} did not include in the data set used the
MiniBooNE results at $E_{\nu} \leq 0.475$ GeV, which show an excess
of events over the estimated background \cite{MBooNEexcess}. The
nature of this excess is not well understood at present. For the
best values of
 $\sin^2\theta_{14}$ and $\Dmq_{41}$ the authors
\cite{GiuntiTalk:2013vaa} find:
 $\sin^2\theta_{14}= 0.028$ and $\Dmq_{41} = 1.60~{\rm eV^2}$,
which are close to the best fit values found in \cite{Kopp:2013vaa}
in the analysis of the $\nu_e$ and $\bar{\nu}_e$ disappearance data
(see Table \ref{tab:1nu}). Actually, in what concerns the problem we
are going to investigate, these two sets of values of
$\sin^2\theta_{14}$ and $\Dmq_{41}$ lead practically to the same
results.
\begin{table}[h]
  \centering
    \begin{tabular}{lcc}
      \toprule
      & $\sin^22\theta_{14}$ & $\Dmq_{41} \, {\rm eV}$ \\ 
      \hline
      SBL rates only            & 0.13 & 0.44 \\ 
      SBL incl.\ Bugey3 spectr. & 0.10 & 1.75\\ 
      SBL + Gallium             & 0.11 & 1.80 \\ 
      SBL + LBL                 & 0.09 & 1.78 \\ 
      global $\nu_e$ disapp.\   & 0.09 & 1.78 \\ 
      \midrule
       & $\sin \theta_{14}$ & $\Dmq_{41} \, {\rm eV}$ \\ 
     \midrule
       global data & 0.15 & 0.93 \\ 
      \bottomrule
  \end{tabular}
  \caption{\label{tab:1nu} The best fit values of the
oscillation parameters  $\sin^22\theta_{14}$ and $\Delta m^2_{41}$
obtained in the $3+1$ scheme in \cite{Kopp:2013vaa}
using different data sets.
The values in the last row are obtained from the
global fit of all available data
and are reported in Table 8 in \cite{Kopp:2013vaa}.
}
\end{table}
%

The authors of ref. \cite{GiuntiTalk:2013vaa} give also 
the allowed intervals of values of $\Delta m^2_{41}$ and 
$\sin^2\theta_{14}$ at 95\% C.L., which are correlated.
The two values of $\Delta m^2_{41}$ correspond approximately 
to the two extreme points of the $\Delta m^2_{41}$ interval.
For $\Delta m^2_{41} = 0.9~{\rm eV^2}$, the $2\sigma$ interval of 
allowed values of  $\sin^2\theta_{14}$ reads:
$0.022 \leq \sin^2\theta_{14} \leq 0.028$. This interval 
is very narrow. Varying  $\sin^2\theta_{14}$ in it in our 
analysis leads practically to the same results as those obtained for 
$\sin^2\theta_{14} = 0.0225$ and we will present results 
only for $\sin^2\theta_{14} = 0.0225$. In the case of 
$\Delta m^2_{41} = 1.78~{\rm eV^2}$, the corresponding 
$2\sigma$ interval of  $\sin^2\theta_{14}$ is:
\be
\Delta m^2_{41} = 1.78~{\rm eV^2}:~~~
0.017 \leq \sin^2\theta_{14} \leq 0.047\,,~~95\%~{\rm C.L.}\,.
\label{th142sigma} 
\ee
%
In this case the value we are using 
$\sin^2\theta_{14} = 0.023$ is approximately 
by a factor 1.35 bigger (a factor of 2.04 smaller) then 
the $2\sigma$ minimal (maximal) allowed value of 
$\sin^2\theta_{14}$. In what follows we will present results for 
the best fit values 
$\Delta m^2_{41} = 1.78~{\rm eV^2}$ and $\sin^2\theta_{14} = 0.023$ 
and will comment how the results change if one varies 
 $\sin^2\theta_{14}$ in the $2\sigma$ interval (\ref{th142sigma}).

\subsection{The case of 3+1 Scheme with NO Neutrino Mass Spectrum}

\begin{figure}
\unitlength=1mm
\begin{center}
\begin{picture}(100,40)
\put(20,5){\line(1,0){20}}
\put(20,8){\line(1,0){20}}
\put(20,13){\line(1,0){20}}
\put(20,35){\line(1,0){20}}
\put(30,5){\line(0,1){30}}
\put(45,4.5){\vector(0,1){3.5}}
\put(45,8){\vector(0,-1){3.5}}
\put(45,8){\vector(0,1){5}}
\put(72,5){\vector(0,1){30}}
\put(72,35){\vector(0,-1){30}}
\put(48,4){$\Delta m^2_{21}$}
\put(48,9.5){$\Delta m^2_{31}$}
\put(75,18){$\Delta m^2_{41}$}
\put(17,4){1}
\put(17,8){2}
\put(17,12.3){3}
\put(17,34.2){4}
\end{picture}
\end{center}
\caption{\label{specNH3plus1} The Mass spectrum in the $3 + 1$ NO (NH)
model.}
\end{figure}
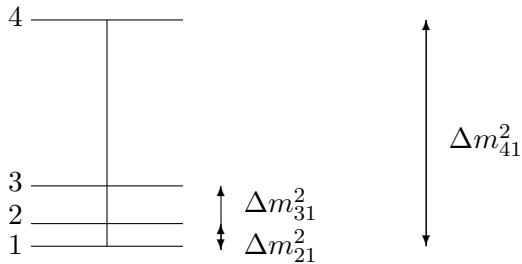

In the case of the $3+1$ scheme with NO neutrino mass spectrum,
$m_1 < m_2 < m_3 < m_4$, we have:
\be
\meff = | m_1 c_{12}^2 c_{13}^2 c_{14}^2 + m_2 e^{i\alpha}
c_{13}^2 c_{14}^2 s_{12}^2 +   m_3 e^{i\beta} c_{14}^2 s_{13}^2
 + m_4 e^{i\gamma}  s_{14}^2 |\,,
\label{meffNO}
\ee
%
where we have used eq. (\ref{Uek31}). The masses $m_{2,3,4}$ can be
expressed in terms of the lightest neutrino mass $m_1$ and the three
neutrino mass squared differences $\Delta m^2_{21} > 0$, $\Delta
m^2_{31} > 0$ and $\Delta m^2_{41} >0$:
\be
\begin{split}
m_1 &\equiv m_{min},
\quad m_2  = \sqrt{m_{min}^2 + \Delta m^2_{21}},
\qquad m_3 = \sqrt{m_{min}^2 + \Delta m^2_{31}}
\quad \text{and} \quad m_4 = \sqrt{m_{min}^2 + \Delta m^2_{41}}\,.
\end{split}
\ee
%
The mass spectrum of the  $3 + 1$ NO (NH) model is shown
schematically in Fig. \ref{specNH3plus1}.
%

\begin{figure}[h!]
  \begin{center}
 \subfigure
 {\includegraphics[width=7cm]{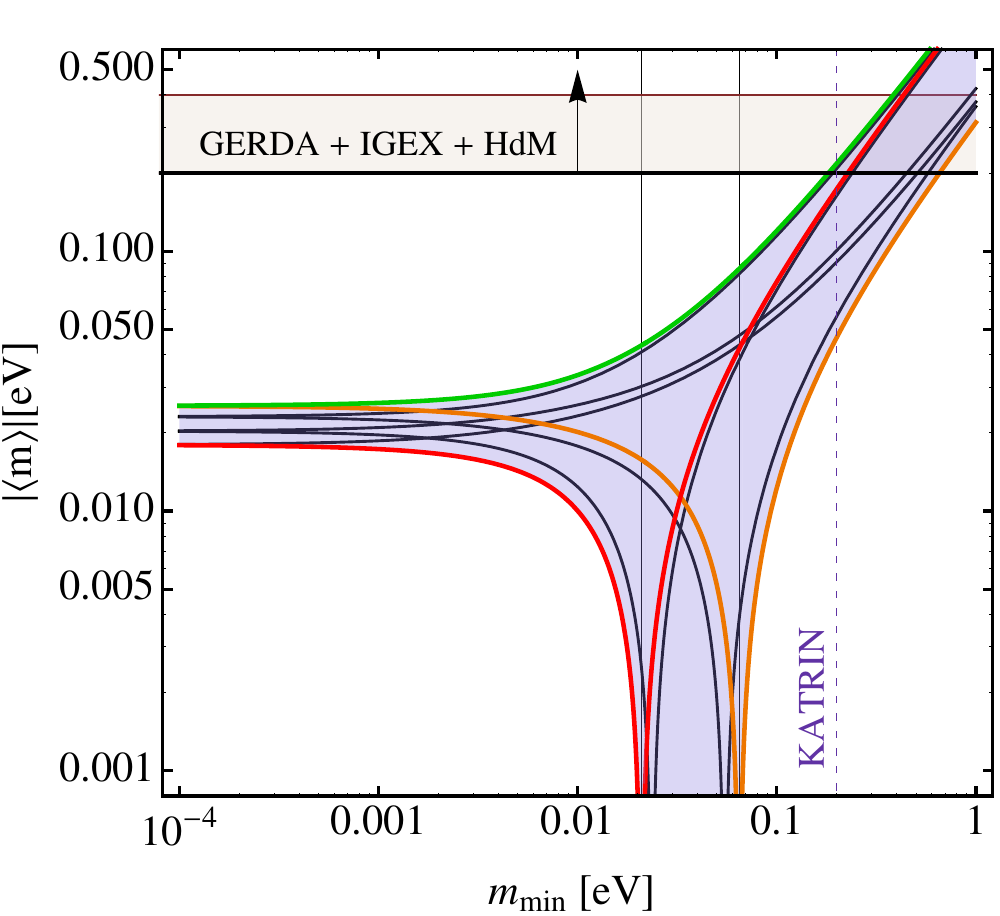}}
 \vspace{5mm}
 \subfigure
   {\includegraphics[width=7cm]{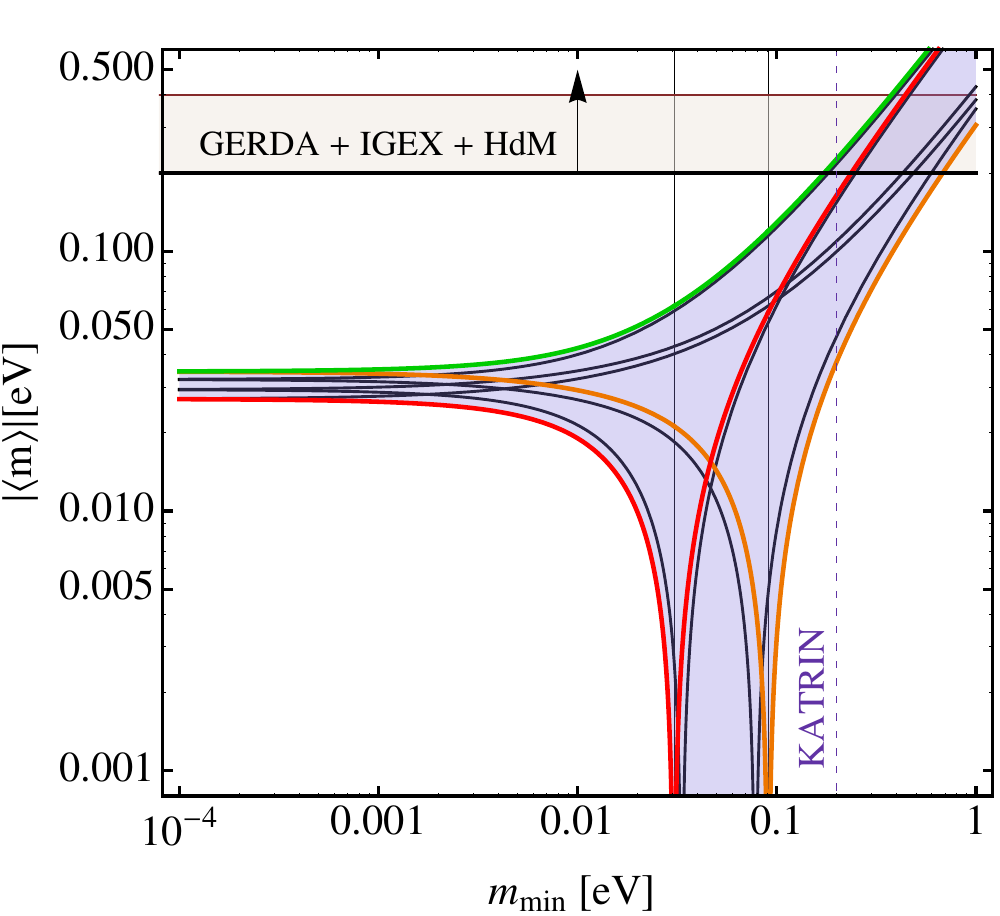}}
     \end{center}
\vspace{-1.0cm} \caption{\label{figMinNH} Left Panel. The value of
\meff as a function of $m_{min}\equiv m_1$ in the NO case for
$\Dmq_{41} = 0.93 \, {\rm eV^2}$, $\sin \theta_{14} = 0.15$ and the
best fit values of the oscillation parameters given in Table
\ref{tabNudata}. The green, red and orange lines correspond
respectively to values of the three CPV Majorana phases
$(\alpha,\beta,\gamma)= (0,0,0),(0,0,\pi),(\pi,\pi,\pi)$.
 The five gray curves show \meff
computed for the other five sets of CP conserving values of the
phases. The interval between the vertical left and right solid
lines,
corresponding to $m_1 = \un{m}_1   \simeq 0.021\, {\rm eV}$ and
$m_1 = \ov m_1\simeq 0.065\, {\rm eV} $, 
indicates the region where $\meff_{min}=0$ for specific choices of
$(\alpha,\beta,\gamma)$.
Right Panel. The same as in the left panel but for
$\Dmq_{41} \equiv 1.78 \, {\rm eV^2} $. The vertical solid lines
correspond to $m_1 = \un{ m}_1 \simeq 0.030\, {\rm eV}$ and $m_1 =
\ov m_1\simeq 0.091 \, {\rm eV}$.
The horizontal band indicates the upper bound $\meff  \sim0.2-0.4$
eV obtained using the 90 \% C.L. limit on the half-life of $^{76}$Ge
reported in \cite{Agostini:2013mba}. See text for further details.}
\end{figure}

 In Fig. \ref{figMinNH}  we show $\meff$ as a function of the lightest
neutrino mass $m_{min} = m_1$. As we have already indicated
and was noticed in \cite{Barry:2011}
(see also \cite{BPP2,SGWRode05}), for the two sets of
values of $\nu_4$ oscillation parameters (\ref{A4}) and (\ref{B4})
and NH 3-neutrino spectrum (i.e., $m_1 \ll m_{2,3,4}$) we have,
depending on the values of the Majorana phases,
$\meff \cong (0.018-0.025)$ eV and  $\meff \cong (0.027-0.034)$ eV,
 respectively. This is in contrast with the prediction for
$|\mefff^{(3\nu)}| \ltap 0.005$ eV. Another important feature of the
dependence of $\meff$ on  $m_{min}$, which is prominent in Fig.
\ref{figMinNH}, is the possibility of a strong suppression of
$\meff$ \cite{Barry:2011,Li:2012,Goswami:2013}. Such a suppression
can take place also for  $|\mefff^{(3\nu)}|$ and the conditions
under which it occurs have been studied in detail in
\cite{Pascoli:2007qh}. In what follows we perform a similar study
for $\meff$. The aim is to determine the range of values of
$m_{min}$ and the Majorana phases $\alpha$, $\beta$ and $\gamma$ for
which $\meff \geq 0.01$ eV.

%

 It proves convenient for the purposes of our analysis to work
with the quantity  $\meff^2$ rather than with  $\meff$, and
to write $\meff^2$ as
\be
\meff^2 = |a + e^{i \alpha }b + e^{i \beta }c + e^{i \gamma }d|^2\,.
\label{meff2abcd}
\ee
%
In the NO case under study the parameters
$a$, $b$, $c$, and $d$ read:
\be
\begin{split}
a &= m_{min} c_{12}^2 c_{13}^2 c_{14}^2\\
b & =  \sqrt{m_{min}^2+\Delta m^2_{21}} c_{13}^2 c_{14}^2 s_{12}^2 \\
c & =  \sqrt{m_{min}^2+\Delta m^2_{31}} c_{14}^2 s_{13}^2 \\
d & = \sqrt{m_{min}^2+ \Delta m^2_{41}}  s_{14}^2\,.
\label{abcd}
\end{split}
\ee
%
The first derivative of $\meff^2$
with respect of $\alpha$, $\beta$ and $\gamma$ leads to
the following system of three coupled equations:
\be
\begin{split}
-& 2b [a \sin (\alpha )+c \sin (\alpha -\beta )+d \sin (\alpha -\gamma )] = 0,\\
-& 2c [a \sin (\beta )-b \sin (\alpha -\beta )+d \sin (\beta -\gamma )] = 0,\\
& 2d [-a \sin (\gamma )+b \sin (\alpha -\gamma )+c \sin (\beta -\gamma )] = 0.\\
\end{split}
\label{genminmax}
\ee
%
 It is possible to solve this system of equations using
the set of variables $v = \tan(\alpha/2)$, $t =
\tan(\beta/2)$, $u = \tan(\gamma/2)$ with $\alpha$, $\beta$, $\gamma
\neq \pi + 2k\pi$. We will give here only the basic formulas
and will describe the results of such minimization
using the best fit values given in Table \ref{tabNudata} and
eqs. (\ref{A4}) and (\ref{B4}).
In  Appendix \ref{AppendixA} we describe in detail
the minimization procedure of the general expression of
$\meff$ in the 3+1 scheme and the 16 solutions found.
We give explicit expressions for the solutions
and derive the domain of each of the 16 solutions.
Eight of these solutions correspond to all possible
combinations of the phases having values $0$ or $\pi$.
Obviously, the solution $(\alpha,\beta,\gamma)=(0,0,0)$ corresponds to
an absolute maximum of the effective Majorana mass $\meff$.
As we show in Appendix  \ref{AppendixA}, the domains of
the solutions of interest are determined by the
properties of the functions $f_i$, $i=1,...,8$:
\be
\begin{split}
f_1 & = a - b - c - d \,,  \qquad f_2  = a + b - c - d \,, \\
f_3 & = a + b - c + d \,, \qquad f_4  = - a + b + c - d \,, \\
f_5 & = a + b + c + d \,, \qquad f_6  = a - b + c + d \,, \\
f_7 & = a - b + c - d \,, \qquad f_8  = a + b + c - d \,, \\
\end{split}
\label{fi}
\ee
%
where $a$, $b$, $c$  and $d$ for the NO case
are defined in eq. (\ref{abcd}).

\begin{figure}[h!]
  \begin{center}
 {\includegraphics[width=7.55cm]{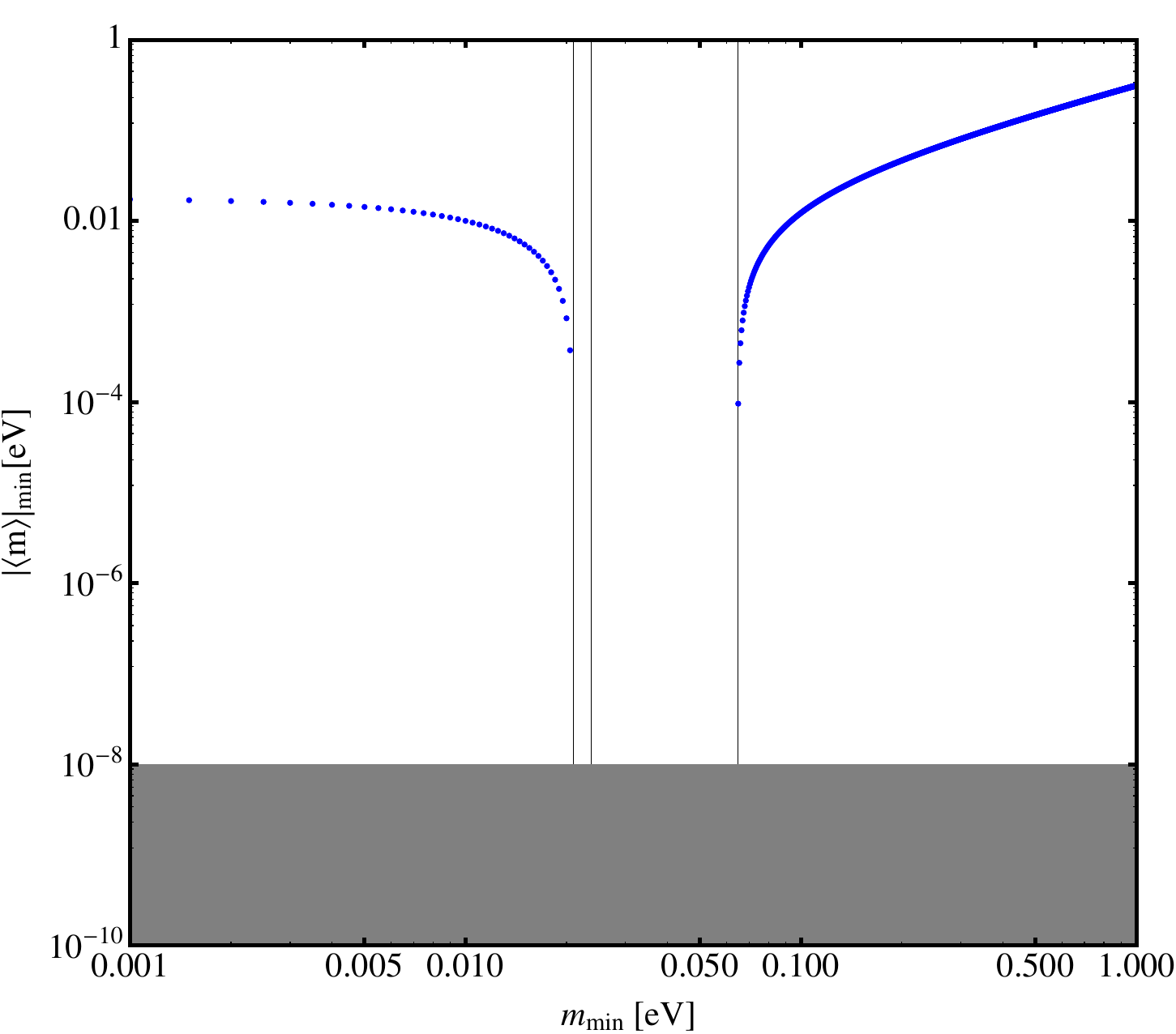}}
     \end{center}
\caption{\label{fig:minnumNH} Minimum
\meff as function of $m_{min} \equiv  m_1$
for $\Dmq_{41} =  0.93 \, {\rm eV^2} $ and $\sin \theta_{14} = 0.15$.
The plot has been obtained numerically
by performing a scan over a sufficiently large sets
of values of $m_{min}$ and of each of
the CPV phases  $(\alpha,\beta,\gamma)$
in the interval $[0,\pi]$.
The black vertical lines correspond respectively to
$m_{min} = \un m_1 \simeq 0.021$ eV,
$m_{min} = \hat m_1\simeq 0.024$ eV and
$m_{min} = \ov m_1  \simeq 0.065$ eV.
}
\end{figure}

 We will focus first on the solutions which
minimize $\meff$ such that the minimum value is exactly zero.
As is shown in Appendix \ref{AppendixA1},
there are six
physical solutions for which $\meff_{min}=0$:
$(u_{\pm},v_{\pm},t_{\pm})$, $(u^{\pm}_3,v^{\pm}_3,t^{\pm}_3)$,
$(v^{\pm}_{4}(u),t^{\pm}_{4}(u))$. In order to study the domain
of existence of these solutions
we define the following points
$\un m_1<\hat m_1<\tilde m_1<\ov m_1$ as the zeros
of the functions $f_8$, $f_2$, $f_7$, $f_1$, respectively:
\be
f_8(\un m_1) = 0 , \quad f_2(\hat m_1) = 0,
\quad f_7(\tilde m_1) = 0, \quad f_1(\ov m_1) = 0.
\label{f82710}
\ee
%
We find from the numerical analysis performed in
Appendix \ref{AppendixA1} that
i) the solution
$(u_{\pm},v_{\pm},t_{\pm})$ is valid
between the zeros of the function $f_2$ and $f_1$
(the region in which $f_1 f_2 f_3 f_4 > 0$),
i.e. in the interval $\hat m_1 < m_1 < \ov m_1$;
ii) the solution $(u^{\pm}_3,v^{\pm}_3,t^{\pm}_3)$ is valid
between the zeros of the function $f_7$ and $f_1$, i.e.
in the interval $\tilde m_1 < m_1 < \ov m_1$;
and finally
iii) the solution $(v^{\pm}_{4}(u),t^{\pm}_{4}(u))$ is valid
in the interval between the zero of the function
$f_8$ and $f_1$, i.e. for $\un m_1 < m_1 < \ov m_1$.
In  Appendix \ref{AppendixA1} we give the numerical ranges
that define the domains of the solutions discussed above.
Using the best fit values of the neutrino oscillation parameters
given in Table  \ref{tabNudata} and eqs. (\ref{A4}) and (\ref{B4}),
we get the following numerical values of
\footnote{Although it will not be specified further,
whenever we present numerical results in the text or in
graphical form of figures in what follows,
we will use the best fit values of
the neutrino oscillation parameters reported
in Table \ref{tabNudata} to obtain them.}
$\un m_1$, $\hat m_1$, $\tilde m_1$, $\ov m_1$:
\begin{itemize}
\item for $\Dmq_{41} = 0.93~{\rm eV^2}$ we have
$( \un{ m}_1, \hat m_1,\tilde m_1, \ov m_1)
\simeq (0.021,0.024,0.055,0.065)$ eV;
\item if $\Dmq_{41} = 1.78~{\rm  eV^2}$ we get
$(\un{ m}_1,\hat m_1,\tilde m_1, \ov m_1)
\simeq (0.030,0.033,0.078,0.091)$ eV.
\end{itemize}
%
For $m_1= \un m_1$ and $m_1 = \ov m_1$ the value of $\meff$
is exactly zero for the CP conserving values of the phases
$(\alpha,\beta,\gamma)= (0,0,\pi)$ and $(\pi,\pi,\pi)$,
respectively. In the intervals described above
(excluding the extrema), it is possible to have $\meff_{min}=0$
for specific values of $(\alpha,\beta,\gamma)$,
which are not necessarily CP conserving.
This can be seen
in Fig. \ref{fig:minnumNH}.
The numerical minima depicted in Fig. \ref{fig:minnumNH}
are obtained minimizing $\meff$ by performing a scan for
different values of $m_1$ and the CPV Majorana phases.

The grey horizontal band in  Fig. \ref{fig:minnumNH}
corresponds to $\meff_{min}\leq 10^{-8}$ eV  and reflects the
precision of the numerical calculation of
 $\meff_{min} = 0$.
%
The minima are reached at specific values of the
phases $(\alpha,\beta,\gamma)$ that can have either
CP conserving or CP nonconserving values.
For $\Delta m^2_{41} = 0.93 \, {\rm eV^2} $ and
$m_1 = 0.03 \, {\rm eV}$, for instance, we have $\meff = 0$
if the three CPV phases have the following CP
nonconserving values:
$(\alpha,\beta,\gamma) = (1.731,0.023,-2.711)$.
Similarly, if
$\Delta m^2_{41} = 1.78~{\rm eV^2}$
and, e.g., $m_1 = 0.04\, {\rm eV}$, we find that $\meff = 0$ for
 $(\alpha,\beta,\gamma) = (1.511, -0.365, -2.761)$.
%

 Combining the results described above we
can conclude that the effective Majorana
mass $\meff$ can be zero only for values of
$m_{min}$ from the following interval:
\be
\un{ m}_1  \leq m_{min} \leq
\ov m_1 \,\,.
\label{condmin31}
\ee
%

 We will derive next simple approximate analytical expressions for
$\ov m_1$ and $\un m_1$. We note first that
for values of $m_{min}$ in the range $ 0.02 - 0.10$ eV,
the term proportional to $ \sqrt{m_{min}^2 + \Delta m^2_{31}}c^2_{14}s^2_{13}$
is approximately by an order of magnitude smaller than the
other three terms in $\meff$ (the terms with the factors
$a,b,d$ in eq. (\ref{meff2abcd})). Neglecting it
as well as  $\Delta m^2_{21}\ll \Delta m^2_{31},\Delta m^2_{41}$,
we find the following rather simple
analytic expressions for $\ov m_1$ and $\un m_1$,
which are valid up to an error of about $10\%$:
\be
\begin{split}
\ov m_1 & \approx \sqrt{\frac{ \Dmq_{41} \sin ^4\theta_{14} }{ \cos ^4\theta_{13} \left( - \cos ^2\theta_{12} \cos^2\theta_{14}+\sin ^2\theta_{12} \right )^2 - \sin ^4\theta_{14} } } \,\, ,\\
\un m_1 & \approx \sqrt{\frac{ \Dmq_{41} \sin ^4\theta_{14} }{ \cos ^4\theta_{13} \left(  \cos ^2\theta_{12} \cos^2\theta_{14}+\sin ^2\theta_{12} \right )^2 - \sin ^4\theta_{14} } } \,\, . \\
\end{split}
\label{barm1unm1approx}
\ee
%
Using these expressions we get
$(\un m_1, \ov m_1) \simeq (0.023,0.060)$ eV
for $\Delta m^2_{41} = 0.93~{\rm  eV^2}$ instead of
$(0.021,0.065)$ eV,
and $(\un m_1, \ov m_1) \simeq (0.032,0.085)$ eV for
 $\Delta m^2_{41} = 1.78~{\rm  eV^2}$ instead of
 $(0.030,0.091)$ eV.

In Fig. \ref{figMinNH} we show two plots of $\meff$
as function of the lightest neutrino mass $m_{min}\equiv m_1$
using  $\Delta m^2_{41} = 0.93~{\rm  eV^2}$ and
$\Delta m^2_{41} = 1.78~{\rm  eV^2}$.
The shaded area is the region of allowed values of
$\meff$. One can see that the green curve represents
the possible maximum value for \meff
corresponding to $(\alpha,\beta,\gamma)= (0,0,0)$.
We notice also that in the limit of $m_{min}\rightarrow 0$
the minimum value of $\meff > 0.01$ eV.
This limit  will be analyzed in detail below.
We also show in Fig. \ref{figMinNH} the prospective
sensitivity to $m_{min}$ of the $\beta$-decay experiment
KATRIN \cite{MainzKATRIN}, which is under preparation.

 As is well known, in the case of 3-neutrino mixing and IH (IO)
neutrino mass spectrum we have
$|\mefff^{(3\nu)}| > 0.01$ eV (see, e.g., \cite{PDG2012}).
We find that in the $3+1$ scheme under discussion
and NO neutrino mass spectrum we have always
$\meff > 0.01$ eV for the following ranges of $m_{min}$:
\begin{itemize}
\item  $m_{min} < 0.010$ eV and  $m_{min} > 0.093$ eV, if
 $\Delta m^2_{41} = 0.93~{\rm  eV^2}$;
\item  $m_{min} < 0.020$ eV and  $m_{min} > 0.119$ eV, for
$\Dmq_{41} = 1.78~{\rm  eV^2}$.
\end{itemize}

What would be the changes of our results presented for 
$\Delta m^2_{41} = 1.78~{\rm eV^2}$ presented so far
if instead of $\sin^2\theta_{14} = 0.023$ we used 
the minimal (maximal) value of the 2$\sigma$ 
interval (\ref{th142sigma}) of allowed values
$\sin^2\theta_{14} = 0.017$ ($\sin^2\theta_{14} = 0.047$)
in the analysis? Qualitatively no new features appear and 
the results remain the same. Quantitatively some of the numarical 
values of $|<m>|$, $\overline{m}_1$ and $\underline{m}_1$, 
quoted in the text and obtained for $\sin^2\theta_{14} = 0.023$, 
are just shifted. More specifically,
this will lead to the decreasing (increasing) 
of the values of $|<m>|$ at $m_{min} \lesssim 10^{-3}$ eV 
and of $\overline{m}_1$ and $\underline{m}_1$ 
approximately by the same factor 1.35 (2.04).

 For $0 \leq m_1 < \un{ m}_1 $ and $m_1 >  \ov m_1$, there are no
physical solutions for the phases for which $\meff = 0$.
Moreover, $(u_{\pm},v_{\pm},t_{\pm})$, $(u^{\pm}_3,v^{\pm}_3,t^{\pm}_3)$
%
and $(v^{\pm}_4(u),t^{\pm}_4(u))$ are not well
defined in the indicated intervals.
However, by studying the Hessian of
$\meff^2$, we find that there are physical solutions
(among those listed in eq. (\ref{condmin2})
of  Appendix \ref{AppendixA}) for
which $\meff_{min}\neq 0$. These solutions are
realised for specific values of the phases, i.e., for
$(\alpha, \beta, \gamma)=(0,0,\pi)$ or $(\pi,\pi,\pi)$.
The analysis performed in Appendix \ref{AppendixA}
allowed to find the minima of \meff at  $m_{min} <  \un{m}_1$ for
$(\alpha,\beta,\gamma) = (0,0,\pi)$, and at $m_{min} > \ov m_1$ for
$(\alpha,\beta,\gamma) = (\pi,\pi,\pi)$. The domain of the solution
at  $m_{min} <  \un{m}_1$, corresponding to $(\alpha,\beta,\gamma) =
(0,0,\pi)$, is the common interval of values of
 $m_{min}$ in which the three inequalities
$c < d$, $b < d - c$ and $f_8 = a+b+c -d <0$
hold. The domain of the solution at
$m_{min} > \ov m_1$ with
$(\alpha,\beta,\gamma) = (\pi,\pi,\pi)$
is determined by the inequality
$f_1 = a-b-c-d >0$.
Actually, as it is possible to show,
we have, in particular,  $f_8 <0$ at $m_{min} <  \un{m}_1$,
and $f_1>0$ for $m_{min} > \ov m_1$.
\\
  The analysis of the Hessian of
$\meff^2$ performed in  Appendix \ref{AppendixA}
shows that there can be two more solutions for
which $\meff_{min}\neq 0$.
They correspond two i) $(\alpha,\beta,\gamma) = (0,\pi,0)$
and ii) $(\pi,0,0)$. The domains of these
solutions (following from the
Sylvester's criterion) are determined by
i) $c > d$, $b < c-d$, $-f_3 = -a-b+c-d> 0$, and
ii) $b>c+d$, $-f_6 = -a+b-c-d >0$.
However, it is not difficult to convince oneself
that for the 
values of the neutrino oscillation
parameters, including those of $\Dmq_{41}$
and $\sin^2\theta_{14}$ used by us
in the present analysis,
there are no physical values of
$m_{min}\geq 0$ for which the inequalities
in i) or in ii) are satisfied.
%

\begin{figure}[h!]
  \begin{center}
 \subfigure
 {\includegraphics[width=7cm]{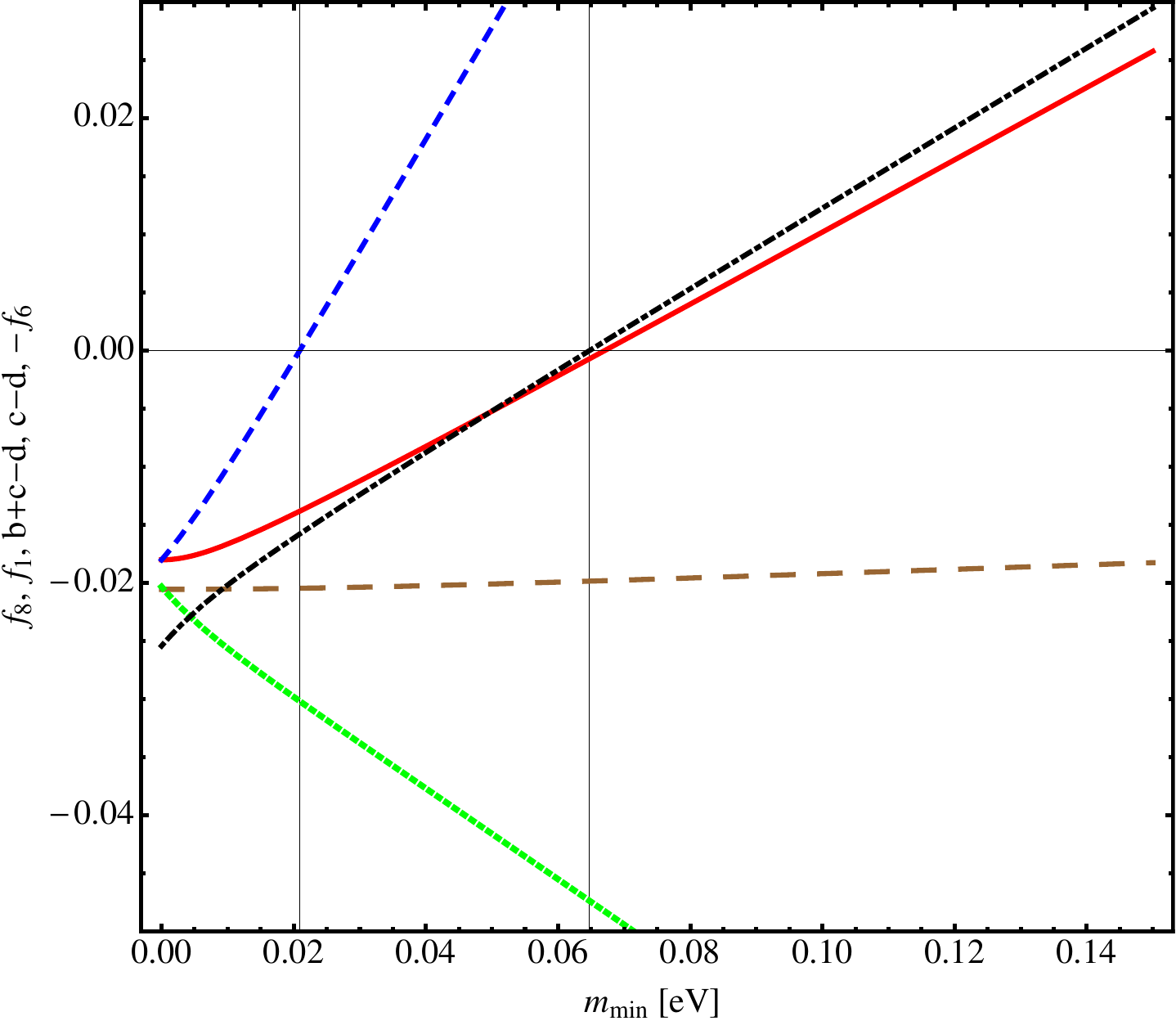}}
 \vspace{5mm}
 \subfigure
   {\includegraphics[width=7cm]{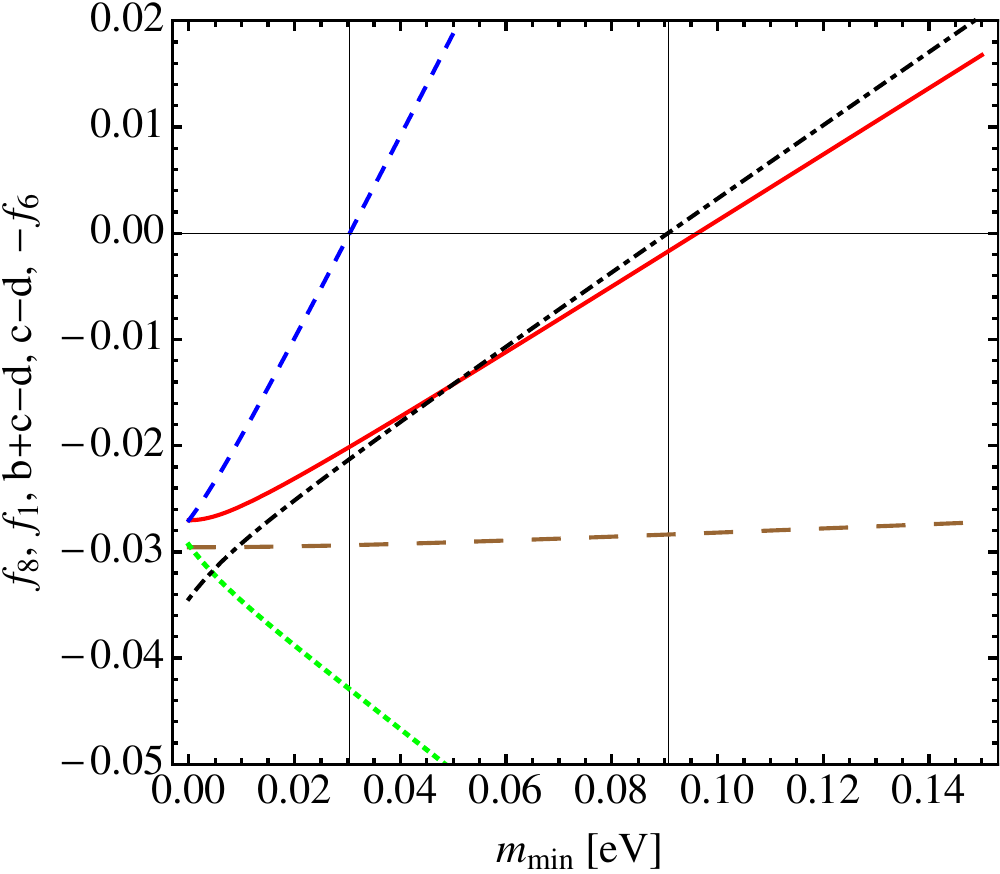}}
     \end{center}
\vspace{-1.0cm} \caption{\label{fig:SylvesterNH}
Left Panel. The functions $f_8$ (short-dashed blue),
$f_1$ (dot-dashed black), $b + c - d$ (solid red),
$c - d$ (large dashed brown), $-f_6$ (dotted green)
versus $m_{min}\equiv  m_1$
for $\Dmq_{41} = 0.93 \, {\rm eV^2}$ and $\sin\theta_{14}=0.15$.
The vertical lines
correspond to $m_{min} = \un{m}_1\simeq 0.021 \,{\rm eV}$
and $m_{min} = \ov m_1 \simeq 0.065 \, {\rm eV}$.
 Right Panel. The same as in the left panel, but for
$\Dmq_{41} = 1.78~{\rm eV^2}$.
The vertical lines
now are at $m_{min} = \un{ m}_1 \simeq 0.030\, {\rm eV}$
and $m_{min} = \ov m_1\simeq 0.091 \,{\rm eV}$.
}
\end{figure}

In Fig. \ref{fig:SylvesterNH} we show
all the relevant functions entering in the
four sets of inequalities listed
above (and in eq. (\ref{condmin2}) of Appendix \ref{AppendixA})
which ensure that the minima  $\meff_{min}\neq 0$.
The figure is obtained for
the best fit values of the oscillation parameters
given in Table \ref{tabNudata} and
for $\Dmq_{41} = 0.93 \, {\rm eV^2} $ (left panel) and
$\Dmq_{41} = 1.78 \, {\rm eV^2} $ (right panel).
One can easily check that only the
two sets of conditions, corresponding
to $(\alpha, \beta, \gamma)=(0,0,\pi)$ or $(\pi,\pi,\pi)$ 
and given above 
\footnote{These are  
 the first two conditions in
eq. (\ref{condmin2}) of Appendix \ref{AppendixA}
following from the Sylvester's criterion for
a minimum.}, are satisfied.

\begin{figure}[h!]
  \begin{center}
 \subfigure
 {\includegraphics[width=7cm]{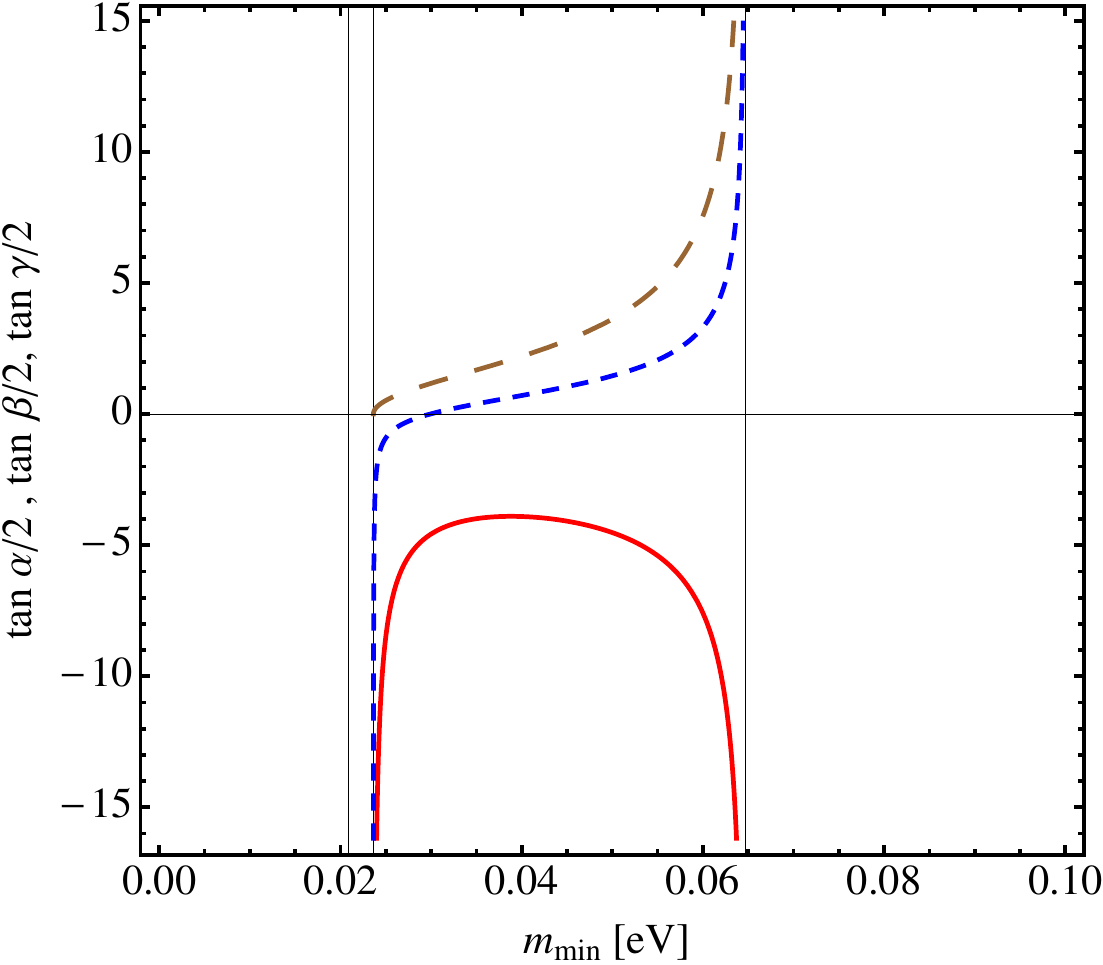}}
 \vspace{5mm}
 \subfigure
   {\includegraphics[width=7cm]{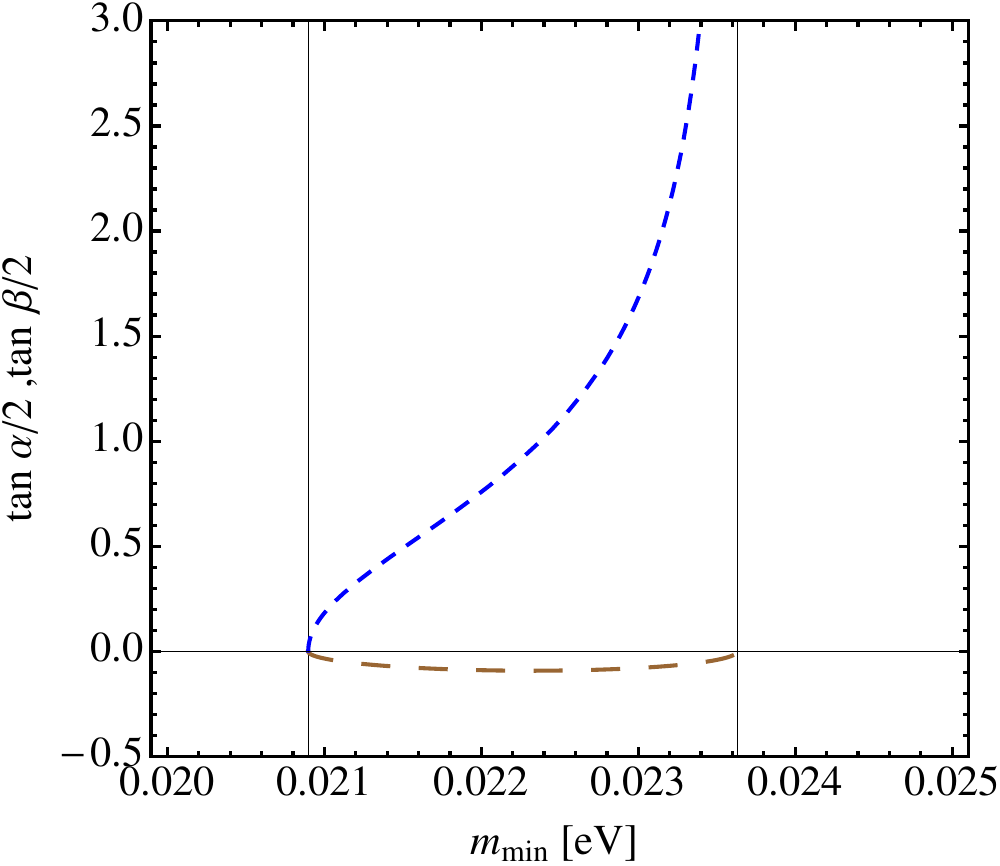}}
     \end{center}
\vspace{-1.0cm} \caption{\label{figMin1}
Left Panel. The values of $(\tan \alpha/2,\tan \beta/2,\tan \gamma/2)$ --
the large dashed (brown), short dashed (blue), solid  (red) lines --
corresponding to the solution $(v_-,t_-,u_-)$
as functions of $m_1$
for $\Dmq_{41} = 0.93~{\rm eV^2}$.
The 2nd and the 3rd vertical
lines from the left are
at $\hat m_1 = 0.024 \, {\rm eV}$ and $\ov m_1 = 0.065 \, {\rm eV}$, and
indicate the domain of existence of this solution.
The 1st vertical line from the left
corresponds to $\un m_1 = 0.021 \, {\rm eV}$.
Right Panel.
The values of $(\tan \alpha/2,\tan \beta/2)$
-- large dashed (brown), solid  (red) lines --
corresponding to the solution $(t^+_4,v^+_4)$
as functions of $m_1$ for $\gamma \simeq \pi$ ($u >> 1$).
The other parameters are the same as in the left panel.
The vertical lines from the left are respectively
at $\un m_1$ and $\hat m_1$.
This solution is well defined only
for $\un m_1 \leq m_1   \leq \hat m_1$.
}
\end{figure}

\begin{figure}[h!]
  \begin{center}
 \subfigure
 {\includegraphics[width=7cm]{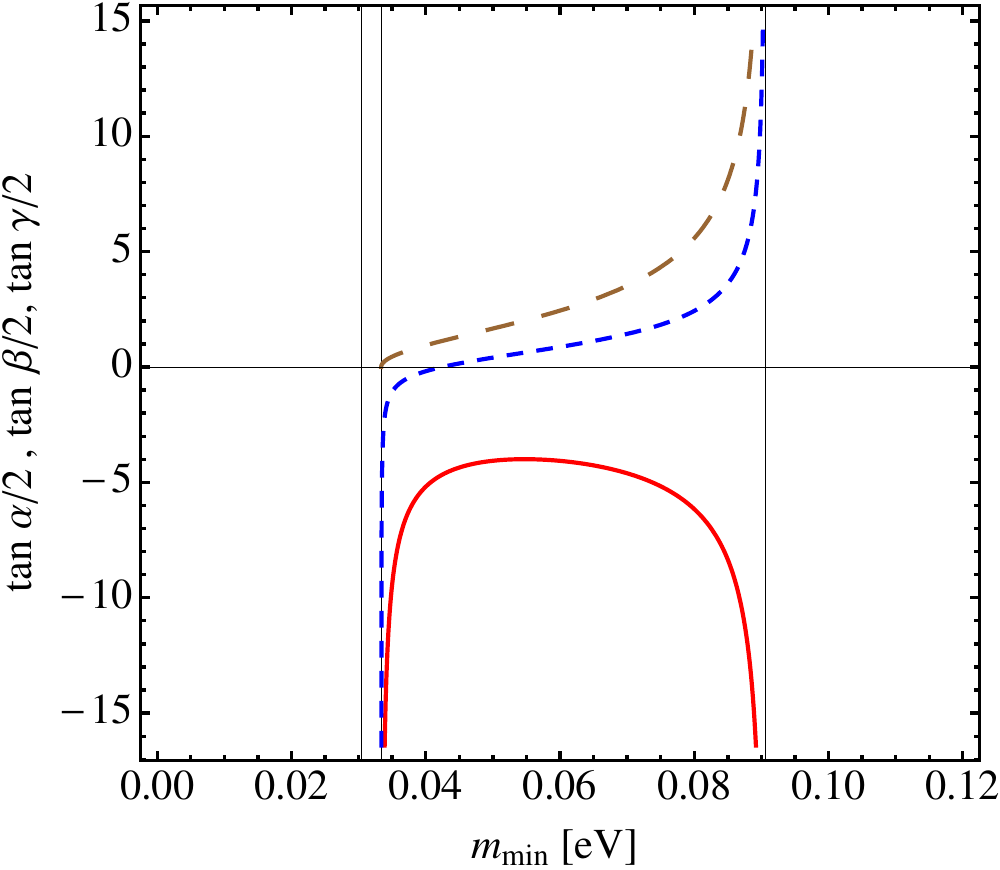}}
 \vspace{5mm}
 \subfigure
   {\includegraphics[width=7cm]{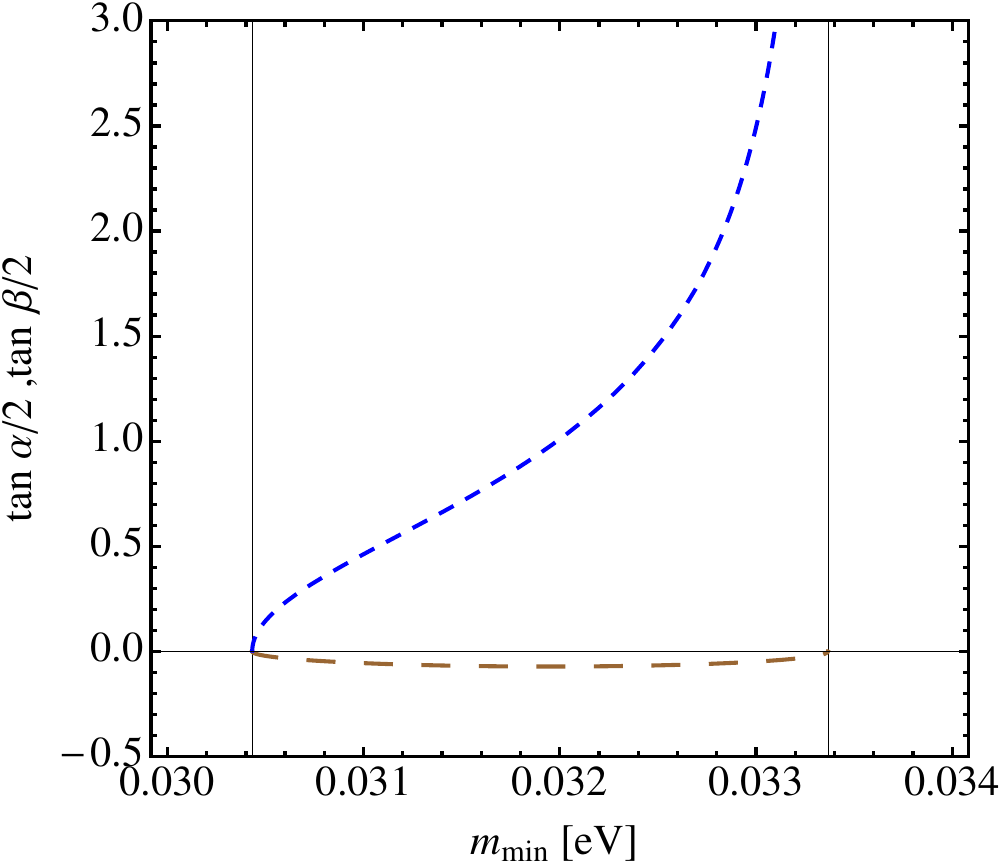}}
     \end{center}
\vspace{-1.0cm} \caption{\label{figMin2}
Left Panel. The same as in Fig. \ref{figMin1}, left panel,
%
for $\Dmq_{41} = 1.78~{\rm eV^2}$.
The 2nd and the 3rd vertical lines from the left
are at $\hat m_1 = 0.033$ eV and $\ov m_1 = 0.091$ eV
and indicate the domain of existence of the solution $(u_-,t_-,v_-)$.
The 1st vertical line from the left corresponds to
$\un m_1 = 0.030 \, {\rm eV}$.
Right Panel.  The same as in Fig. \ref{figMin1}, right panel, for
$\Dmq_{41} = 1.78~{\rm eV^2}$.
The vertical lines from the left are at  $\un m_1$ and $\hat m_1$.
The solution considered $(t^+_4,v^+_4)$ is well defined only
in the interval $\un m_1\leq m_1 \leq \hat m_1$.
}
\end{figure}

In Figs. \ref{figMin1} and \ref{figMin2} we show as
an illustrative examples
$(\tan \alpha/2,\tan \beta/2,\tan \gamma/2)$
as function of $m_{min}$ for two of the
physical solutions, namely, $(u_-,t_-,v_-)$
and  $(v^+_4,t^+_4)$, found in
Appendix \ref{AppendixA}:
\be
\begin{split}
 & \begin{cases}
u_{\pm} & = \pm\, \sqrt{\dfrac{(-a+b+c-d) (a+b-c+d)}{(a-b-c-d) (a+b-c-d)}}, \\
v_{\pm} & = \pm\, \dfrac{(b+c)}{(b-c)}\dfrac{[ (a+b-c)^2 - d^2 ]}{ \sqrt{(-a+b+c-d) (a+b-c+d) (a - d - c -b )(a-d-c+b)}} ,\\
t_{\pm} & = \pm\, \dfrac{a^2+b^2-c^2-d^2}{\sqrt{(a-b-c-d) (a+b-c-d) (-a+b+c-d) (a+b-c+d)}} ,\\
\end{cases}\\
\end{split}
\label{uvt}
\ee
\be
\begin{split}
& \begin{cases}
v^{\pm}_4(u) & =
\dfrac{4 b d u \pm F(a,b,c,d,u) }{-u^2 (a-b-c-d) (a-b+c-d)-(a-b+d)^2+c^2}, \\
t^{\pm}_4(u) & =
\dfrac{-\, 4 c d u \pm F(a,b,c,d,u)}{u^2 (a-b-c-d) (a+b-c-d)+(a-c+d)^2-b^2} ,\\
\end{cases}\\
\end{split}
\label{v4t4}
\ee
%
where $a,b,c$ and $d$ are given in eq. (\ref{abcd}) and
\beq
\begin{split}
F(a,b,c,d,u) & =  \bigg\{  \Big[-u^2 (a+b-c-d) (a-b+c-d)-(a+d)^2+(b-c)^2 \Big] \times\\
& \times \Big[a^2 \left(u^2+1\right)-2 a d \left(u^2-1\right)-\left(u^2+1\right) (b+c-d) (b+c+d) \Big]   \bigg\}^{1/2}.\\
\end{split}
\eeq
%
The corresponding figures for, e.g.,
the solutions $(v_+,t_+,u_+)$ and
$(v^-_4,t^-_4)$ are obtained from
Figs. \ref{figMin1} and \ref{figMin2} by reversing
the $y-$axis.

%
\subsection{The Case of $m_1 =0$}
%

\begin{figure}[h!]
  \begin{center}
 \subfigure
 {\includegraphics[width=8cm]{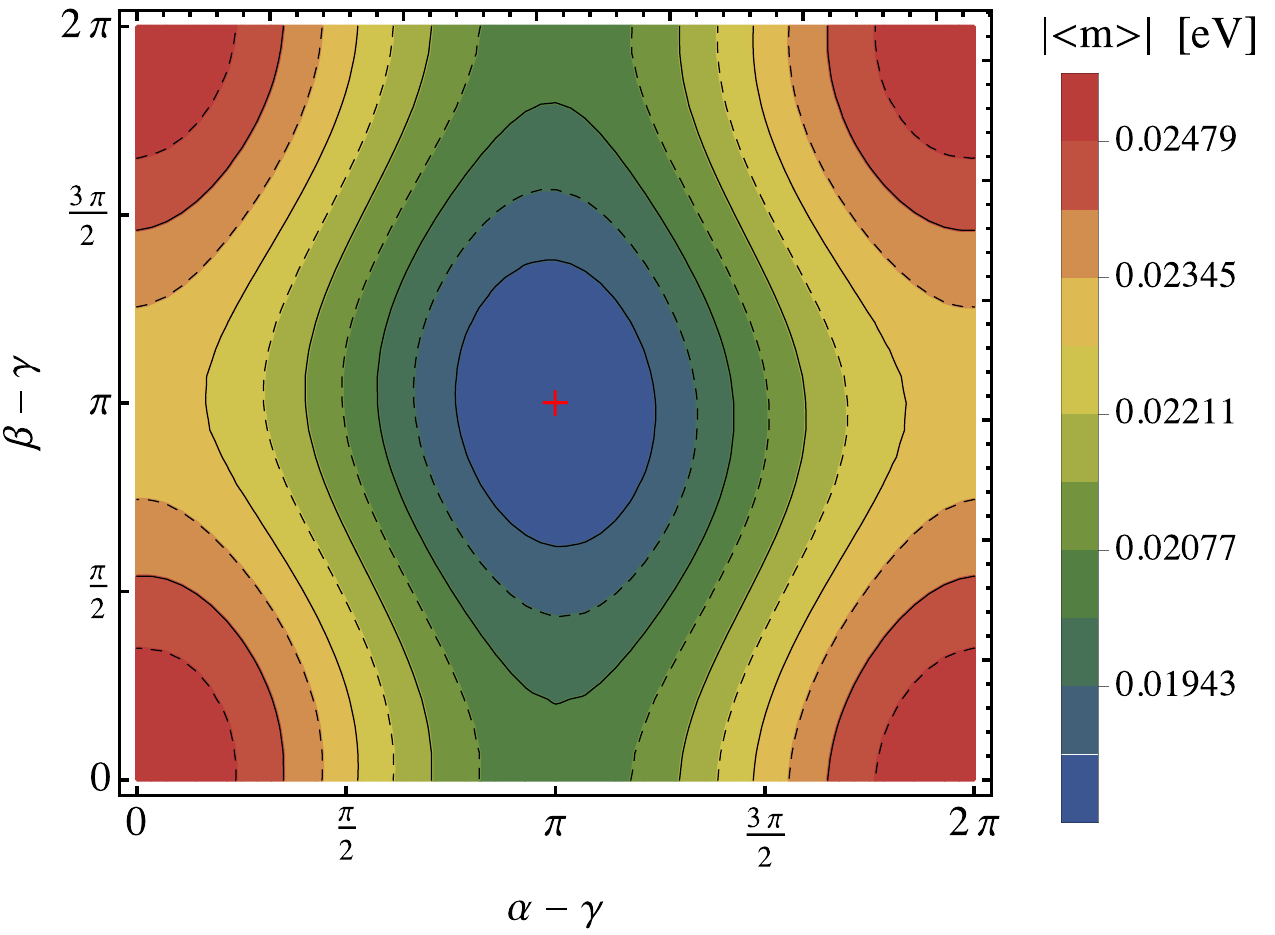}} \qquad \qquad
 \vspace{5mm}
     \end{center}
\vspace{-1.0cm} \caption{\label{fig:minNH} The value of \meff for
 NH spectrum in the 3+1 scheme and
$m_{min}=0$. The minimum corresponds to ($\alpha -
\gamma,\beta - \gamma$) =($\pi,\pi$).
At the minimum (the point with the cross) $\meff = 1.80 \times 10^{-2} \, {\rm eV}$.
The values in the first four contours are, respectively,
$(1.88, 1.94,  2.01, 2.08)  \times 10^{-2} \, {\rm eV}$
and obtained for $\Delta m^2_{41} = 0.93~{\rm eV^2}$.
See text for further details.
}
\end{figure}

  The investigation of the minima of $\meff$ in the NH case
in the limit of $m_1 =0$ and arbitrary values
of the relevant parameters
$b_0,c_0,d_0=b(m_1=0),c(m_1=0),d(m_1=0)$,
can be done following the
general analysis presented in Appendix
\ref{AppendixA} and, more specifically,
using the system eq. (\ref{genminmax})
that can be written as
\be
\begin{split}
&c_0 \sin (\alpha -\beta +\gamma - \gamma)+d_0 \sin (\alpha -\gamma ) = 0\\
-&b_0 \sin (\alpha -\beta +\gamma - \gamma)+d_0 \sin (\beta -\gamma ) = 0\,,\\
\end{split}
\label{minNHa0}
\ee
%
with
\be
b_0
= \sqrt{\Delta m^2_{21}} c_{13}^2 c_{14}^2 s_{12}^2\,,
\quad
c_0
= \sqrt{\Delta m^2_{31}} c_{14}^2 s_{13}^2\,,
\quad
d_0 = \sqrt{ \Delta m^2_{41}}  s_{14}^2\,.
\ee
%
%
 We have solved the system eq. (\ref{minNHa0}) in ($\alpha - \gamma$) and
($\beta - \gamma$) and found the solution $\sin(\alpha - \gamma) =
0$,  $\sin(\beta - \gamma) = 0$. The solution value of $(\alpha -
\gamma,\beta - \gamma) = (0,0)$ is a maximum, while the
second one $(\alpha - \gamma,\beta - \gamma) = (\pi,\pi)$ is a minimum.
In other words, solving the system of two equations
we find a unique minimum at $(\alpha -\gamma,\beta - \gamma) = (\pi,\pi)$
independently of the value of
\footnote{For the specific values of the neutrino
oscillation parameters used in the present analysis, the fact that
the minimum of \meff is reached for just one set of values of
 $(\alpha -\gamma,\beta - \gamma) = (\pi,\pi)$ follows
from the explicit expression for \meff, eqs. (\ref{meffNO}),
in the case of $m_1 = 0$.}
$\Delta m^2_{41}$.
The corresponding minimum value
of \meff is 0.018 (0.027) eV  in the case of
$\Delta m^2_{41} = 0.93~{\rm eV^2}~(1.78~{\rm eV^2})$.
This result is depicted in Fig. \ref{fig:minNH}.
The darkest region in the figure corresponds to the
minimum of \meff and the red cross indicates the precise value of
($\alpha - \gamma ,\beta - \gamma$) at the minimum.

  It follows from the results of our analysis that
for $m_1 = 0$ and any values of the CPV phases
$(\alpha - \gamma, \beta - \gamma)$ we have:
\begin{itemize}
\item  $\meff \geq 0.018 \, eV$ if $\Delta m^2_{41} = 0.93~{\rm  eV^2}$;
\item  $\meff \geq 0.027 \, eV$ for
$\Delta m^2_{41} = 1.78~{\rm  eV^2}$. If instead of 
$\sin^2\theta_{14} = 0.0223$ we use $\sin^2\theta_{14} = 0.017~(0.047)$, 
we get $\meff \geq 0.019~(0.059)\, eV$.

\end{itemize}
%

%
\section{The case of IO Spectrum in the 3+1 Scheme}
%

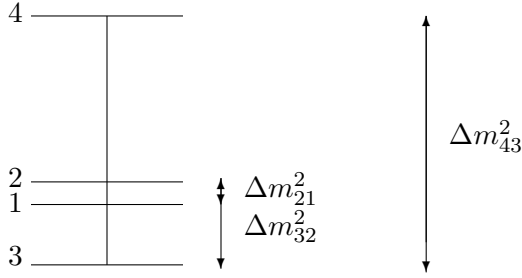
\begin{figure}
\unitlength=1mm
\begin{center}
\begin{picture}(100,40)
\put(20,2){\line(1,0){20}}
\put(20,10){\line(1,0){20}}
\put(20,13){\line(1,0){20}}
\put(20,35){\line(1,0){20}}
\put(30,2){\line(0,1){33}}
\put(45,10){\vector(0,-1){8.5}}
\put(45,10){\vector(0,1){3.5}}
\put(45,13.5){\vector(0,-1){3.5}}
\put(72,5){\vector(0,1){30}}
\put(72,35){\vector(0,-1){34}}
\put(48,11){$\Delta m^2_{21}$}
\put(48,6){$\Delta m^2_{32}$}
\put(75,18){$\Delta m^2_{43}$}
\put(17,2){3}
\put(17,9){1}
\put(17,12.3){2}
\put(17,34.2){4}
\end{picture}
\end{center}
\caption{\label{specIH3plus1} The mass spectrum in the 3+1 IO scheme.}
\end{figure}

In the case of 3+1 scheme with IO 3-neutrino mass spectrum,
$m_3 < m_1 < m_2 < m_4$,
one can write the effective Majorana mass
following the notation in \cite{PDG2012} as:
\be
\meff = | m_1 c_{12}^2 c_{13}^2 c_{14}^2 + m_2 e^{i\alpha}
c_{13}^2 c_{14}^2 s_{12}^2 +   m_3 e^{i\beta} c_{14}^2 s_{13}^2
 + m_4 e^{i\gamma}  s_{14}^2 |\,.
\ee
%
The masses $m_{1,2,4}$ can be expressed in terms of the lightest
neutrino mass $m_{min} = m_3$ and  the neutrino mass squared differences
as follows:
\be
\begin{split}
m_1 = & \sqrt{m_{min}^2 + |\Delta m^2_{32}| - \Delta  m^2_{21} },
\quad
m_2 = \sqrt{m_{min}^2 + |\Delta m^2_{32}|  },
\quad
m_4 = \sqrt{m_{min}^2 + \Delta m^2_{43} }\, ,\\
m_3 = & m_{min} \,, \quad
\Delta m^2_{21} > 0\,, \quad
\Delta m^2_{32} < 0\,, \quad
\Delta m^2_{43} > 0\,.\\
\end{split}
\ee
%
The neutrino mass spectrum of this scheme is depicted
schematically in Fig. \ref{specIH3plus1}.

%
The parameters $a$, $b$, $c$ and $d$ are given by:
\be
\begin{split}
a & = 
\sqrt{m_{min}^2 + |\Delta m^2_{32}| - \Delta  m^2_{21} } \,
 c_{12}^2 c_{13}^2 c_{14}^2\\
b & =
\sqrt{m_{min}^2 + |\Delta m^2_{32}|  } \,\,  c_{13}^2 c_{14}^2 s_{12}^2 \\
c & = m_{min}  \,\, c_{14}^2 s_{13}^2\\
d & = \sqrt{m_{min}^2 + \Delta m^2_{43}} \,\,s_{14}^2 \\
\label{IH31abcd}
\end{split}
\ee

In this case only a few solutions among those found in
Appendix \ref{AppendixA}
are relevant and their existence depends on the numerical
values of the parameters $a$, $b$, $c$ and $d$.
In Appendix \ref{AppendixA1} we list the domain of
existence of all the solutions.
Here we will analyze the solutions $(u_{\pm},v_{\pm},t_{\pm})$,
given in eq. (\ref{uvt}) with the parameters  $a$, $b$, $c$ and $d$
defined in eq. (\ref{IH31abcd}),
 because their domain is the largest
(the numerical details are given in Appendix \ref{AppendixA1}).

\begin{figure}[h!]
  \begin{center}
 {\includegraphics[width=7.55cm]{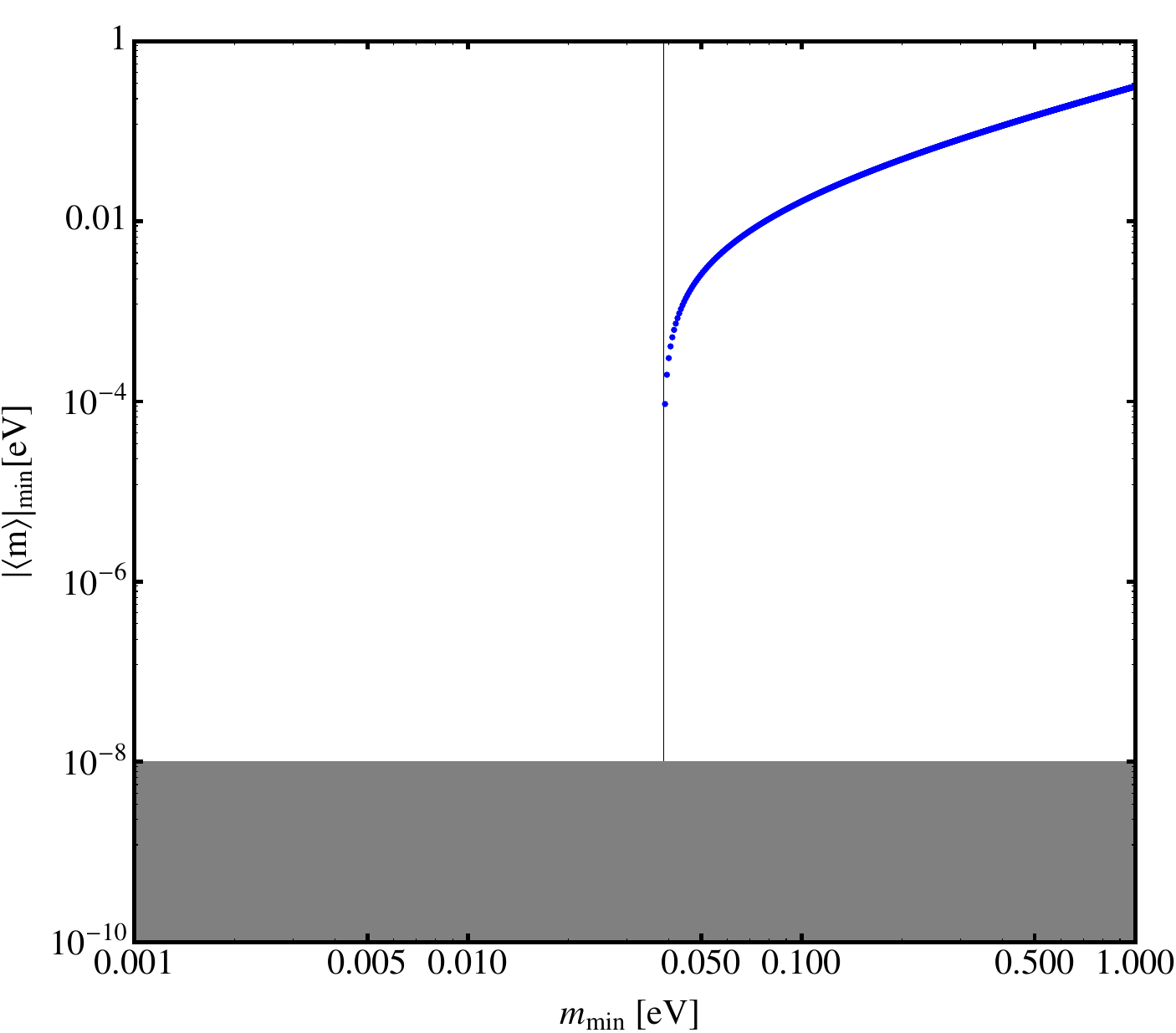}}
      \end{center}
 \caption{\label{fig:minIHa}  Minimum \meff
as function of $ m_{min}\equiv  m_3$. The figure has been
obtained numerically
for  $\Dmq_{43} =  0.93 \, {\rm eV^2} $, $\sin \theta_{14} = 0.15$.
The vertical line corresponds to $m_{min} = \ov m_{3} \simeq
0.038 \, {\rm eV}$. See text for details.
 }
\end{figure}

 We observe that the solutions $(u_{\pm},v_{\pm},t_{\pm})$ are
well defined when the product $f_1f_2 f_3 f_4$ is positive,
where $f_{1,2,3,4}$ are given in eq. (\ref{fi}).
Defining $\ov m_3$ as the zero of the function $f_1$,
$f_1(\ov m_3) = 0$, we find that the
effective Majorana mass can be
zero for $m_3 < \ov m_3$ for specific
CP non-conserving values of the CPV phases
$\alpha$, $\beta$ and $\gamma$.
For $\Dmq_{43} =  0.93~(1.78)~{\rm eV^2}$
and the best fit values of Table \ref{tabNudata}, we find
$\ov m_3 \simeq 0.038~(0.074)\,{\rm eV}$.
These results are presented graphically in Fig. \ref{fig:minIHa},
where we show the numerically calculated $\meff_{min}$
as function of $m_3$.
The numerical minima depicted in Fig.  \ref{fig:minIHa}
are obtained by
performing a scan over the values of $m_3$
and of each of the phases
$(\alpha,\beta,\gamma)$ in the interval $[0,2\pi]$.
The grey horizontal band  
in  Fig. \ref{fig:minIHa}, corresponding to $\meff_{min} < 10^{-8}$
eV, reflect the precision of the numerical calculation of
$\meff_{min} = 0$.
 The minima of \meff under discussion
are reached for values of the phases
$(\alpha,\beta,\gamma)$ that can be either $CP$ conserving or CP
non-conserving.
%

  We will find next an analytical approximation of $\ov m_3$.
We observe that for values $m_3$ in the range
$ m_3 \approx  0.05 - 0.10 \, eV$ the term
$\propto m_3 \cos^2 \theta_{14} \sin^2 \theta_{13}$
is by approximately an order of magnitude smaller then the
other three terms in \meff.
Neglecting it as well the term  $\propto \Dmq_{21}$,
we find the following expression for $\ov m_3$,
which is valid up to an error of about the $15\%$:
\be
\ov m_3 \approx \sqrt{\frac{\Dmq_{43} \sin ^4 \theta_{14}
-\, |\Dmq_{32}| \cos ^2 2 \theta_{12} \cos ^4 \theta_{13} \cos ^4 \theta_{14}}
{\cos ^2 2 \theta_{12}  \cos ^4 \theta_{13} \cos ^4 \theta_{14}
-\, \sin ^4 \theta_{14}}} \,.
\ee
%
Using this approximation we get $\ov m_3 \simeq 0.032 \, {\rm eV}$ for
$\Dmq_{43} = 0.93 \, {\rm eV^2}$,
and $\ov m_3 \simeq 0.068 \, {\rm eV}$ for
$\Dmq_{43} = 1.78 \, {\rm eV^2}$,
instead of $0.038 \, {\rm eV}$ and $0.074 \, {\rm eV}$ found numerically.

\begin{figure}[h!]
  \begin{center}
 \subfigure
 {\includegraphics[width=7cm]{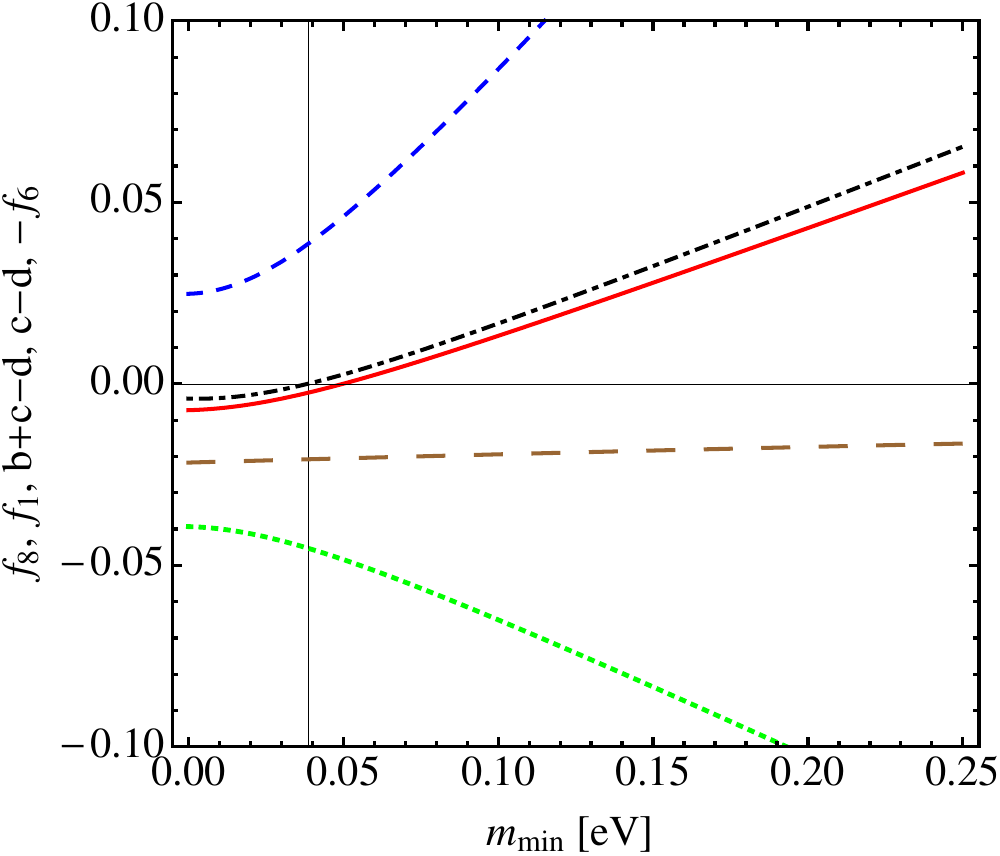}}
 \vspace{5mm}
 \subfigure
   {\includegraphics[width=7cm]{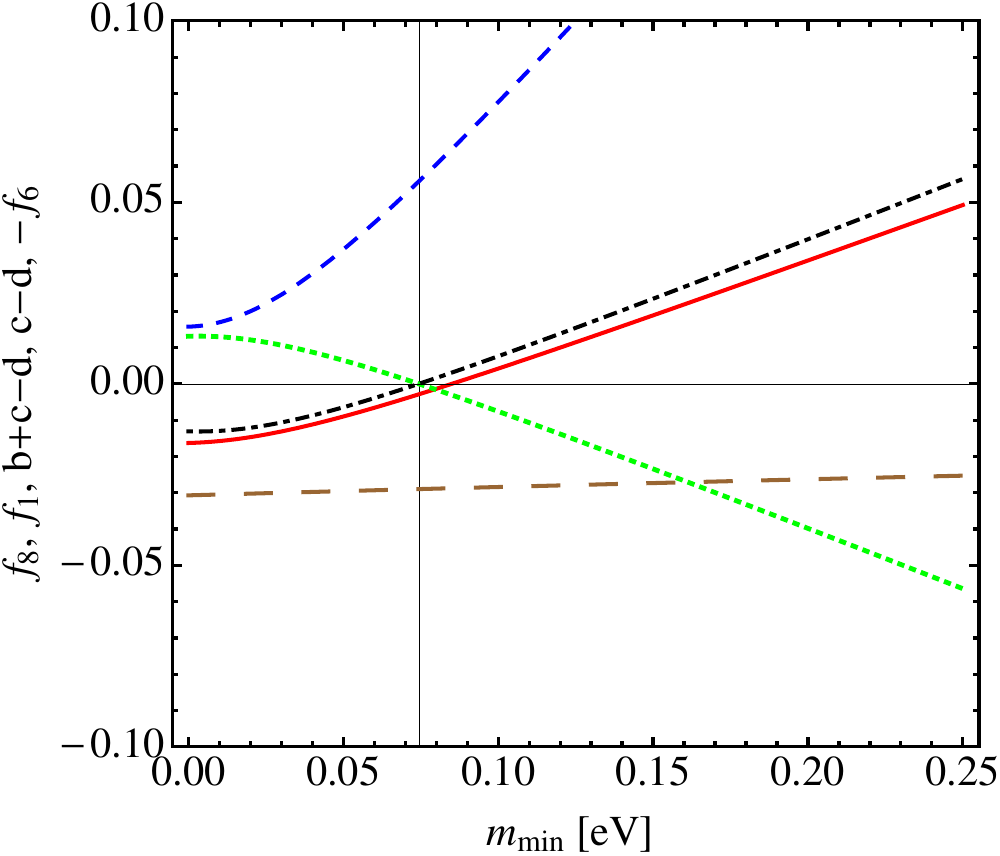}}
     \end{center}
\vspace{-1.0cm} \caption{\label{fig:SylvesterIH}
Left Panel. The functions $f_8$ (short-dashed blue),
$f_1$ (dot-dashed black),
$b + c - d$ (solid red),
$c - d$ (large dashed brown),
$-f_6$ (dotted green) versus $ m_{min}\equiv  m_3$
for $\Dmq_{43} = 0.93 \, {\rm eV^2}$, $\sin\theta_{14} = 0.15$.
The vertical line corresponds to $m_{min} = \ov m_3\simeq 0.038\,{\rm eV}$.
Right Panel. The same as in the left panel, but for
 $\Dmq_{43} = 1.78\, {\rm eV^2}$.
The vertical line corresponds to
$m_{min} = \ov m_3 \simeq 0.074 \,{\rm eV}$.
}
\end{figure}

  To find the minima of \meff for values of
$m_3 > \ov m_3$ we have to study the Hessian of $\meff$.
From the analysis in  Appendix \ref{AppendixA} it follows that in the
region in which $f_1 > 0$ (corresponding to the region $m_3 > \ov m_3$),
the minimum of \meff (according to the Sylvester's criterion)
takes place at $(\alpha,\beta,\gamma) = (\pi,\pi,\pi)$.
In Fig. \ref{fig:SylvesterIH} we show all  the
relevant functions entering in the conditions determining the minima,
which are listed in eq. (\ref{fi})
(and eq. (\ref{condmin2})),  with the
parameters  $a$, $b$, $c$ and $d$ defined in
eq. (\ref{IH31abcd}).

\begin{figure}[h!]
  \begin{center}
 \subfigure
 {\includegraphics[width=7cm]{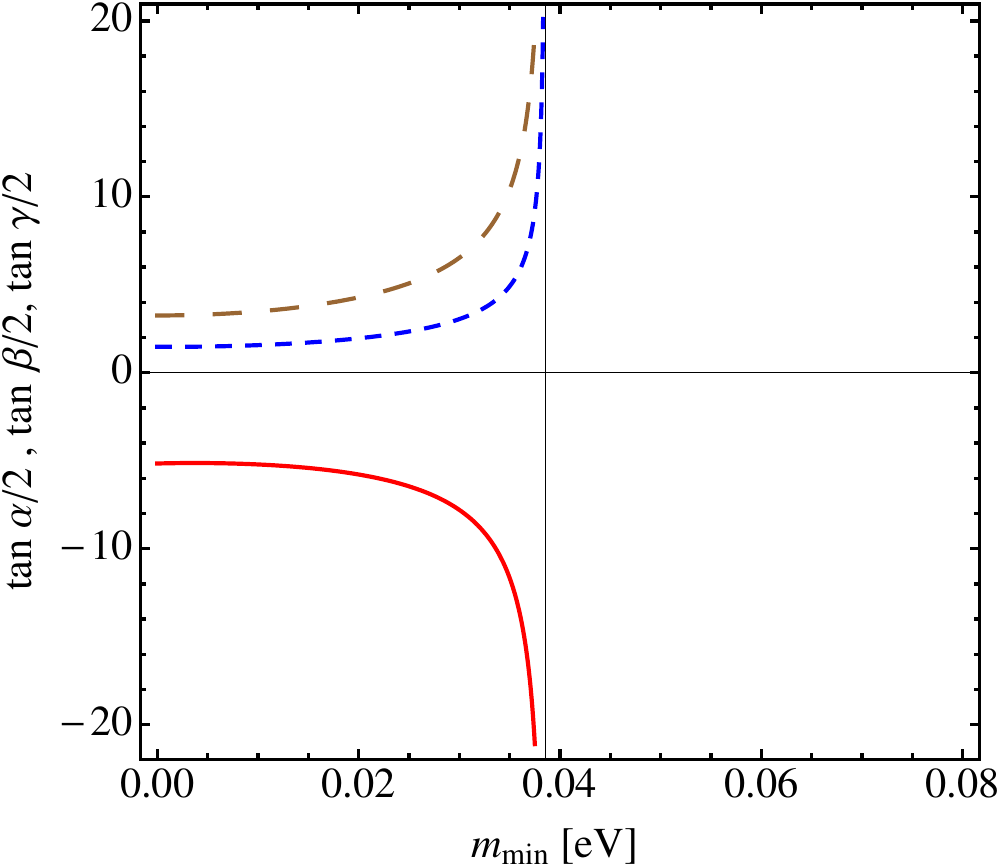}}
 \vspace{5mm}
 \subfigure
   {\includegraphics[width=7cm]{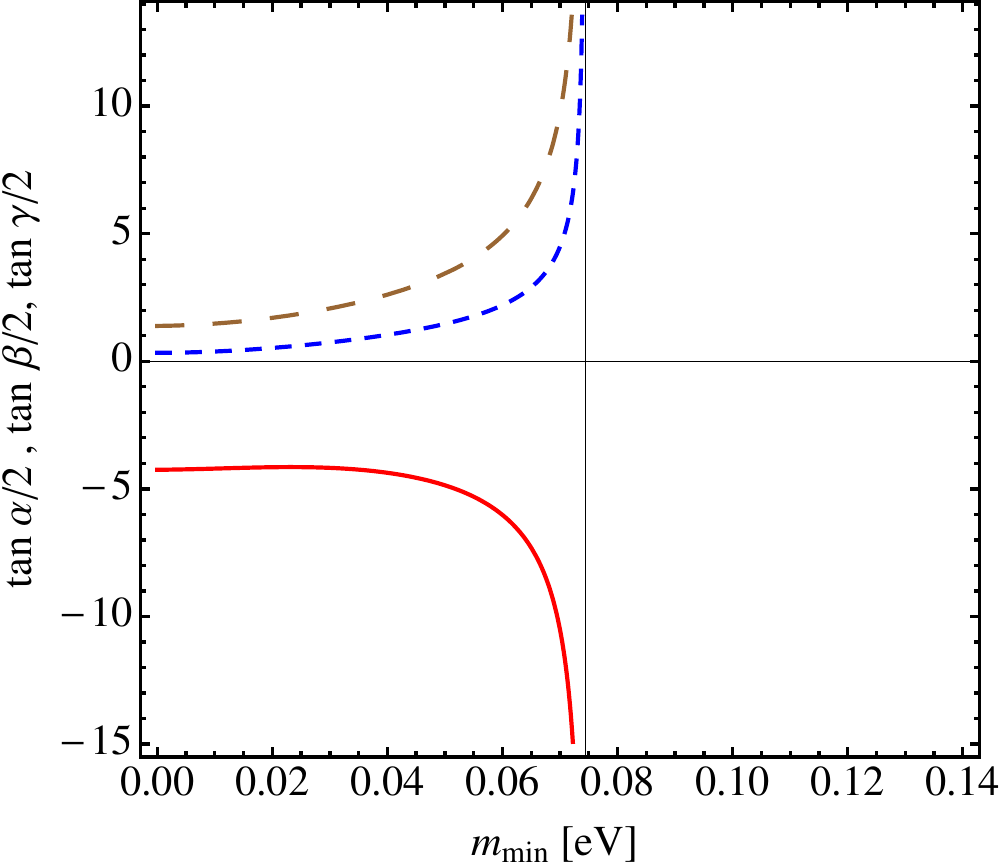}}
     \end{center}
\vspace{-1.0cm} \caption{\label{fig:MinIHphase}
Left Panel.
The values of $(\tan \alpha/2,\tan \beta/2,\tan \gamma/2)$
 -- large dashed (brown), short dashed (blue), solid  (red) lines --
corresponding to the solution $(v_-,t_-,u_-)$,
eq. (\ref{uvt}), as functions of $m_3$
for  $\Dmq_{43} = 0.93 \, {\rm eV^2}$, $\sin\theta_{14} = 0.15$.
The vertical line is at $\ov m_3 = 0.038 \, {\rm eV}$,
indicating the domain of existence of this solution,
$m_3 \leq \ov m_3$.
Right Panel. The same as in the left panel, but for
$\Dmq_{43} = 1.78 \, {\rm eV^2}$.
The vertical line
is at $\ov m_3 = 0.074 \, {\rm eV}$, indicating the domain of
existence of the solution $(v_-,t_-,u_-)$, $m_3 \leq \ov m_3$.
%
}
\end{figure}

In Fig. \ref{fig:MinIHphase} we show as an example
the values of the three phases versus $m_{min}$
for the solution  $(u_{-},v_{-},t_{-})$.
The analogous figure for the solution $(u_{+},v_{+},t_{+})$
is obtained formally from  Fig. \ref{fig:MinIHphase}
by reversing the $y-$axis.

\begin{figure}[h!]
  \begin{center}
 \subfigure
 {\includegraphics[width=7cm]{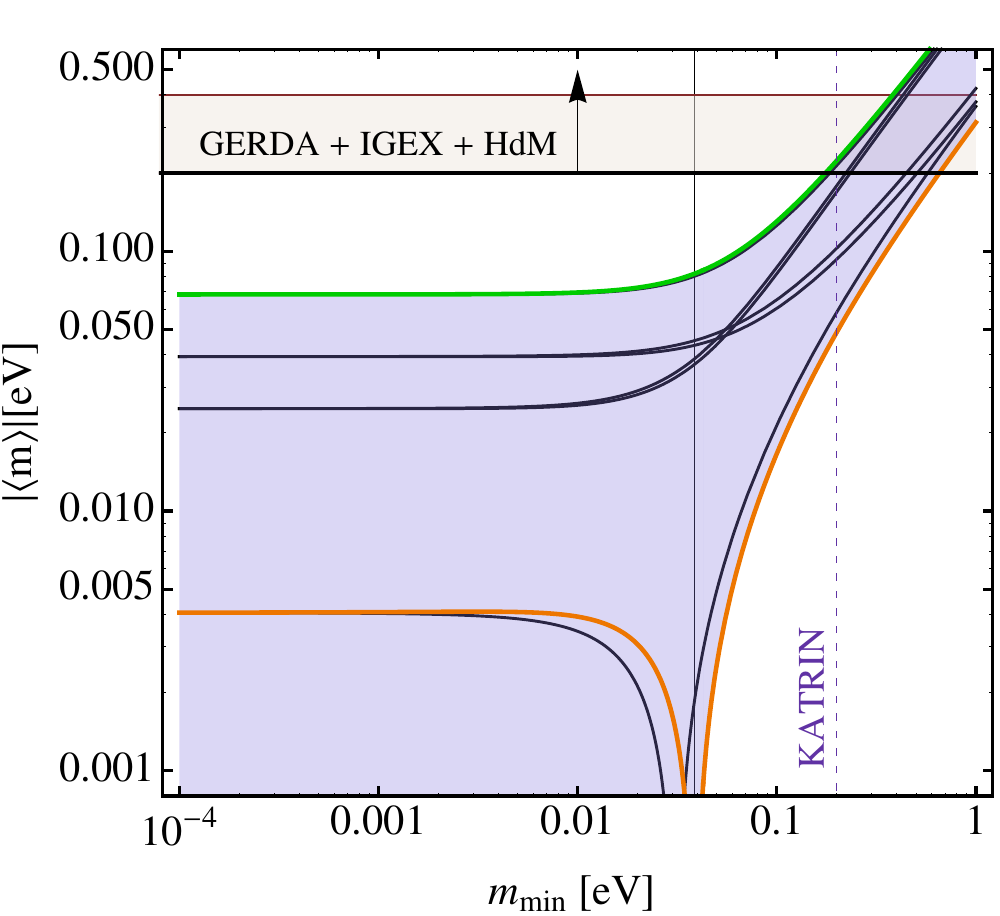}}
 \vspace{5mm}
 \subfigure
   {\includegraphics[width=7cm]{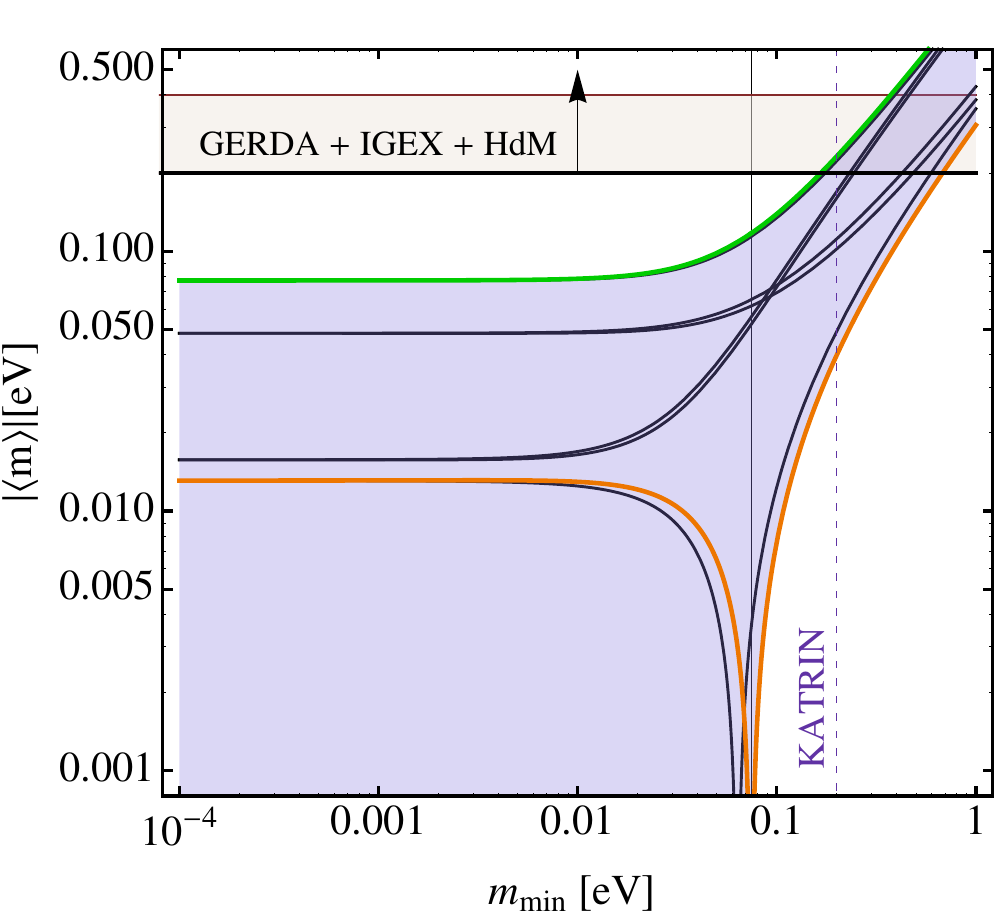}}
     \end{center}
\vspace{-1.0cm}\caption{\label{fig:Min}
 Left Panel. The value of \meff as function of $m_{min} = m_3$
 for $\Dmq_{43} = 0.93 \, {\rm eV^2}$, $\sin \theta_{14} = 0.15$.
The green and orange lines correspond
respectively  to $(\alpha,\beta,\gamma)= (0,0,0)$  and $(\pi,\pi,\pi)$.
 The six gray curves correspond to
the other possible sets of CP conserving
values $0$ or $\pi$ of the CPV phases  $(\alpha,\beta,\gamma)$.
The vertical line is at $m_3 = \ov m_3\simeq
0.038 \,{\rm eV}$. If $m_3 \leq \ov m_3$,  we can have
$\meff_{min}=0$ at any fixed $m_3$
for specific values of $(\alpha,\beta,\gamma)$,
while for  $m_3 > \ov m_3$, the minimum
of \meff is realized at $(\alpha,\beta,\gamma) = (\pi,\pi,\pi)$
and $\meff_{min}\neq 0$.
Right Panel. The same as
in the left panel, but for
 $\Delta m^2_{43} \equiv 1.78 \, {\rm eV^2}$.
The vertical
line is at $m_3 = \ov m_3\simeq 0.074 \,{\rm eV}$.
The horizontal band indicates the upper bound
$\meff  \sim0.2-0.4$ eV obtained using the 90 \% C.L.
limit on the half-life of $^{76}$Ge
reported in \cite{Agostini:2013mba}.
}
\end{figure}

%
Finally, we show in Fig. \ref{fig:Min} \meff as function of the
lightest neutrinos mass, $m_{min}$. In this case the region of
allowed values of \meff (the shaded area) is larger than in the NO
case since \meff can reach zero for any $m_{min}\leq \ov m_3$. This
is due to the fact that, depending on the values of the CPV phases
$\alpha$, $\beta$ and $\gamma$, a complete cancellation among the
terms in the expression for \meff can occur.

 The results of the analysis performed in this section show that
we always have  $\meff > 0.01$ eV for:
\begin{itemize}
\item  
$m_{min} > 0.078$ eV, if
 $\Delta m^2_{43} = 0.93~{\rm  eV^2}$;
\item
$m_{min} > 0.108$ eV, if
$\Delta m^2_{43} = 1.78~{\rm  eV^2}$.
\end{itemize}

If in the case of
 $\Delta m^2_{41} = 1.78~{\rm eV^2}$, instead of 
$\sin^2\theta_{14} = 0.023$ we used the extreme value 
of the $2\sigma$ allowed interval quoted in eq. (\ref{th142sigma}),  
$\sin^2\theta_{14} = 0.017$ ($\sin^2\theta_{14} = 0.047$), 
this will lead  to the decreasing (increasing) 
of the numerical values of  $|<m>|$ at 
$m_{min} \lesssim 10^{-3}$ eV and of 
$\overline{m}_3$, obtained for $\sin^2\theta_{14} = 0.023$,  
approximately by the factors 
1.1 (1.4) and 2.0 (2.4), respectively.

\subsection{The Case of $m_3 =0$}

The effective Majorana mass in this case is
\be
\meff = \left|\sqrt{|\Delta m^2_{32}| -\Delta m^2_{21}}
\,(c_{12}c_{13}c_{14})^2
+ \sqrt{|\Delta m^2_{32}|}\, (c_{13}c_{14}s_{12})^2 e^{i \alpha}
+ \sqrt{\Delta m^2_{43}}\, s_{14}^2 e^{i \gamma} \right|\,.
\label{eq:IHm0}
\ee
%
Now only two phases enter into the expression of
\meff: $\alpha$ and  $\gamma$.
%
%
In this case the minima of \meff can be obtained from
the general solutions derived  in Appendix \ref{AppendixA}
and take place for
\be
\begin{split}
\sin \gamma_\pm & =\mp \frac{\sqrt{-[(a_0-d_0)^2 - b_0^2][(a_0+d_0)^2 - b_0^2]}}{2a_0 d_0} \\
\sin \alpha_\pm & = \pm\frac{\sqrt{-[(a_0-d_0)^2 - b_0^2][(a_0+d_0)^2 - b_0^2]}}{2a_0 b_0}\\
\end{split}
\ee
%
%
%
where
\be
\begin{split}
a_0 & =
\sqrt{|\Delta m^2_{32}| - \Delta  m^2_{21} } \,\ c_{12}^2 c_{13}^2 c_{14}^2\,,\\
b_0 & =
\sqrt{|\Delta m^2_{32}|  } \,\,  c_{13}^2 c_{14}^2 s_{12}^2\,, \\
d_0 & = \sqrt{\Delta m^2_{43}} \,\,s_{14}^2\,. \\
\end{split}
\ee
%
%
%

\begin{figure}[h!]
  \begin{center}
      \subfigure
   {\includegraphics[width=8cm]{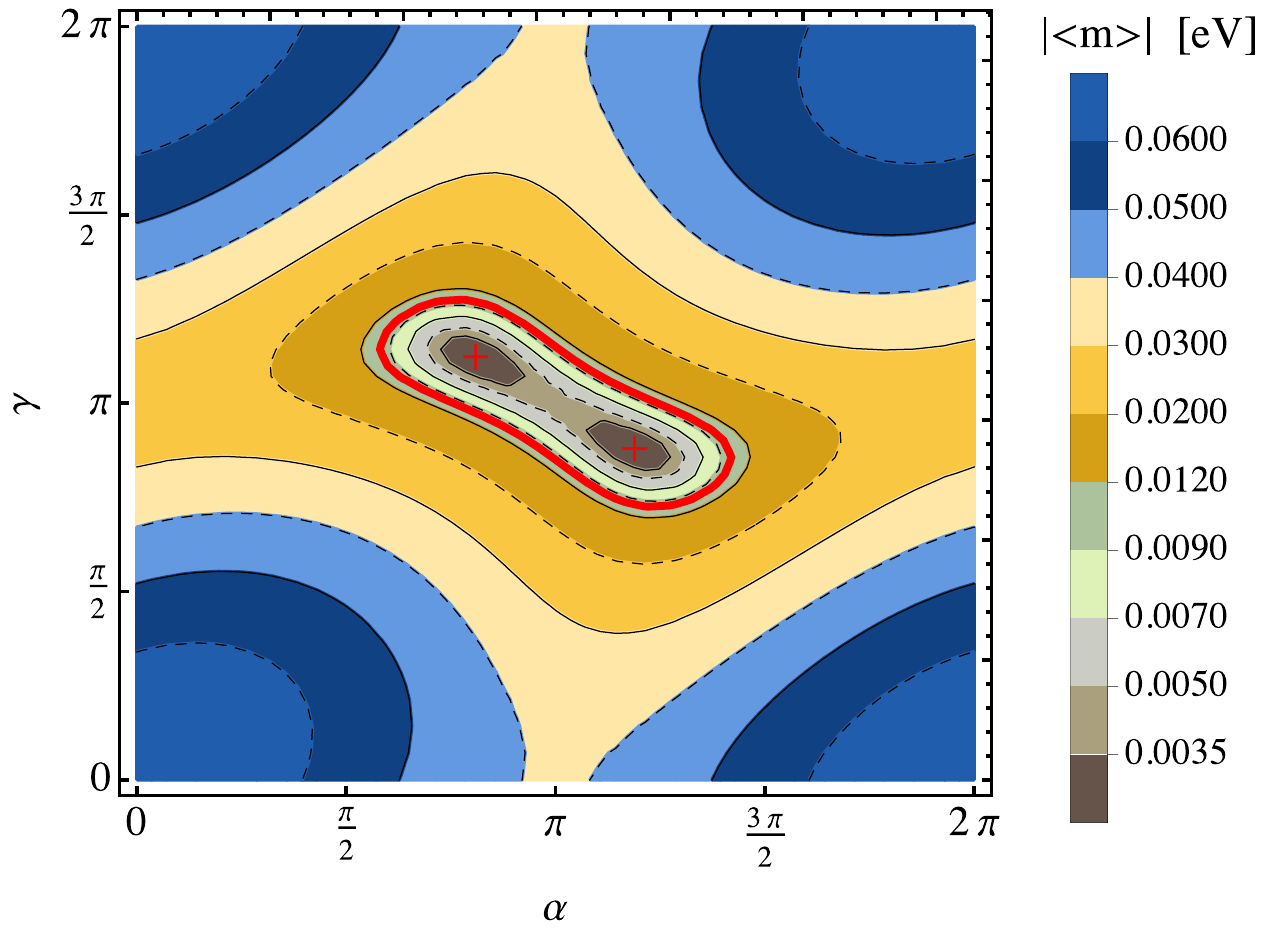}} \qquad \qquad
 \vspace{5mm}
     \end{center}
     \vspace{-1.0cm}
 \caption{\label{fig:minIH}
The values of \meff for the IH spectrum in the 3+1 scheme versus
$\alpha$ and $\gamma$ in the case of $m_{min}= m_3=0$,
$\Dmq_{43} =  0.93 \, {\rm eV^2}$ and $\sin \theta_{14} = 0.15$.
In this case there are two minima in the crossed points
at ($\sin \alpha, \sin \gamma) = (\mp0.562, \pm0.373)$,
and \meff in these minima is exactly zero.
The red line corresponds to $\meff = 0.01$ eV.
See text for details.
 }
\end{figure}

Both minima correspond to $\meff = 0$
independently of the value of $\Delta m^2_{43}$.
However, the location of the minima on the $\alpha - \gamma$ plane
depends on $\Delta m^2_{43}$.
For instance,  if we use $\Delta m^2_{43}= 0.93~{\rm eV^2}$, the
minima are at $(\sin \alpha, \sin \gamma) = (\mp 0.562, \pm0.373)$.
This result is shown in Fig. \ref{fig:minIH}. We notice
that the existence of solutions for $(\alpha, \gamma)$ such that
$\meff \sim 0$ is clear from the expression in eq. (\ref{eq:IHm0})
since for the values of the oscillation parameters used 
in the present study a complete cancellation among the 
three terms in \meff can take place. 
Indeed, in the case of the best fit values, for instance,

the first term
$\propto\sqrt{|\Delta m^2_{32}|-\Delta m^2_{21}}\,
(c_{12}c_{13}c_{14})^2 \approx 0.032$, can
be compensated completely by the sum of the other 
two terms which are of the order of
$\sqrt{|\Delta m^2_{32}|}\, (c_{13}c_{14}s_{12})^2 \approx0.014$ and
$\sqrt{\Delta m^2_{43}}\, s_{14}^2\approx 0.022$, respectively.

It follows from our analysis that in the case of $m_3 = 0$ we have
 $\meff > 0.01$ eV for values of the phases $\alpha$ and $\gamma$ outside the
 region delimited by the red line in Figure \ref{fig:minIH}.

We note finally that in the limit
$m_3 \rightarrow 0$ (or equivalently $c \rightarrow 0$)
there are four  out of the
nine solutions determined analytically, which admit $\meff = 0$ (this can
be seen in Table  \ref{DomainIH} in the Appendix \ref{AppendixA1}).
The four solutions are $(u_{\pm}, v_{\pm},t_{\pm})$ and
$(v^{\pm}_4(u),t^{\pm}_4(u))$. If the solutions
$(v^{\pm}_4(u),t^{\pm}_4(u))$ are evaluated at
$u^{\pm}$, i.e., $v^{\pm}_4(u^{\pm})\rightarrow v_{\pm}$, in this case
the two minima of the first solution coincide with the two minima
of the second one.

%
\section{The 3+2 Scheme: Two Sterile Neutrinos}
%

In this Section we analyze the case of two extra sterile neutrino
states. In this case  the PMNS mixing matrix is a $5\times5$ unitary
matrix. Following the parametrization used in  \cite{Kopp:2013vaa}
it can be written as:
\be
 U = V_{35} O_{34} V_{25} V_{24} O_{23}
O_{15} O_{14} V_{13} V_{12} \,\, diag
(1,e^{i\alpha/2},e^{i\beta/2},e^{i\gamma/2}, e^{i\eta/2})\,,
\ee
%
where $\eta$ is an additional Majorana CPV phase. As in the case of
the 3+1 scheme, we can set to zero the phases in the matrices $
V_{25}$, $V_{24}$, $V_{13}$ and $ V_{12}$ without loss of
generality. In this case the elements of the first row of the PMNS
matrix of interest for our analysis are given by:
\be
\begin{split}
U_{e1} & = c_{12} c_{13} c_{14} c_{15}\\
U_{e2} & = e^{i \alpha/2} c_{13} c_{14} c_{15} s_{12} \\
U_{e3} & = e^{i\beta/2} c_{14} c_{15} s_{13} \\
U_{e4} & = e^{i\gamma/2} c_{15} s_{14}\\
U_{e5} & = e^{i\eta/2}  s_{15}
\end{split}
\ee
%
The \betabeta-decay effective Majorana mass reads:
\be
\meff =\left|  m_1 |U_{e1}|^2 +  m_2 |U_{e2}|^2 e^{i \alpha}
+ m_3 |U_{e3}|^2e^{i \beta} + m_4 |U_{e4}|^2 e^{i\gamma}
+ m_5 |U_{e5}|^2 e^{i
\eta}  \right|.
\ee
%
The values for $\theta_{14}$,
$\Dmq_{41(43)}$, $\theta_{15}$ and $\Dmq_{51(53)}$ --- for NO
(IO)---, obtained in the global analysis performed in
\cite{Kopp:2013vaa}, are summarized in Table \ref{tab:nu2}.
\begin{table}[H]
  \centering
  \begin{tabular}{lccc}
    \toprule
    $\Dmq_{41(43)} \,\, [{\rm eV^2}]$& $\Dmq_{51(53)} \,\, [{\rm eV^2}]$ & $\theta_{14}$ & $\theta_{15}$
    \\
    \midrule
   0.47 & 0.87 & 0.13 & 0.14     \\
    \bottomrule
  \end{tabular}
  \caption{\label{tabNudata3plus2} Best global fit values of
the sterile neutrino oscillation parameters in the
 3+2 scheme with NO (IO) neutrino mass spectrum
(from \cite{Kopp:2013vaa}). The
    relation to the mixing matrix elements is $|U_{e4}| =
    \cos\theta_{15}\sin\theta_{14}$ and $|U_{e5}| =
    \sin\theta_{15}$.
\label{tab:nu2}}
\end{table}
%

%
\section{The 3+2 Scheme with NO  Spectrum}
%

\begin{figure}[h!]
\unitlength=1mm
\begin{center}
\begin{picture}(100,40)
\put(20,5){\line(1,0){20}}
\put(20,8){\line(1,0){20}}
\put(20,13){\line(1,0){20}}
\put(20,25){\line(1,0){20}}
\put(20,35){\line(1,0){20}}
\put(30,5){\line(0,1){30}}
\put(45,4.5){\vector(0,1){3.5}}
\put(45,8){\vector(0,-1){3.5}}
\put(45,8){\vector(0,1){5}}
\put(82,5){\vector(0,1){30}}
\put(82,35){\vector(0,-1){31}}
\put(65,9){\vector(0,1){15}}
\put(65,13){\vector(0,-1){9}}
\put(48,4){$\Delta m^2_{21}$}
\put(48,9.5){$\Delta m^2_{31}$}
\put(67,12){$\Delta m^2_{41}$}
\put(84,18){$\Delta m^2_{51}$}
\put(17,4){1}
\put(17,8){2}
\put(17,12.3){3}
\put(17,24.2){4}
\put(17,34.2){5}
\end{picture}
\end{center}
\caption{\label{specNH3plus2} The neutrino mass spectrum
in the 3+2 NO scheme.}
\end{figure}
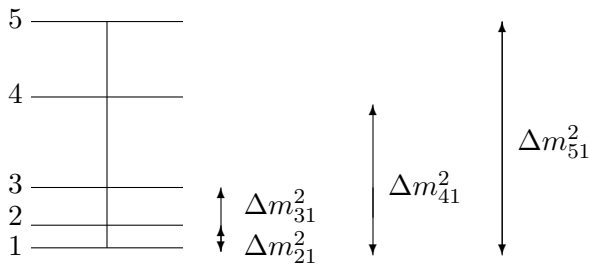

In the case of the 3+2 scheme with NO spectrum,
$m_1 < m_2 < m_3 < m_4 < m_5$,
one can write the effective Majorana mass as:
\be
\meff  = | m_1 c_{12}^2 c_{13}^2 c_{14}^2 c_{15}^2 +
m_2  e^{i\alpha} c_{13}^2 c_{14}^2 c_{15}^2 s_{12}^2 +
m_3  e^{i\beta} c_{14}^2 c_{15}^2 s_{13}^2 +
m_4 e^{i\gamma}  c_{15}^2 s_{14}^2  + m_5 e^{i\eta}  s_{15}^2 | \\
\ee
%
As in the case of the 3+1 scheme, it proves convenient
to express the masses $m_{2,3,4,5}$ in terms of
the lightest neutrino mass $m_1$ and the neutrino mass squared
differences:
\be
\begin{split}
 m_{min} &\equiv m_{1}, \quad m_2 = \sqrt{m_1^2 + \Delta m^2_{21}},
\quad m_3 = \sqrt{m_1^2 + \Delta m^2_{31} }, \quad
m_4 = \sqrt{m_1^2 + \Delta m^2_{41} }, \\
m_5 & = \sqrt{m_1^2 + \Delta m^2_{51} }, \quad
\Delta m^2_{21}
> 0, \ \
\Delta m^2_{31}
> 0, \  \
\Delta m^2_{41} > 0  \,\,\,\mbox{and}\,\, \,\Delta m^2_{51} > 0 \,\,.\\
\end{split}
\ee
%
The neutrino mass spectrum in 3+2 NO scheme is shown
in Fig. \ref{specNH3plus2}.
%

In what follows we will analyze the conditions
for minimization of \meff. As in the 3+1 case, we will work
with $\meff^2$ rather than with \meff:
 \be
 \meff^2 = |a+e^{i \alpha } b+e^{i
\beta } c+e^{i \gamma } d + e^{i \eta} e |^2  \,,
\ee
%
where
\be
\begin{split}
a &= m_{min} c_{12}^2 c_{13}^2 c_{14}^2 c_{15}^2 \,,\\
b & =  \sqrt{m_{min}^2+\Delta m^2_{21}} c_{13}^2 c_{14}^2 c_{15}^2 s_{12}^2 \,, \\
c & =  \sqrt{m_{min}^2+\Delta m^2_{31}} c_{14}^2 c_{15}^2 s_{13}^2 \,,\\
d & = \sqrt{m_{min}^2+ \Delta m^2_{41}}  c_{15}^2 s_{14}^2 \,, \\
e & =  \sqrt{m_{min}^2+ \Delta m^2_{51}}  s_{15}^2 \,.\\
\end{split}
\ee
%

\begin{figure}[h!]
  \begin{center}
 \subfigure
 {\includegraphics[width=7cm]{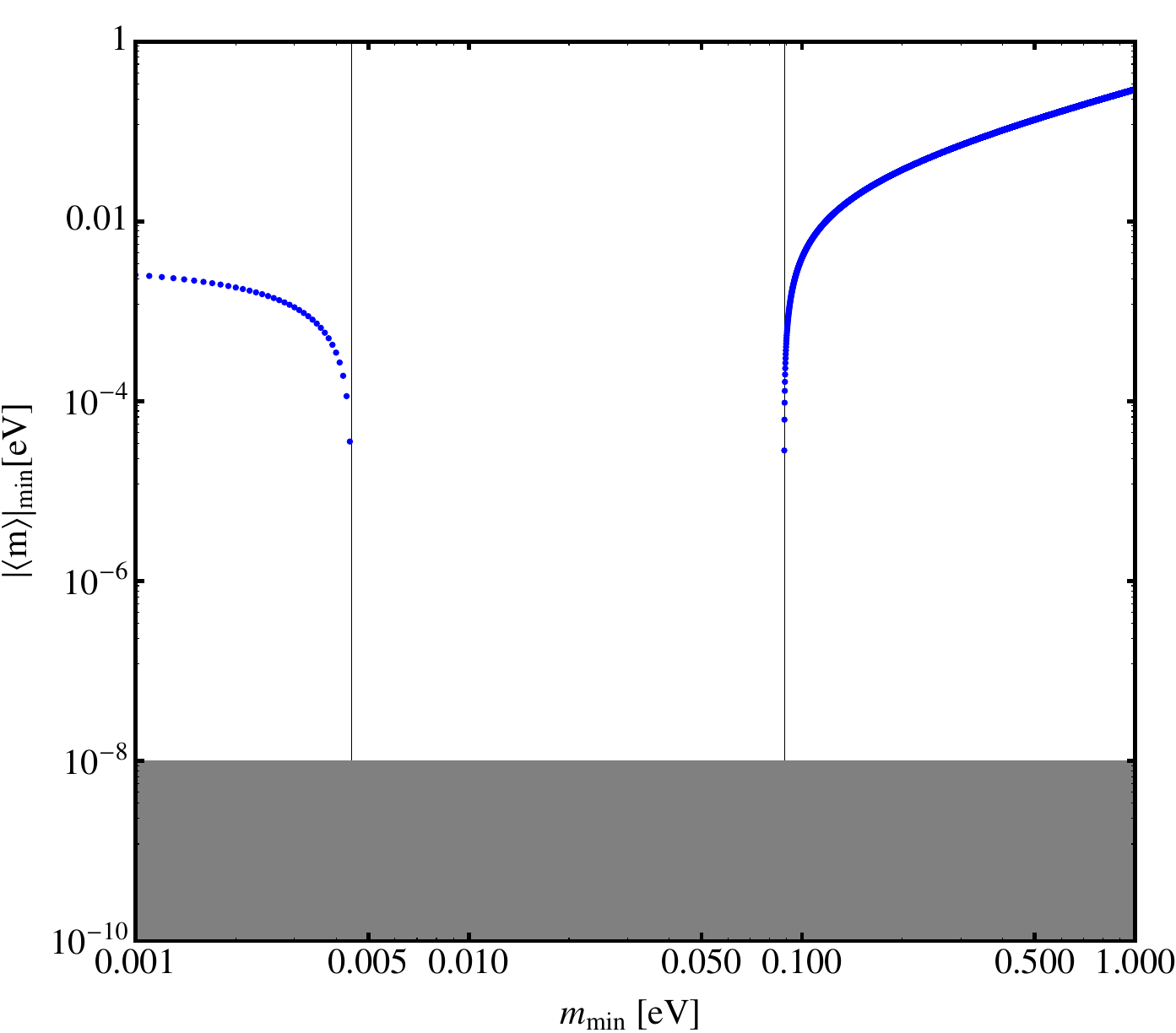}}
 \vspace{5mm}
 \subfigure
   {\includegraphics[width=7cm]{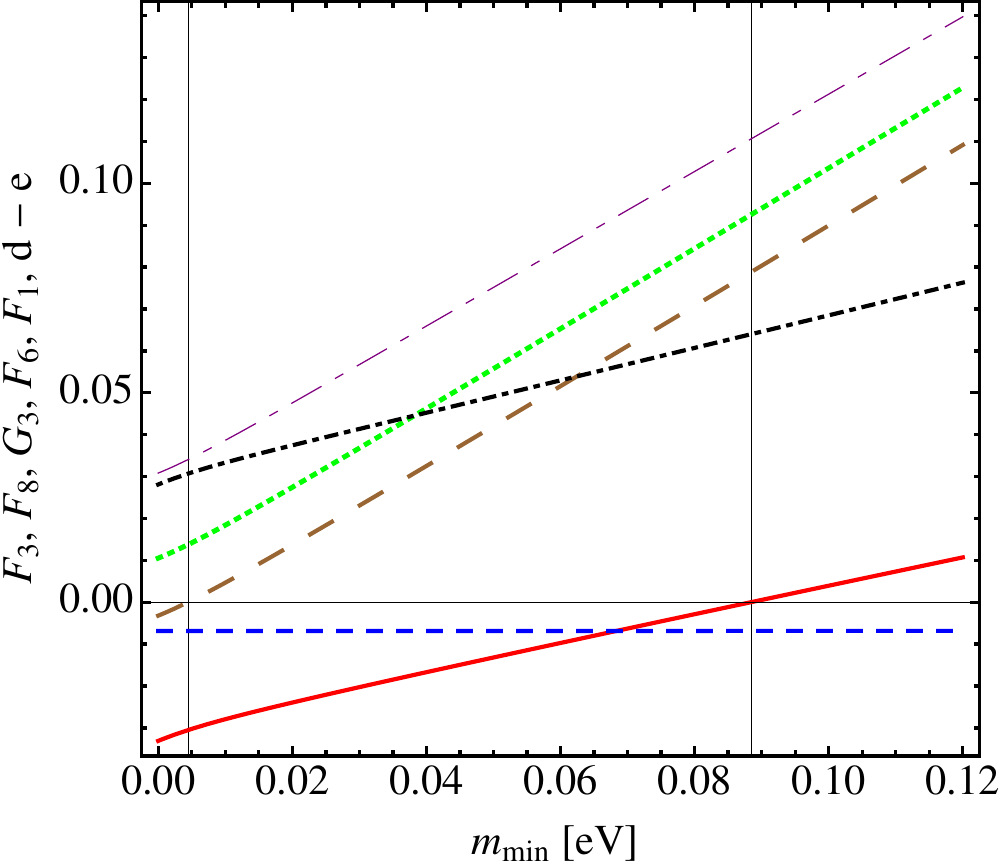}}
     \end{center}
\vspace{-1.0cm} \caption{\label{fig:SylvesterNH3plus2} Left Panel.
Minimum \meff as function of $m_{min}\equiv m_{1}$. The figure has
been obtained numerically for $\Dmq_{41} =  0.47 \, {\rm eV^2}$,
$\Dmq_{51} =  0.87 \, {\rm eV^2}$, $\sin \theta_{14} = 0.13$, $\sin
\theta_{15} = 0.14$ (see Table \ref{tab:nu2}) and performing a scan
over a sufficiently large sets of values of $m_{min}$.  The vertical
lines correspond to $\ov m_1 \simeq   8.84\times 10^{-2}$ eV and
$\un{m}_1 \simeq 4.44\times 10^{-3}$ eV. In the interval $\un{m}_1
\leq m_1 \leq \ov m_1$ we have min(\meff) = 0.  Right Panel. The
functions $F_3$ (dotted green), $F_8$ (large dashed brown), $G_3$
(short-large dashed purple), $F_6$ (dot-dashed black), $F_1$ (solid
red), $d - e$ (short-dashed blue), defined in eq. (\ref{eq:F1F8}) as
function of $m_{min}$ for the best fit values of Table
\ref{tabNudata3plus2}. The vertical lines are at $\ov m_1 \simeq
8.84\times 10^{-2}$ eV
 and $\un{m}_1 \simeq 4.44\times 10^{-3}$ eV.
}
\end{figure}

The analytical study of the minima of $\meff^2$ in this case is a
non-trivial task since four phases are involved and the
non-linearity of the system of the first derivatives of $\meff^2$
with respect to the four phases makes the analysis rather
complicated. Therefore finding all possible solutions of the
minimization procedure in analytical form is a complex problem.
Thus, we have performed the general analysis of the minimization of
\meff numerically. It is possible, however, to perform analytically
the analysis of the minima of \meff, corresponding to the 16 sets of
CP conserving values (either  $0$ or $\pi$) of the  four phases $
\alpha$, $\beta$, $\gamma$ and $\eta$. This analysis is described in
Appendix \ref{AppendixB}. It follows from the results found in
Appendix \ref{AppendixB} that only
$(\alpha,\beta,\gamma,\eta)=(\pi,\pi,\pi,\pi)$, $(0,0,0,\pi)$,
$(0,0,\pi,0)$, $(0,\pi,0,0)$ and $(\pi,0,0,0)$ can correspond to
minima of \meff. These minima take place in intervals of values of
$m_1$ which are determined by the following sets of inequalities:
\be
\begin{split}
 (\alpha,\beta,\gamma,\eta) & = (\pi,\pi,\pi,\pi) \quad if
\quad  F_1 = a - b - c - d - e > 0,\\
 (\alpha,\beta,\gamma,\eta) &= (0,0,0,\pi)  \quad if
\quad  ( d < e) \land (c < e - d) \land (b < - c - d + e)\land\\
 & \phantom{,aaaaaaaaaaaaaaa}\land  (F_8 = a + b + c + d - e  < 0), \\
(\alpha,\beta,\gamma,\eta) &= (0,0,\pi,0)  \quad if \quad   (d > e) \land  (c < d - e)  \land  (b  < - c + d - e) \land\\
 &  \phantom{,aaaaaaaaaaaaaaa}\land  (F_3 = a + b + c - d + e < 0),\\
 (\alpha,\beta,\gamma,\eta) &= (0,\pi,0,0)  \quad if \quad  (c > d + e)  \land  (b  <  c - d - e) \land \\
 & \phantom{,aaaaaaaaaaaaaaa} \land  (G_3 = a + b - c + d + e < 0),\\
 (\alpha,\beta,\gamma,\eta) &= (\pi,0,0,0)  \quad if \quad   (b > c + d + e) \land  (F_6 = a - b + c + d + e < 0)\,.
\label{cond3+2}
\end{split}
\ee
%
The dependence of $F_1$, $F_8$, $F_3$, $G_3$, $F_6$ and $(d-e)$
on $m_1$ is shown in the right panel of
Fig. \ref{fig:SylvesterNH3plus2}.

  It is not difficult to check that for the values of the
oscillation parameters quoted in Tables \ref{tabNudata} and
\ref{tabNudata3plus2}, the sets of inequalities
listed above in each of the cases of
$(\alpha,\beta,\gamma,\eta)=(0,0,\pi,0)$, $(0,\pi,0,0)$ and
$(\pi,0,0,0)$ cannot simultaneously be fulfilled
for $m_1 \geq 0$. Thus, only
$(\alpha,\beta,\gamma,\eta)=(\pi,\pi,\pi,\pi)$ and
$(0,0,0,\pi)$ correspond to true minima of \meff.
Defining $\ov m_1$ and $\un{m}_1$ as the zero of
the functions $F_1$ and  $F_8$,
\be
F_1(\ov m_1) = a - b - c - d - e = 0, \qquad
F_8(\un{m}_1) = a + b + c + d - e = 0\,,
\label{eq:F1F8}
\ee
%
we find that the minima of \meff for $m_1 > \ov m_1$ take place
only at $(\alpha,\beta,\gamma,\eta)=(\pi,\pi,\pi,\pi)$, while for
$m_1 <\un{m}_1$ they occur at $(\alpha,\beta,\gamma,\eta)=(0,0,0,\pi)$.
Further, the numerical analysis performed by us shows that
in the interval of $\ov m_1<m_{min}<\un{m}_1$,
the minimum value of \meff is exactly zero and is reached,
in general,  for CP nonconserving values of the phases
$(\alpha,\beta,\gamma,\eta)$.
These results are presented graphically in the left panel of
Fig. \ref{fig:SylvesterNH3plus2}.
Figure \ref{fig:SylvesterNH3plus2} shows also,
in particular, that at $m_{min}\rightarrow 0$
we have $\meff \neq 0$.
%

\begin{figure}[H]
\begin{center}
\includegraphics[scale=0.8]{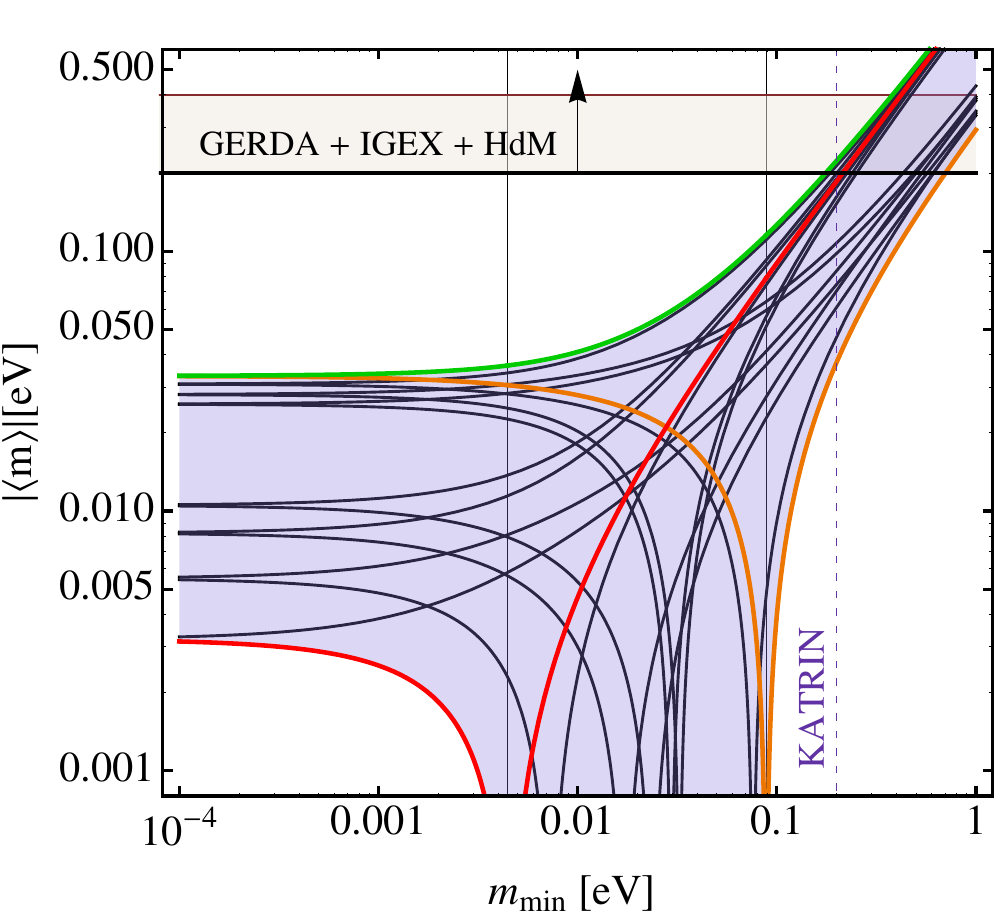}
\caption{The value of \meff versus the lightest neutrino mass
$m_1$ for $\Dmq_{41} =  0.47 \, {\rm eV^2}$, $\Dmq_{51} =  0.87 \, {\rm eV^2}$,
$\sin \theta_{14} = 0.13$, $\sin \theta_{15} = 0.14$.
The green, red and orange lines correspond to
$(\alpha,\beta,\gamma,\eta) =
(0,0,0,0),(0,0,0,\pi),(\pi,\pi,\pi,\pi)$,
while the blue lines  are obtained for the other
13 sets of CP conserving values (0 or $\pi$)
of the four CPV phases.
The vertical solid lines
are at $m_1 = \un{m}_1 \simeq 0.004\, {\rm eV}$
and $m_1 = \ov m_1\simeq 0.088 \,{\rm eV}$.
The minimum of \meff in the interval
$ \un{m}_1 \leq m_1 \leq \ov m_1$ is exactly zero.
The horizontal band indicates the upper bound
$\meff  \sim0.2-0.4$ eV obtained using the 90 \% C.L.
limit on the half-life of $^{76}$Ge
reported in \cite{Agostini:2013mba}.
See text for further details.
}
\label{NH3+2}
\end{center}
\end{figure}

 In Fig. \ref{NH3+2} we show
\meff as a function of the lightest neutrino mass $m_{min}$.
The shaded area indicates the allowed values for \meff. The red,
orange, green and gray lines correspond to the different sets of CP
conserving values (0 or $\pi$) of the CPV phases
$(\alpha,\beta,\gamma)$. The vertical solid lines are at $m_1 =
\un{m}_1 \simeq 4.44\times 10^{-3}\, {\rm eV}$ (and
$(\alpha,\beta,\gamma,\eta)=(0,0,0,\pi)$) and $m_1 = \ov m_1\simeq
8.84\times 10^{-2} \,{\rm eV}$ (and
$(\alpha,\beta,\gamma,\eta)=(\pi,\pi,\pi,\pi)$). It is clear from
the figure that \meff can be zero in the interval $\un{m}_1\leq
m_{min} \leq \ov m_1$, while for $m_{min}\rightarrow 0$ we have
$\meff_{min} \rightarrow 3.21\times 10^{-3}$ eV and ${\rm
max}(\meff) = 0.033$ eV. The indicated $\meff_{min}$ and ${\rm
max}(\meff)$ values at $m_{min}= 0$ in Fig. \ref{NH3+2} are reached
for $(\alpha,\beta,\gamma,\eta)=(0,0,0,\pi)$ and
$(\alpha,\beta,\gamma,\eta)=(0,0,0,0)$ (corresponding to the red and
green lines). At  $m_1 = \ov m_1$ and $m_1 = \un{m}_1$, we have
$\meff_{min}=0$: at $m_1 = \un{m}_1$ the first four terms in the
expression for \meff are positive and their sum is compensated by
the last term, $\sqrt{\Dmq_{51}}\sin^2\theta_{15}$, while at $m_1 =
\ov m_1$ a cancellation occurs between the first term proportional
to $m_{min}$ and the sum of all the other terms. We have also
indicated in the figure with a dotted vertical line the prospective
constraint on $m_{min}$ that might be obtained in the $\beta$-decay
experiment KATRIN \cite{MainzKATRIN}. We find that in 3+2 NO scheme
under discussion one always has
\begin{itemize}
\item
$\meff > 0.01$ eV for $m_{min} > 0.118$ eV.
\end{itemize}

%
\subsection{The Case of $m_{1} =0$}
%

In the case of $m_{min}\equiv m_1=0$,
the expression of \meff symplifies to:
\be
 \meff^2 \bigg|_{m_1 = 0} = |e^{i \alpha } b_0+  e^{i \beta } c_0+ e^{i \gamma } d_0 + e^{i \eta} e_0 |^2  \,\,,
\ee
%
where the parameters $b_0,c_0,d_0$ and $e_0$ read:
\be
\begin{split}
b_0 & =
\sqrt{\Delta m^2_{21}} c_{13}^2 c_{14}^2 s_{12}^2\,,\\
c_0& =   \sqrt{\Delta m^2_{31}} c_{14}^2 c_{15}^2 s_{13}^2 \,,\\
d_0 & = \sqrt{ \Delta m^2_{41}}  c_{15}^2 s_{14}^2 \,, \\
e_0 & =  \sqrt{ \Delta m^2_{51}}  s_{15}^2 \,.\\
\end{split}
\ee
%

\begin{figure}[h!]
       \begin{center}
      \subfigure
    {\includegraphics[width=8cm]{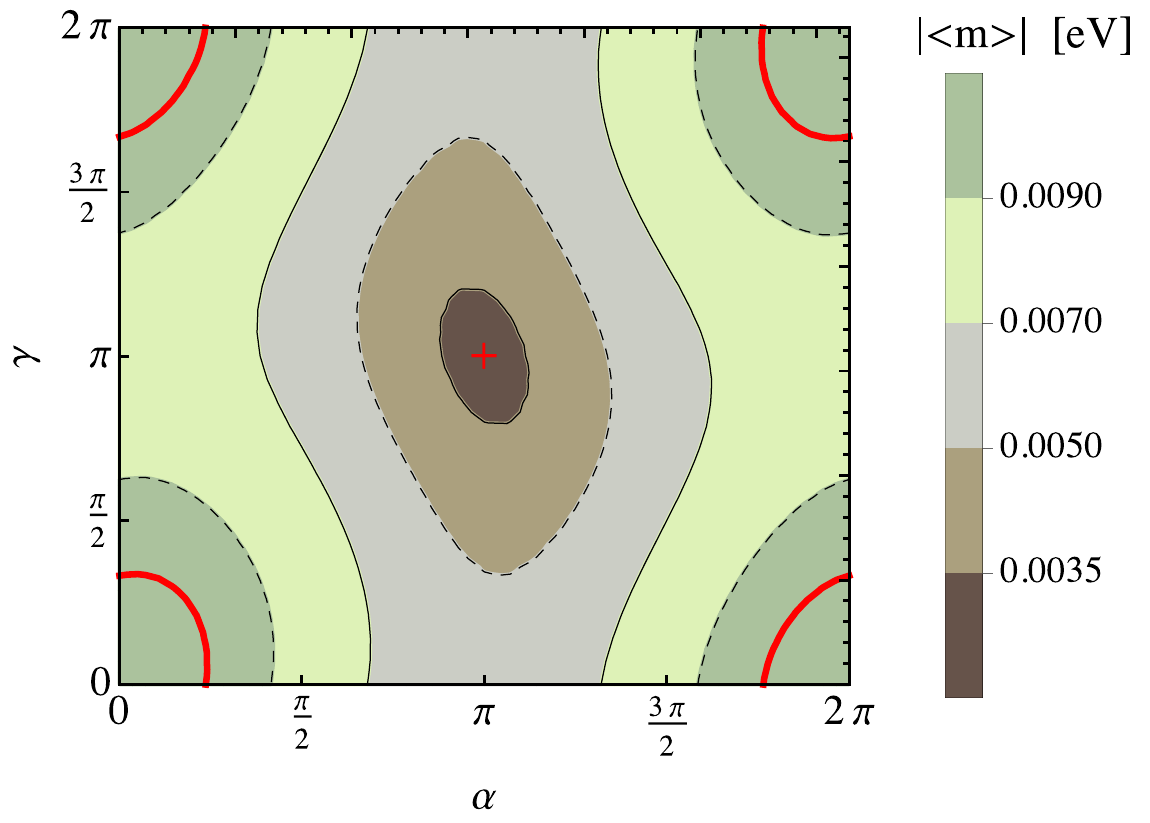}} \qquad \qquad
 \vspace{5mm}
     \end{center}
\vspace{-1.0cm} \caption{\label{fig:minNH2plus2}
The value of \meff in the 3+2 scheme with NO spectrum
at $m_{min}=0$ for $\gamma - \eta = \pi$.
The minimum of \meff corresponds to ($\alpha -
\eta,\beta - \eta$) =($\pi,\pi$) (the crossed point).
The values of \meff at this minimum is $3.21\,  \times 10^{-3}$ eV.
The red line corresponds to $\meff = 0.01$ eV.
 }
\end{figure}

The minimum of the effective Majorana mass is reached in this case
for $(\alpha,\beta,\gamma,\eta) = (0,0,0,\pi)$ and at the minimum
$\meff \neq 0$. Indeed, numerically we have
$b_0 \simeq 2.51 \times 10^{-3}$ eV,
$ c_0 \simeq 1.14 \times 10^{-3}$ eV,
$d_0\simeq 1.13 \times 10^{-2}$ eV and
$ e_0 \simeq 1.82 \times 10^{-2}$ eV, and it is
clear that the four terms in the expression for \meff
cannot compensate each other completely.
For the minimum value of \meff in the case under study we get
$\meff = 0.0032$ eV.
In Fig. \ref{fig:minNH2plus2}
we show the values of \meff
versus  $\alpha -\eta$ and $\beta - \eta$,
fixing for convenience
$\gamma - \eta = \pi$. The minimum is at the
crossed point corresponding to
at $\alpha -\eta = \pi$, $\beta - \eta = \pi$.
It follows from our analysis that in the case of $m_1 = 0$ and $\gamma - \eta = \pi$ we have
 $\meff > 0.01$ eV for values of the phases $\alpha - \eta$ and $\beta -\eta$ in the
 region delimited by the red lines in Figure \ref{fig:minNH2plus2}.

%
\section{The 3+2 Scheme with IO Spectrum}
%

\begin{figure}[h!]
\unitlength=1mm
\begin{center}
\begin{picture}(100,40)
\put(20,2){\line(1,0){20}}
\put(20,10){\line(1,0){20}}
\put(20,13){\line(1,0){20}}
\put(20,25){\line(1,0){20}}
\put(20,35){\line(1,0){20}}
\put(30,2){\line(0,1){33}}
\put(45,10){\vector(0,-1){8.5}}
\put(45,10){\vector(0,1){3.5}}
\put(45,13.5){\vector(0,-1){4.2}}
\put(82,5){\vector(0,1){30}}
\put(82,35){\vector(0,-1){34}}
\put(65,9){\vector(0,1){15}}
\put(65,13){\vector(0,-1){12}}
\put(48,11){$\Delta m^2_{21}$}
\put(48,6){$|\Delta m^2_{32}|$}
\put(67,12){$\Delta m^2_{43}$}
\put(84,18){$\Delta m^2_{53}$}
\put(17,2){3}
\put(17,9){1}
\put(17,12.3){2}
\put(17,24.2){4}
\put(17,34.2){5}
\end{picture}
\end{center}
\caption{\label{specIH3plus2} The neutrino mass spectrum in the 3+2 IO scheme.}
\end{figure}
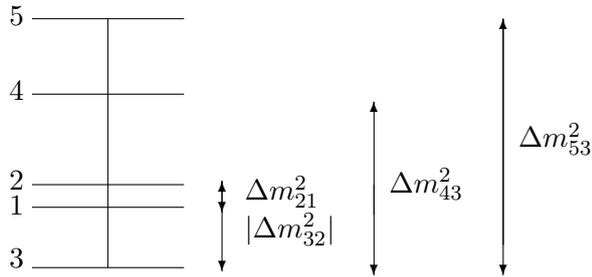

In the case of the 3+2 scheme with IO spectrum,
$m_3 < m_1 < m_2 < m_4 < m_5$, \meff can be written as:
\be
\meff = | m_1 c_{12}^2 c_{13}^2 c_{14}^2 c_{15}^2 + m_2  e^{i\alpha} c_{13}^2 c_{14}^2 c_{15}^2 s_{12}^2 +   m_3  e^{i\beta} c_{14}^2 c_{15}^2 s_{13}^2 +  m_4 e^{i\gamma}  c_{15}^2 s_{14}^2  + m_5 e^{i\eta}  s_{15}^2 |\,. \\
\label{meffIOr}
\ee
%
We have:
\be
\begin{split}
m_1 & = \sqrt{m_3^2 - \Delta m^2_{32} - \Delta  m^2_{21} },
\quad
m_2 = \sqrt{m_3^2 - \Delta m^2_{32}  },
\quad
m_3 \equiv m_{min},
\quad
m_{4} = \sqrt{m_3^2 + \Delta m^2_{43} }, \\
m_{5} & = \sqrt{m_3^2 + \Delta m^2_{53} },
\quad
\Delta m^2_{21} > 0\,, \ \
\Delta m^2_{32} < 0\,, \ \
\Delta m^2_{43} > 0\,, \qquad    \Delta  m^2_{53} > 0 \,. \\
\end{split}
\ee
%
The neutrino mass spectrum in 3+2 IO scheme is
shown schematically in Fig. \ref{specIH3plus2}.
%

We define:
\be
 \meff^2 = |a+e^{i \alpha } b+e^{i
\beta } c+e^{i \gamma } d + e^{i \eta} e |^2  \,\,,
\label{eqIOabcde}
\ee
%
where the parameters $a$, $b$, $c$, $d$ and $e$ in this case read:
\be
\begin{split}
a &= \sqrt{m_{min}^2 + |\Delta m^2_{32}| - \Delta  m^2_{21} } c_{12}^2 c_{13}^2 c_{14}^2 c_{15}^2 \,,\\
b & = \sqrt{m_{min}^2 + |\Delta m^2_{32}|  }  c_{13}^2 c_{14}^2 c_{15}^2 s_{12}^2 \,, \\
c & =  m_{min} c_{14}^2 c_{15}^2 s_{13}^2 \,,\\
d & = \sqrt{m_{min}^2 + \Delta m^2_{43} }  c_{15}^2 s_{14}^2 \,, \\
e & = \sqrt{m_{min}^2 + \Delta m^2_{53} } s_{15}^2 \,.\\
\end{split}
\ee
%

\begin{figure}[h!]
  \begin{center}
 \includegraphics[width=7.55cm]{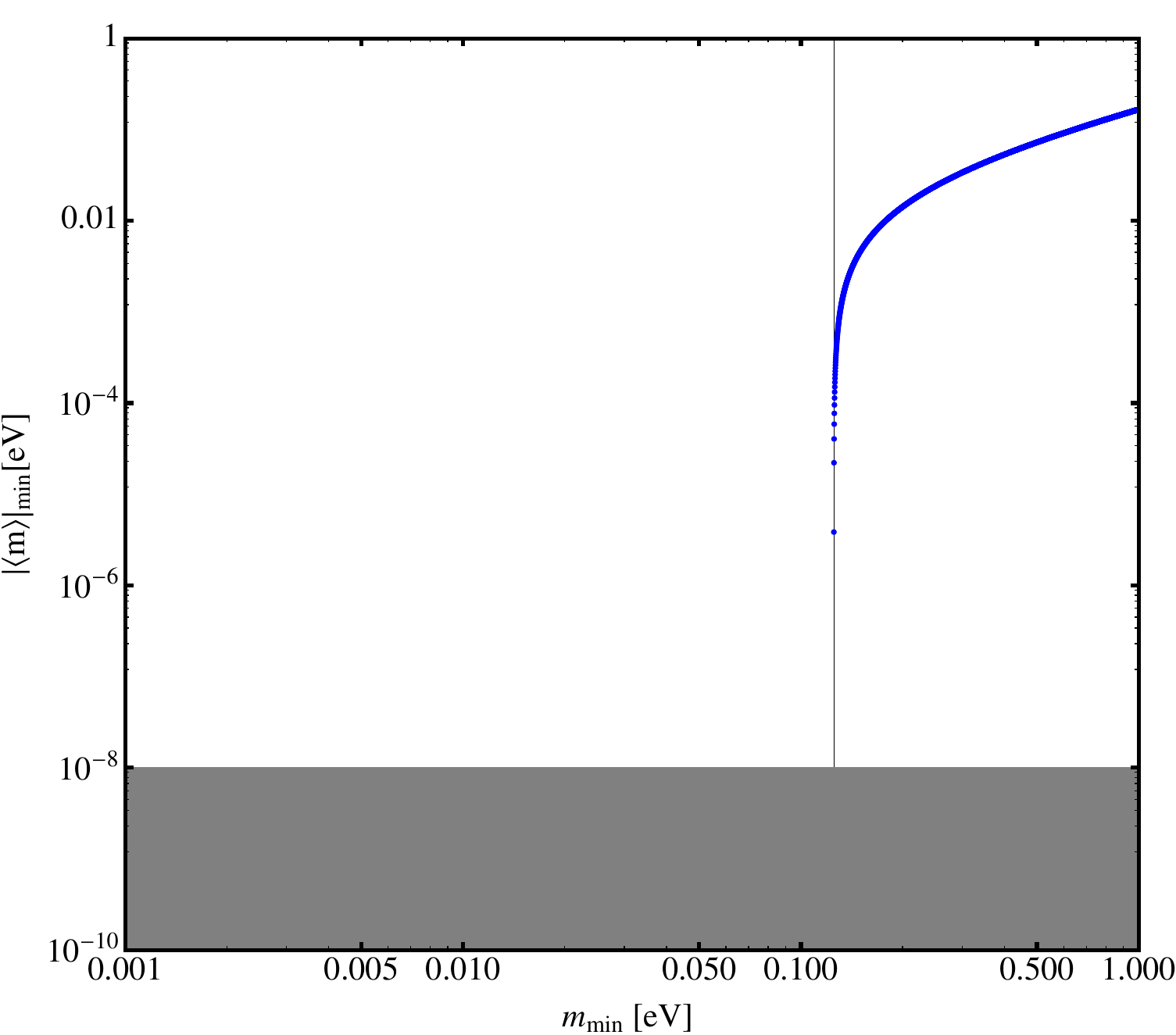} 
     \end{center}
\caption{\label{fig:SylvesterIH3plus2A}
Minimum \meff as a function of $m_{min}
\equiv m_{3}$. The plot has been obtained numerically for $\Dmq_{43}
=  0.47 \, {\rm eV^2}$, $\Dmq_{53} =  0.87 \, {\rm eV^2}$,
$\sin\theta_{14} = 0.13$, $\sin\theta_{15} = 0.14$ (see Table
\ref{tab:nu2}) by varying $m_{min}$ in the interval [0.001, 1.000]
eV and each of the four CPV  phases in the interval $[0,2\pi]$.
The vertical line corresponds to $\ov m_3 \simeq 1.25 \times 10^{-1} $ eV
($F_1(\ov m_3)=0$).
For any given $m_3 \leq \ov m_3$ we have  $\meff_{min} = 0$.
For the different  $m_3 < \ov m_3$, the
minima of \meff occur at different sets of
CP nonconserving values of   $(\alpha,\beta,\gamma,\eta)$
(see text for further details).
 }
\end{figure}

 As in the case of NO spectrum, we have performed the
general  analysis of minimization of $\meff$ numerically. Analytical
results have been obtained only for the CP conserving values (0 or
$\pi$) of the four CPV  phases. As it follows from the analysis
performed in Appendix \ref{AppendixB}, only one set of
 CP conserving values of the phases corresponds
to a minimum of \meff, namely,
 $(\alpha,\beta,\gamma,\eta) = (\pi,\pi,\pi,\pi)$.
The domain of this minimization solution is determined by the
inequality $F_1(m_3) \equiv (a - b - c - d - e) > 0$. Let us define
by  $\ov m_3$ the zero of $F_1$: $F_1(\ov m_3) = 0$. As can be
shown (and is seen also in Fig.
\ref{fig:SylvesterIH3plus2} in Appendix \ref{AppendixB}),
the inequality of interest  $F_1(m_3) >
0$ is satisfied for $m_3 > \ov m_3$. Thus, for $m_3 > \ov m_3$,
\meff takes minimum values only for the values of the phases
$(\alpha,\beta,\gamma,\eta) = (\pi,\pi,\pi,\pi)$. Moreover, the
minima of \meff  at $m_3 > \ov m_3$ are different from zero. This
follows from the fact that the minima under discussion correspond to
the contribution of the first term $\propto \sqrt{m_{min}^2 +
|\Delta m^2_{32}| - \Delta  m^2_{21} }$ in the expression for \meff,
eq. (\ref{meffIOr}), being compensated by the sum of the other terms
in \meff, and that for the values of the oscillation parameters used
in the present analysis the compensation cannot be complete. For the
indicated values of the phases, a complete compensation leading to
$\meff_{min} = 0$ is possible only in the point $m_3 = \ov m_3$. At
any given $m_3 < \ov m_3$, as our numerical analysis shows, we have
$\meff_{min} =0$ and the minimum takes place, in general, for CP
nonconserving values of $(\alpha,\beta,\gamma,\eta)$. These results
are illustrated in Fig.
\ref{fig:SylvesterIH3plus2A}, where we show $\meff_{min}$ as function
of the lightest neutrino mass.
%

\begin{figure}[h!]
\begin{center}
\includegraphics[scale=0.8]{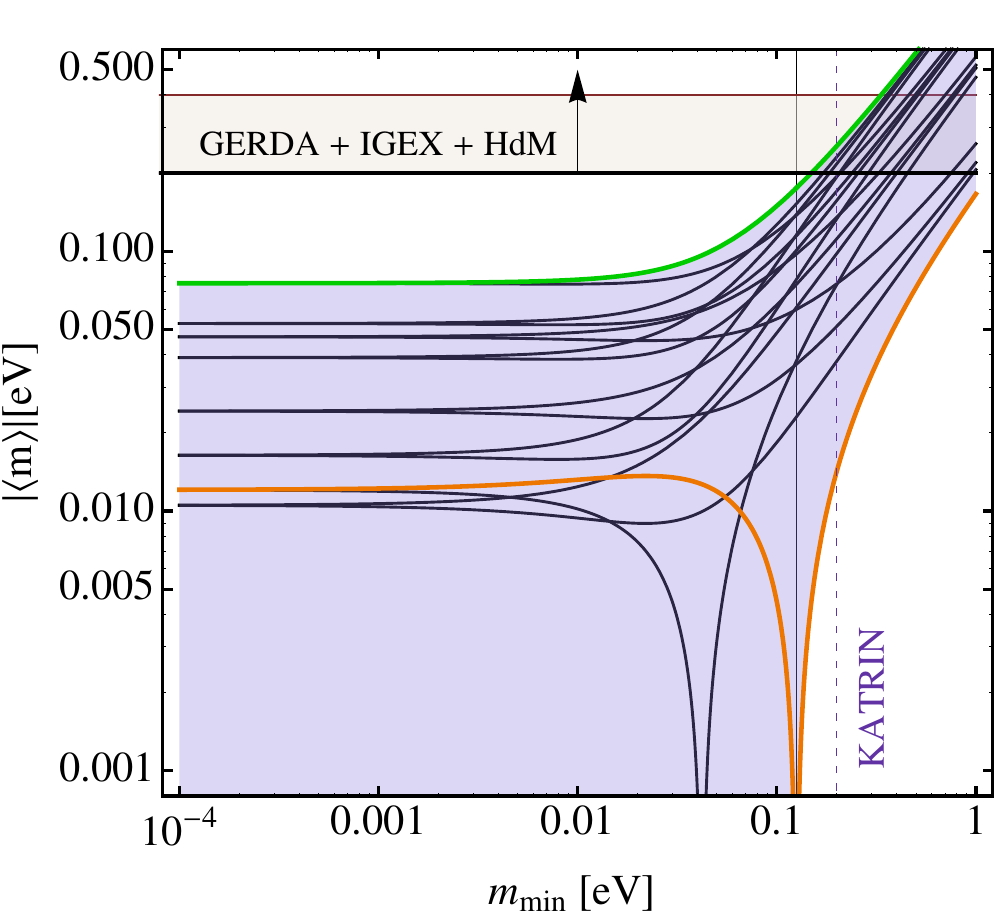}
\caption{The value of \meff as function of
$m_{min} = m_3$ for
$\Dmq_{43} =  0.47~{\rm eV^2}$,
$\Dmq_{53} =  0.87~{\rm eV^2}$,
$\sin\theta_{14} = 0.13$, $\sin\theta_{15} = 0.14$ (see Table \ref{tab:nu2}).
The green and orange lines correspond to
$(\alpha,\beta,\gamma) = (0,0,0,0)$ and $(\pi,\pi,\pi,\pi)$,
while the gray lines corresponding to the other
14 sets of CP conserving values (0 or $\pi$)
of the four CPV phases. The vertical solid line corresponds
to $m_{min}  = \ov m_3\simeq 0.125$ eV.
At $m_{min}\leq \ov m_3$ we can have $\meff_{min}=0$.
The  dotted line represents
the prospective upper limit from
the $\beta$-decay experiment
KATRIN \cite{MainzKATRIN}.
The horizontal band indicates the upper bound
$\meff  \sim0.2-0.4$ eV obtained using the 90 \% C.L.
limit on the half-life of $^{76}$Ge given
in \cite{Agostini:2013mba}.
}
\label{IH3+2}
\end{center}
\end{figure}

 In Fig. \ref{IH3+2} we show \meff as function of the
lightest neutrino mass $m_{min}= m_3$.
The gray lines correspond to \meff computed for
$CP$ conserving values of the phases
$(\alpha,\beta,\gamma,\eta)$ (either 0 or $\pi$).
The shaded area indicates the possible allowed
values of  \meff and is obtained for
the values of the oscillation parameters
quoted in Tables \ref{tabNudata} and \ref{tab:nu2}.
The vertical solid line corresponds to
$m_{min} = \ov m_3\simeq 0.125$ eV
and  $(\alpha,\beta,\gamma,\eta) = (\pi,\pi,\pi,\pi)$.
At  $m_{min} \leq \ov m_3$, we can have $\meff_{min} = 0$
for specific, in general CP nonconserving, values of
the phases $(\alpha,\beta,\gamma,\eta)$.
This behaviour of $\meff_{min}$
is very different from the behavior in
the case of  NO spectrum discussed in
the previous Section, where $\meff_{min}$
can be zero only in a limited interval
of values of  $m_{min}$.

We find also that in the 3+2 IO scheme under discussion
and the values of the neutrino oscillation parameters used in
the present analysis one always has
\begin{itemize}
\item 
$\meff > 0.01$ eV for $m_{min} > 0.178$ eV.
\end{itemize}
%
\subsection{The case of $m_3 =0$}
%

  In the limit $m_{min}\equiv m_3 = 0$,
(which implies $c = 0$ in eq. (\ref{eqIOabcde})), the analysis of
the minimization of the effective Majorana mass is exactly the same
as in the case of the 3+1 scheme. This becomes clear after a
redefinition of the phases and the coefficients involved. For
$m_{min}\equiv m_3 = 0$, $\meff^2$ can be written as:
\be
\meff^2 \bigg|_{m_3 = 0} =
|a_0+e^{i \alpha } b_0+e^{i \gamma } d_0 + e^{i \eta} e_0 |^2  \,\,,
\ee
%
where
\be
\begin{split}
a_0 &= \sqrt{|\Delta m^2_{32}| - \Delta  m^2_{21} }\,
c_{12}^2 c_{13}^2 c_{14}^2 c_{15}^2 \,,\\
b_0 & = \sqrt{|\Delta m^2_{32}|  }\, c_{13}^2 c_{14}^2 c_{15}^2 s_{12}^2 \,, \\
d_0 & = \sqrt{\Delta m^2_{43} } \,  c_{15}^2 s_{14}^2 \,, \\
e_0 & =  \sqrt{\Delta m^2_{53} } \, s_{15}^2 \,.\\
\end{split}
\label{32IOm30abcd}
\ee
%
Using the analysis performed in  Appendix \ref{AppendixA}
for the 3+1 scheme we find that the solutions which minimize \meff,
such that $\meff_{min}$ is exactly zero, are:
$(u_{\pm},v_{\pm},t_{\pm} )$$\equiv$
$(\tan (\gamma_{\pm}/2),$ $\tan (\alpha_{\pm}/2),$ $\tan (\eta_{\pm}/2) )$,
$(u^{\pm}_3,v^{\pm}_{3},t^{\pm}_{3})$ $\equiv$
$(\tan (\gamma^{\pm}_3/2) ,\tan (\alpha^{\pm}_3/2) ,\tan (\eta^{\pm}_3/2) )$,
and $(v^{\pm}_{4}(u),t^{\pm}_{4}(u))
\equiv(\tan (\alpha^{\pm}_{4}/2 ),\tan (\eta^{\pm}_{4}/2))$.
The solutions $(u_{\pm},v_{\pm},t_{\pm} )$ and
$(v^{\pm}_{4}(u),t^{\pm}_{4}(u))$ can be obtained formally
from eqs. (\ref{uvt}) and (\ref{v4t4}) by replacing,
respectively,  $a$, $b$, $c$ and $d$ with
$a_0$, $b_0$, $e_0$ and $d_0$
defined in eq. (\ref{32IOm30abcd}),
while the solution $(u^{\pm}_3,v^{\pm}_{3},t^{\pm}_{3})$ is given by:
%
\be
\begin{split}
\begin{cases}
u^{\pm}_3 & = \pm \dfrac{\sqrt{(-a_0+b_0-d_0+e_0) (a_0-b_0+d_0+e_0)}}{\sqrt{(a_0-b_0-d_0-e_0) (a_0-b_0-d_0+e_0)}} \,,\\
v^{\pm}_3 & = \pm \dfrac{a_0^2-b_0^2-d_0^2+e_0^2}{\sqrt{(a_0-b_0-d_0-e_0) (a_0-b_0-d_0+e_0)} \sqrt{(-a_0+b_0-d_0+e_0) (a_0-b_0+d_0+e_0)}} \,,\\
t^{\pm}_3 & = \pm \dfrac{(b_0+e_0) (-a_0+b_0+d_0-e_0) \sqrt{(-a_0+b_0-d_0+e_0) (a_0-b_0+d_0+e_0)}}{(b_0-e_0) \sqrt{(a_0-b_0-d_0-e_0) (a_0-b_0-d_0+e_0)} (-a_0+b_0-d_0+e_0)} \,. \\
\end{cases}
\label{sol2}
\end{split}
\ee
%
Using the best fit values $\Dmq_{43} =  0.47 \, {\rm eV^2}$,
$\Dmq_{53} =  0.87 \, {\rm eV^2}$, $\sin\theta_{14} = 0.13$,
$\sin\theta_{15} = 0.14$ we find that the minima corresponding to
$(u_{\pm},v_{\pm},t_{\pm} )$ and to eq. (\ref{sol2}) are given
numerically by:
 \be u_{\pm} \simeq \pm 1.44, \quad
v_{\pm} \simeq \pm 18.5, \quad
t_{\pm} \simeq \mp 2.61\label{solnum1}\ee
%
and
 \be u^{\pm}_3 \simeq \pm 1.30, \quad
v^{\pm}_3 \simeq \mp 2.63, \quad
t^{\pm}_3 \simeq \mp 24.7\,.\label{solnum2}
\ee
%
The third minimum
corresponding to the solution $(v^{\pm}_{4}(u),t^{\pm}_{4}(u))$ is
not determined uniquely since it depends on $u$. However, one can
define the minimum for a specific choice of $u$, or equivalently,
for a value for one of the other phases, because the expressions of
this solution are invertible. In order to check numerically whether
the three solutions are minima we plot the dependence of \meff on
two of the CPV phases  $(\alpha, \gamma, \eta)$, fixing the value of
the third phase. It proves convenient to set the value of $\eta$,
i.e. of $t$, equal to the values  of the solution
$(u_{\pm},v_{\pm},t_{\pm} )$.
One can, of course, do the same using the solutions
$(u^{\pm}_3,v^{\pm}_{3},t^{\pm}_{3})$, or choosing an arbitrary
value of $\eta$. Our aim is  to show that in the 3+2 IO scheme with
$m_3 =0$, the two solutions $(u^{+},v^{+},t^{+})$ and
$(v^{+}_{4}(u),t^{+}_{4}(u))$ represent two different minima, in
contrast to the case of 3+1 IO scheme with $m_3=0$.

\begin{figure}[h!]
  \begin{center}
 \subfigure
 {\includegraphics[width=7.65cm]{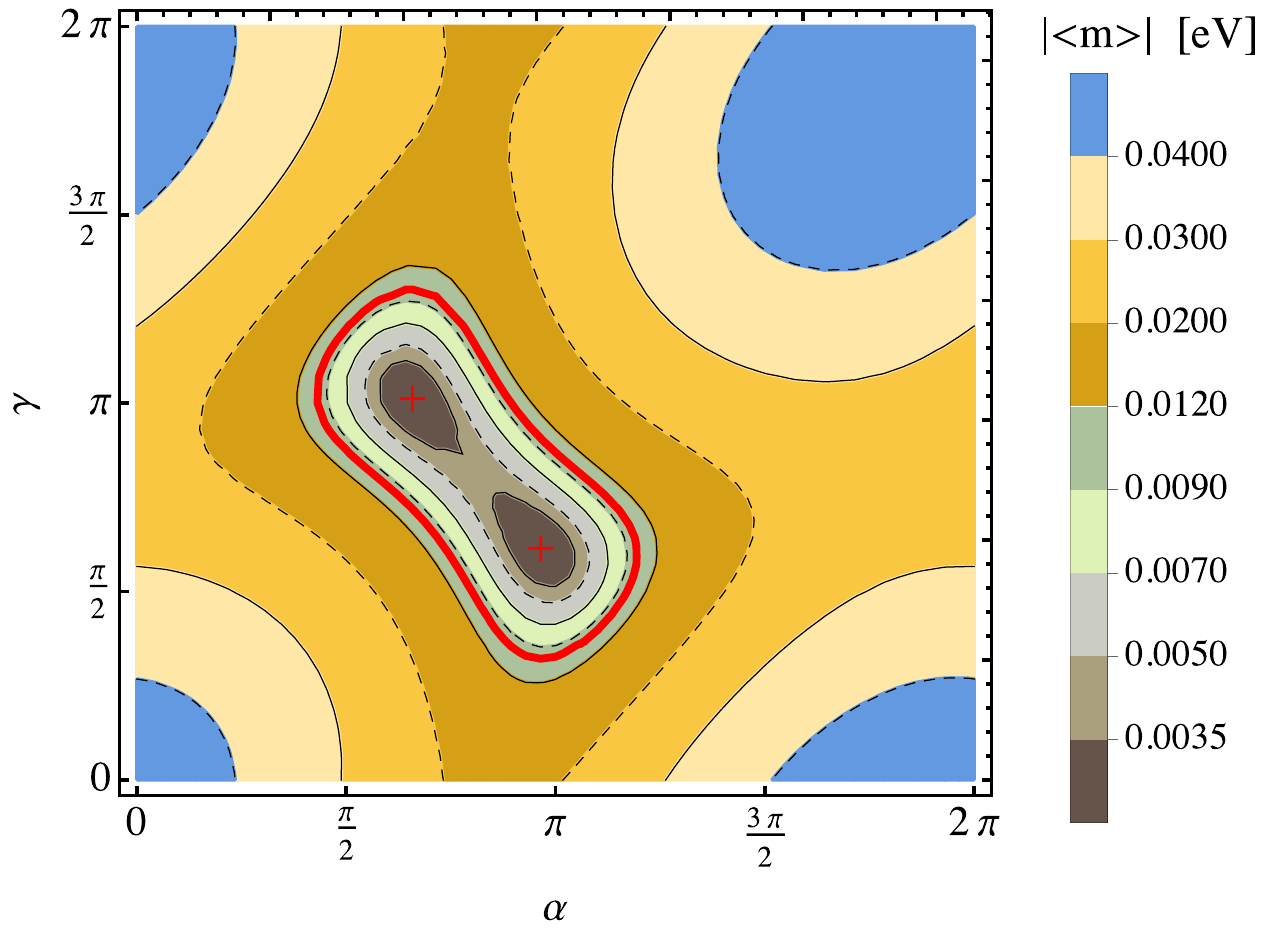}}
 \vspace{5mm}
 \subfigure
   {\includegraphics[width=7.65cm]{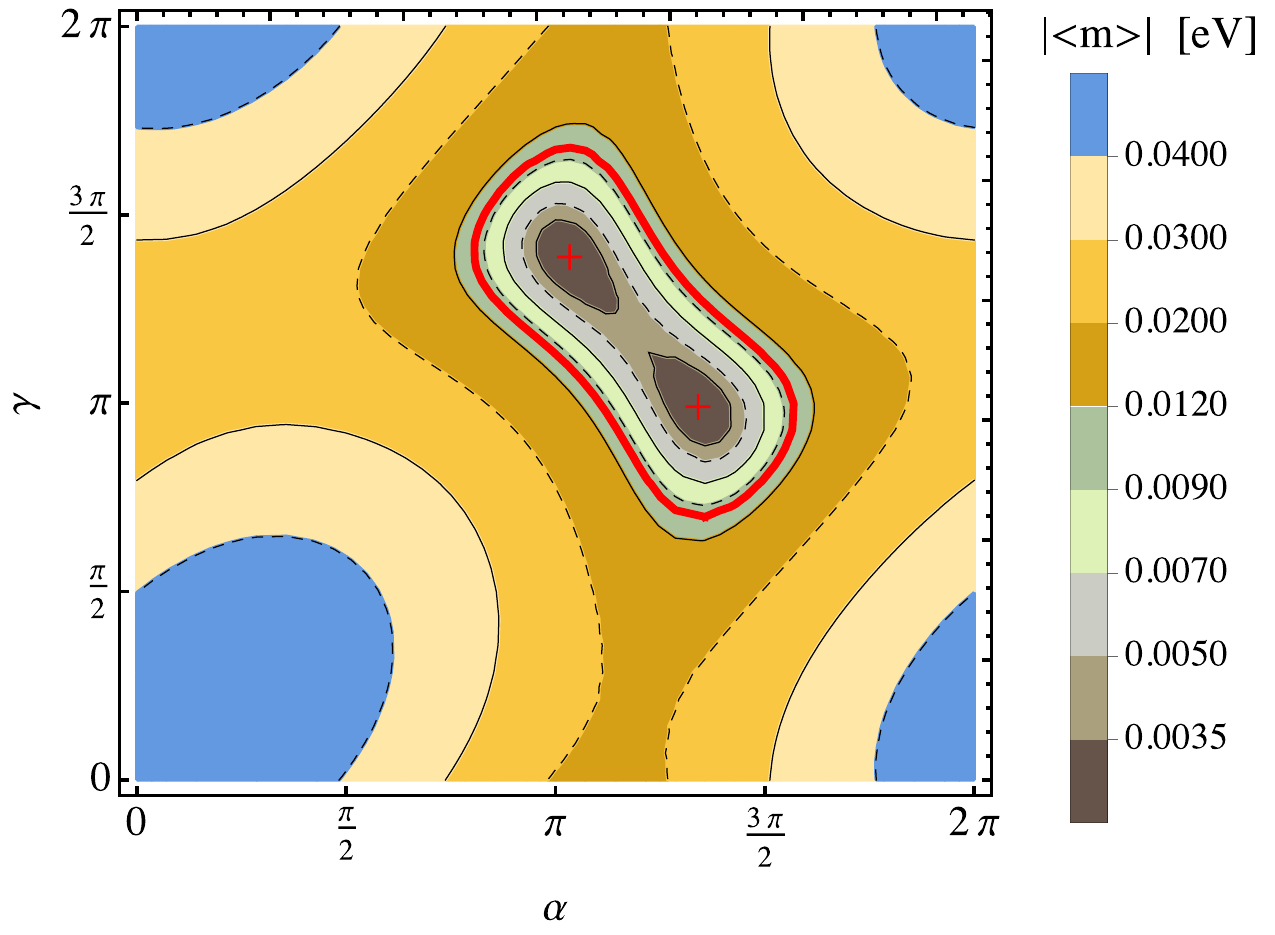}}
    \vspace{5mm}
     \end{center}
\vspace{-1.0cm} \caption{\label{fig:minIH3plus2}
Left Panel. The value of \meff in the
3+2 IO scheme for $m_3 \equiv m_{min} = 0$ and
the best fit values of
Table \ref{tabNudata3plus2}.
The phase $\eta$ is set to the value $\eta_+$,
$\tan \eta_+/2 \simeq -2.61$.
The values of $\meff$ in the marked points
$(\alpha, \gamma) = (3.03,1.93),(2.07,3.17)$,
are zero. 
The red contour line corresponds to $\meff = 0.01$ eV.
Right Panel. The same as in the left panel,
but setting $\eta$ to the value $\eta_-$,
$\tan \eta_-/2 \simeq 2.61$.
The values of $\meff$ in the marked points
$(\alpha, \gamma) =  (3.25,4.35),(4.21,3.11)$,
are zero. The red contour line corresponds to
$\meff = 0.01$ eV. See text for details.
 }
\end{figure}

More specifically, if we fix $t \simeq -2.61$,
we find $v^{+}_{4} \simeq 1.68$ at a value of $u^
{+}_4  \simeq -59.3$. The Left Panel of
Fig. \ref{fig:minIH3plus2} shows the values of
\meff with $t = t_+ \simeq -2.61$. The marked points correspond
to the two different minima: the first corresponds
to the solution $(u_+,v_+,t_+)$ and
takes place at $(\alpha, \gamma) = (3.03,1.93)$,
while the second one is associated with the
solution $(v^{+}_4(u),t^{+}_4(u))$ and occurs at
$(\alpha,\gamma) = (2.07,3.17)$.
In these two minima the effective Majorana mass is exactly zero.
Repeating the same analysis with $t  \simeq 2.61$, we
find $v^{-}_{4} \simeq -1.68$ at the value of $u^{-}_4 \simeq 59.3$.
The Right Panel of Fig. \ref{fig:minIH3plus2}
shows the values of $\meff$ with $t = t_- \simeq 2.61$.
The points marked with a cross correspond to the
two different minima,
one evaluated from the function $(u_-,v_-,t_-)$ and
corresponding to $(\alpha, \gamma) = (3.25,4.35)$,
and the second evaluated from the function
$(v^-_4(u),t^{-}_4(u))$ and corresponding to
$(\alpha,\gamma) = (4.21,3.11)$. As in the previous case,
in these two minima the effective Majorana mass is exactly zero.
The existence of two minima in the 3+2 scheme in the
limit of $m_3=0$ is very different from the
3+1 case where the two minima
coincide.

 Finally, it follows from our analysis that for $m_3 = 0$
in the cases we have considered and which are illustrated in
 Fig. \ref{fig:minIH3plus2} we have
 $\meff > 0.01$ eV for values of the phases $\alpha$ and $\gamma$
 outside the region delimited by the red line in Fig. \ref{fig:minIH3plus2}.

\section{Conclusions}

In the present article we have investigated the predictions for
neutrinoless double beta (\betabeta-) decay effective Majorana mass
$\meff$ in the $3+1$ and $3+2$ schemes with one and two additional
sterile neutrinos with masses at the eV scale. These two schemes are
widely used in the interpretation of the reactor neutrino and
Gallium anomalies as well as of the data of the LSND and MiniBooNE
experiments in terms of active-sterile neutrino oscillations. Due to
the assumed active-sterile neutrino mixing,
the ``$3 + 1$'' and ``$3 + 2$'' models have altogether 4 and 5 light
massive neutrinos $\nu_j$ coupled to the electron and muon in the
weak charged lepton current. In the minimal versions of these models
the massive neutrinos are Majorana particles. The additional
neutrinos $\nu_4$ and $\nu_4$, $\nu_5$, should have masses $m_4$ and
$m_4$, $m_5$ at the eV scale. It follows from the data that if
$\nu_4$ or $\nu_4$, $\nu_5$ exist, they should couple to the
electron and muon in the weak charged lepton current with couplings
$U_{e k}\sim 0.1$ and $U_{\mu k}\sim 0.1$, $k=4;~4,5$.

 As was shown in \cite{BPP2,SGWRode05} and more recently in
\cite{Barry:2011,Li:2012,Goswami:2013}, the contribution of the
additional light Majorana neutrinos $\nu_4$ or $\nu_{4,5}$ to the
$\betabeta$-decay amplitude, and thus to the $\betabeta$-decay
effective Majorana mass $\meff$, can change drastically the
predictions for $\meff$ obtained in the reference 3-flavour neutrino
mixing scheme, $|\mefff^{(3\nu)}|$. Using the values of the neutrino
oscillation parameters of the  $3+1$ and $3+2$ schemes, obtained in
the global analyses of the data relevant for the active-sterile
neutrino oscillation hypothesis (positive evidence and negative
results), performed in \cite{GiuntiTalk:2013vaa,Kopp:2013vaa} (see
Tables \ref{tabNudata}, \ref{tab:1nu} and \ref{tab:nu2}), we have
investigated in detail in the present article the possibility of a
complete or partial cancellation among the different terms in
$\meff$, leading to a strong suppression of $\meff$. This was done
in the $3+1$ and $3+2$ schemes both in the cases of 3-neutrino mass
spectra with normal ordering (NO) and inverted ordering (IO), as well
as in the cases of normal hierarchical (NH) and inverted
hierarchical (IH) spectra with ${\rm min}(m_j) = 0$, where
$j=1,2,3,4$ $(j=1,2,3,4,5)$ for the $3+1$ ($3+2$) scheme. In this
type of analysis the free parameters are the CP violation (CPV)
Majorana phases and the lightest neutrino mass. In the case of the
$3+1$ scheme, in which there are three physical CPV Majorana phases,
we have found all the solutions of the system of equations which
determine the minima of $\meff$ as well as their domains (i.e., the
regions of their validity), in analytic form. This was done for all
types of neutrino mass spectra we have considered. In the more
complicated case of $3+2$ scheme with four physical CPV Majorana
phases, the non-linearity of the system of four equations which
determine the extrema of \meff makes the analytical study of the
extrema of interest a complicated problem. Thus, in this case we have
performed the general analysis of the minimization of \meff
numerically. It was possible, however, to perform analytically the
analysis of the minima of \meff, corresponding to the 16 sets of CP
conserving values (either  $0$ or $\pi$) of the  four phases.

 We have found that if the neutrino mass spectrum is of the NO type,
we can have $\meff =0$, and thus strongly suppressed \meff, in a
specific interval of values of ${\rm min}(m_j) \equiv m_{min}$,
$\un{ m}_1  \leq m_{min} \leq \ov m_1$. This results is valid both
for the $3+1$ and $3+2$ schemes. The specific values of $\un{ m}_1$
and $\ov m_1$ depend on the scheme: they are determined by the
values of the oscillation parameters in each of the two schemes. For
the best fit values reported in Tables 1, 2 and 3, in the $3+1$ with
$\Dmq_{41} = 0.93 \, {\rm eV^2}$, $3+1$ with $\Dmq_{41} = 1.78 \,
{\rm eV^2}$ and $3+2$ schemes they read, respectively: $(\un{ m}_1,
\ov m_1) = (0.021,0.065)$ eV, $(\un{ m}_1, \ov m_1) =
(0.030,0.091)$ eV and $(0.004,0.088)$ eV.
For the different values of $m_{min}$ from the indicated interval,
the minimum $\meff =0$ is reached for different sets of CP
nonconserving, in general, values of the CPV Majorana phases.

  For the best fit values reported in Tables 1, 2 and 3, we find that
we always have $\meff>0.01$ eV,
\begin{itemize}
\item in the $3+1$ scheme with $\Delta m^2_{41} = 0.93~{\rm  eV^2}$ --
for  $m_{min} < 0.010$ eV and $m_{min} > 0.093$ eV;
\item in the $3+1$ scheme with  $\Dmq_{41} = 1.78~{\rm  eV^2}$ --
for $m_{min} < 0.020$ eV and  $m_{min} > 0.119$ eV;
\item in the $3+2$ scheme -- for $m_{min} > 0.118$ eV.
\end{itemize}

 The results we have obtained for IO spectrum are different. In this
case one can have $\meff =0$ in the interval  $m_{min} \leq \ov
m_3$, where  $\ov m_3$ is determined by the values of neutrino
oscillation parameters. For a given   $m_{min}$ from the indicated
interval, $\meff =0$ takes place for specific, in general, CP
nonconserving values of the relevant Majorana phases. The values of
$\ov m_3$ in the two schemes, $3+1$ and $3+2$, differ. Using the
values of the oscillation parameters given in Tables 1, 2 and 3, we
find: $\ov m_3 = 0.038\,(0.074)$ eV for
$\Dmq_{41} = 0.93\,(1.78) \, {\rm eV^2}$
in the $3+1$ scheme, and  $\ov m_3 = 0.125$ in the  $3+2$ scheme.

Using the values of the oscillation parameters given in Tables 1, 2
and 3,  we find also that one has always $\meff>0.01$ eV,
\begin{itemize}
\item in the $3+1$ scheme with $\Delta m^2_{43} = 0.93~{\rm  eV^2}$ --
for $m_{min} > 0.078$ eV;
\item in the $3+1$ scheme with $\Delta m^2_{43} = 1.78~{\rm  eV^2}$ --
for $m_{min} > 0.108$ eV;
\item in the $3+2$ scheme -- for $m_{min} > 0.178$ eV.
\end{itemize}

 We have investigated also the specific cases of NH and IH spectra
in the limit  $m_{min}=0$, which present certain
peculiarities both in the $3+1$ and $3+2$ schemes.

 The analysis performed by us allowed to derive the general
conditions under which the effective Majorana mass satisfies $\meff
> 0.01$ eV, and thus to determine the regions of values $m_{min}$
for which \meff is predicted to lie in  the range planned to be
explored by the next generation of $\betabeta$-experiments. The
results of these experiments will provide additional tests of the
hypothesis of existence of sterile neutrinos with masses at the eV
scale, and couplings $\sim 0.1$ to the electron and muon in the weak
charged lepton current.

\section*{Acknowledgements}

This work was supported in part by the INFN program on
``Astroparticle Physics'',
the World Premier International Research Center
Initiative (WPI Initiative), MEXT, Japan  (S.T.P.) and
by the European Union FP7-ITN INVISIBLES and UNILHC
(Marie Curie Action, PITAN-GA-2011-289442 and PITN-GA-2009-23792).

\clearpage

\appendix
\section{\texorpdfstring{The Extrema of \meff in the 3+1 Scheme
with NO or IO Neutrino Mass Spectrum}{The Extrema of the effective Majorana mass in the 3+1 Scheme
with NO or IO Neutrino Mass Spectrum}}
\label{AppendixA}

We are interested in the minima and the maxima of \meff. It turns
out to be somewhat simpler to study the extrema of $\meff^2$ which
obviously coincide with those of \meff. The expression for
$\meff^2$ in both the cases of NO and IO spectra can be written as:
\be \meff ^2= |a+e^{i \alpha } b+
e^{i\beta } c+e^{i \gamma } d|^2.
\ee
%
The zeros of the first derivatives of
$\meff^2$ with respect to the phases $\alpha$, $\beta$ and
$\gamma$ are  given by the following system of three
equations:
\be
\begin{split}
-& 2b [a \sin (\alpha )+c \sin (\alpha -\beta )+d \sin (\alpha -\gamma )] = 0,\\
-& 2c [a \sin (\beta )-b \sin (\alpha -\beta )+d \sin (\beta -\gamma )] = 0,\\
& 2d [-a \sin (\gamma )+b \sin (\alpha -\gamma )+c \sin (\beta -\gamma )] = 0.\\
\end{split}
\label{genminmax}
\ee
%
In order to solve this system we use the
following parametrization:
\be
\begin{split}
& \sin \alpha = \frac{2v}{1 + v^2}, \qquad
\cos \alpha = \frac{1-v^2}{1+v^2},\\
& \sin \beta = \frac{2t}{1 + t^2}, \qquad
\cos \beta = \frac{1-t^2}{1+t^2},\\
& \sin \gamma = \frac{2u}{1 + u^2}, \qquad
\cos \gamma = \frac{1-u^2}{1+u^2}.\\
\end{split}
\ee
%
where, respectively, $v \equiv \tan(\alpha/2)$, $t \equiv\tan(\beta/2)$, $u
\equiv \tan(\gamma/2)$ with $\alpha$, $\beta$, $\gamma \neq \pi + 2k\pi$.
In terms of the new variables the system in eq.(\ref{genminmax})
can be written as:
\be
\begin{split}
-& \frac{2b}{ 1 + v^2} \bigg [ 2 a v-\frac{2 c t
\left(1-v^2\right)}{t^2+1}+\frac{2 c \left(1-t^2\right)
v}{t^2+1}-\frac{2 d u \left(1-v^2\right)}{u^2+1}+\frac{2
   d \left(1-u^2\right) v}{u^2+1} \bigg] = 0,\\
-& \frac{2c}{ 1 + t^2} \bigg [ 2 a t-\frac{2 b \left(1-t^2\right)
v}{v^2+1}+\frac{2 b t \left(1-v^2\right)}{v^2+1}-\frac{2 d
\left(1-t^2\right) u}{u^2+1}+\frac{2
   d t \left(1-u^2\right)}{u^2+1} \bigg] = 0,\\
 &\frac{2d}{ 1 + u^2} \bigg[ -2 a u+\frac{2 b \left(1-u^2\right) v}{v^2+1}
-\frac{2 b u \left(1-v^2\right)}{v^2+1}
+\frac{2 c t \left(1-u^2\right)}{t^2+1}-
\frac{2c \left(1-t^2\right) u}{t^2+1} \bigg] = 0.\\
\end{split}
\label{utvgen}
\ee
%
The new coordinates $u$, $v$ and $t$ are  singular if at least
one angle $\alpha, \beta, \gamma$ is equal to $\pi$. We observe that
seven  solutions of the system in eq. (\ref{genminmax})
are given for one of the three phases equal to $\pi$, i.e.,
for $u$ or $v$ or $t$ going to $\infty$:
\be
\begin{split}
\alpha & = \beta = \pi(0) \quad \mbox{and} \quad \gamma = 0(\pi), \\
\alpha & = \gamma = \pi(0) \quad \mbox{and} \quad \beta = 0(\pi), \\
\gamma & = \beta = \pi(0) \quad \mbox{and} \quad \alpha = 0(\pi), \\
 \alpha & = \beta = \gamma = \pi \,\,.\\\label{solpi}
\end{split}
\ee
%
We can recover this type of solutions as a limit of the system in
eq. (\ref{utvgen}) when a pair of variables $u$, $v$, $t$ are equal.
For example, in the limit in which $t = v = 0$, the system
in eq. (\ref{utvgen}) is reduced to
\be
\begin{split}
\begin{cases}
& (4 b d ) \dfrac{u}{u^2+1} = 0, \\
& (4 c d) \dfrac{ u}{u^2+1} = 0,\\
& [4 d  (a+b+c)] \dfrac{u}{u^2+1} = 0\,.\\
\end{cases}
\end{split}
\ee
%
Evidently, we have a solution in the limit
$u \rightarrow \infty$. This is equivalent to say that
the solutions under discussion
can be found as a limit of the system (\ref{utvgen})
when the variables $u$, $t$, and $v$ are
sent to $\infty$. \\

The solutions of the system in eq. (\ref{utvgen}),
assuming $\alpha, \beta,
\gamma \neq \pi$ and $b, c, d \neq 0$,
are the zeros of the following system of equations:
\be
\begin{split}
\begin{cases}
&  2 a v-\dfrac{2 c t \left(1-v^2\right)}{t^2+1}+\dfrac{2 c
\left(1-t^2\right) v}{t^2+1}-\dfrac{2 d u
\left(1-v^2\right)}{u^2+1}+\dfrac{2
   d \left(1-u^2\right) v}{u^2+1} = 0,\\
& 2 a t-\dfrac{2 b \left(1-t^2\right) v}{v^2+1}+\dfrac{2 b t
\left(1-v^2\right)}{v^2+1}-\dfrac{2 d \left(1-t^2\right)
u}{u^2+1}+\dfrac{2
   d t \left(1-u^2\right)}{u^2+1}  = 0,\\
 & -2 a u+\dfrac{2 b \left(1-u^2\right) v}{v^2+1}-\dfrac{2 b u
\left(1-v^2\right)}{v^2+1}+\dfrac{2 c t
\left(1-u^2\right)}{t^2+1}-\dfrac{2
   c \left(1-t^2\right) u}{t^2+1} = 0.\\
\end{cases}
\end{split}
\label{utvgenBis}
\ee
%
The solutions of this system are  nine:
$(u_1,v_1,t_1)$, $(u_{\pm},v_{\pm},t_{\pm})$, $(u^{\pm}_i,v^{\pm}_i,t^{\pm}_i)$ with $i=2, 3$ and $(v^{\pm}_4(u),t^{\pm}_4(u))$.
We found $(u_1,v_1,t_1) = (0,0,0)$ and
\be
\begin{split}
 &\begin{cases}
u_{\pm} & = \pm \sqrt{\dfrac{(-a+b+c-d) (a+b-c+d)}{(a-b-c-d) (a+b-c-d)}}, \\
v_{\pm} & = \pm \dfrac{(b+c)}{(b-c)}\dfrac{[ (a+b-c)^2 - d^2 ]}{ \sqrt{(-a+b+c-d) (a+b-c+d) (a - d - c -b )(a-d-c+b)}} ,\\
t_{\pm} & = \pm \dfrac{a^2+b^2-c^2-d^2}{\sqrt{(a-b-c-d) (a+b-c-d) (-a+b+c-d) (a+b-c+d)}} ,\\
\end{cases}\\
 & \\
& \begin{cases}
u^{\pm}_2 & = 0,\\
v^{\pm}_2 & = \pm \sqrt{\dfrac{(-a-b+c-d) (a+b+c+d)}{(a-b-c+d) (a-b+c+d)}}, \\
t^{\pm}_2& = \pm \dfrac{(a-b+c+d) (a+b+c+d)}{\sqrt{(a-b-c+d) (a-b+c+d) (-a-b+c-d) (a+b+c+d)}}, \\
\end{cases}\\
 & \\
& \begin{cases}
u^{\pm}_3 & = \pm \sqrt{ \dfrac{(-a+b+c-d) (a-b+c+d)}{(a-b-c-d) (a-b+c-d)} }, \\
v^{\pm}_3 & = \pm \dfrac{a^2-b^2+c^2-d^2}{\sqrt{(a-b-c-d) (a-b+c-d) (-a+b+c-d) (a-b+c+d)}}, \\
t^{\pm}_3 & = \pm \dfrac{(b+c) (-a+b-c+d)}{(b-c) (-a+b+c-d)} \sqrt{\dfrac{(-a+b+c-d) (a-b+c+d)}{(a-b-c-d) (a-b+c-d)}} ,\\
\end{cases}\\
 & \\
& \begin{cases}
v^{\pm}_4(u) & = \dfrac{   4 b d u   \pm F(a,b,c,d,u) }  {-u^2 (a-b-c-d) (a-b+c-d)-(a-b+d)^2+c^2}, \\
t^{\pm}_4(u) & = \dfrac{    - 4 c d u \pm F(a,b,c,d,u)  }  {u^2 (a-b-c-d) (a+b-c-d)+(a-c+d)^2-b^2} ,\\
\end{cases}\\
\end{split}
\label{sol3+1}
\ee
%
where
\beq
\begin{split}
F(a,b,c,d,u) & =  \bigg\{  \Big[-u^2 (a+b-c-d) (a-b+c-d)-(a+d)^2+(b-c)^2 \Big] \times\\
& \times \Big[a^2 \left(u^2+1\right)-2 a d \left(u^2-1\right)-\left(u^2+1\right) (b+c-d) (b+c+d) \Big]   \bigg\}^{1/2}.\\
\end{split}
\eeq
%
We observe that in the NH case, the limit $m_1 \rightarrow 0$
corresponds to setting $a \rightarrow 0$ in eqs. (\ref{utvgen}),
while the limit $m_3 \rightarrow 0$ in the IH case
corresponds to $c \rightarrow 0$.
We define the constants $a,b,c,d$ in
these limits respectively as $b_0,c_0,d_0$ and $a_0,b_0,d_0$.
Moreover, we observe that \meff evaluated at the solutions
$(u^{\pm}_i,v^{\pm}_i,t^{\pm}_i)$ with $i = 2,3$, $(v^{\pm}_4,t^{\pm}_4)$ and
$(u_{\pm},v_{\pm},t_{\pm})$ is exactly zero.

In the subsection 6.2 we discuss, in particular, the limiting case of
$m_3\rightarrow 0$. Therefore it is useful to show the solutions
$(u_{\pm},v_{\pm},t_{\pm})$ in terms of
$\sin \gamma_+  = -\sin
\gamma_- $, $\sin \alpha_+ = -\sin\alpha_-$, $\sin \beta_+  = -\sin
\beta_-$, so we can write:
 \beq
\begin{split}
\sin \gamma_- & = \frac{\sqrt{-[(a-d-c)^2 - b^2][(a+d-c)^2 - b^2]}}{2(a-c) d} \,, \\
\sin \alpha_- & = \frac{(b-c) (b+c) (a+b-c-d) \sqrt{(-a+b+c-d)
(a+b-c+d)} (-a+b+c+d)}{2 \left(a b \left(c (a-c)+b^2\right)-b c
d^2\right)
   \sqrt{(-a+c+d)^2-b^2}}\,,\\
   \sin \beta_- & = \frac{\sqrt{(a-b-c-d) (a+b-c-d)} \sqrt{(-a+b+c-d) (a+b-c+d)} \left(-a^2-b^2+c^2+d^2\right)}{2 (a-c) \left[a \left(c
   (a-c)+b^2\right)-c d^2\right]} \,.\\
\end{split}
\eeq
%

Next, in order to study the domains of existence of the solutions
given in eq. (\ref{sol3+1}), which depend on the parameters
$a$ $b$, $c$ and $d$, we need to define the following functions:
\be
\begin{split}
f_1 & = a - b - c - d \,,  \qquad f_2  = a + b - c - d \,, \\
f_3 & = a + b - c + d \,, \qquad f_4  = - a + b + c - d \,, \\
f_5 & = a + b + c + d \,, \qquad f_6  = a - b + c + d \,, \\
f_7 & = a - b + c - d \,, \qquad f_8  = a + b + c - d \,. \\
\end{split}
\ee
%
We notice that the conditions of existence for the solutions
$(u_{\pm},v_{\pm},t_{\pm})$,  $(u^{\pm}_{2},v^{\pm}_{2},t^{\pm}_{2})$ and  $(u^{\pm}_{3},v^{\pm}_{3},t^{\pm}_{3})$
are respectively $ f_3 f_4 f_5 f_6 > 0$, $f_1 f_4 f_6 f_7 > 0$
and $ f_1 f_2 f_3 f_4 > 0$.
We discuss  the domains of the other solutions in
Appendix \ref{AppendixA1} using numerical methods.

Finally, we would like to comment on
the solutions $(v^{\pm}_{4},t^{\pm}_{4})$
because, as can be seen from their explicit expressions,
they depend on a free variable, $u$.
Thus, we would like to provide some details about
the study of the domain of such solutions.
Defining $k= u^2$ and $F(a,b,c,d,k) = \sqrt{ f(k) } $
we find that the function $f(k)$ is a parabola of the form:
\be
f(k) = A k ^2 + B  k + C \,,
\ee
%
with coefficient of the term of maximum degree equal to
\be
A = (a+b-c-d) (a-b+c-d) (a+b+c-d) (-a+b+c+d) \,,
\label{A}
\ee
%
and discriminant $\Delta = B^2 - 4A C= 256 \,a^2 b^2 c^2 d^2$.
The discriminant  $\Delta$
is always positive or equal to zero.
The zeros of the function $f(k)$,
namely $k_1$ and $k_2$, are given by:
\be
\begin{split}
\tilde k_1 & = \frac{(a-b-c+d) (a+b+c+d)}{(a+b+c-d) (-a+b+c+d)}\,,\\
\tilde k_2 & =  -\frac{(a+b-c+d) (a-b+c+d)}{(a+b-c-d) (a-b+c-d)}\,.\\
\end{split}
\ee
%
Depending on the values of the parameters $a,b,c$ and $d$
one can find a range of $m_{min}$ for which this solution
is well defined. We will discuss this in the next
section \ref{AppendixA1}.\\

The method described above cannot be used to determine
the physical domain of the minimization solutions found by us in the
case in which at least one of the phases $\alpha, \beta, \gamma$ is
equal to $\pi$ (eq. (\ref{solpi})). In order to study these cases we
use the Hessian  matrix of $\meff^2$, $H(\alpha,\beta,\gamma)$. The
determinant of the Hessian, evaluated for the phases either  $0$ or
$\pi$ and assuming $a,b,c,d > 0$, can be positive only for
$(\alpha,\beta,\gamma) =
(\pi,0,0),(0,\pi,0),(0,0,\pi),(\pi,\pi,\pi)$. Therefore the local
minima and maxima can correspond only to these configurations. We
derive the relations among the coefficients $a,b,c,d$ in order to
have a minimum using the Sylvester's criterion. We assume that
$a,b,c,d$ are real and positive, $a,b,c,d > 0$.
\\
We have a minima at
\be
\begin{split}
(\alpha,\beta,\gamma) &= (\pi,\pi,\pi) \quad if \quad f_1 = a - b - c - d > 0;\\
(\alpha,\beta,\gamma) &= (0,0,\pi) \quad if \quad (c < d) \land (b < d - c ) \land (f_8 = a + b + c - d < 0) ;\\
(\alpha,\beta,\gamma) &= (0,\pi,0) \quad if \quad (c > d) \land ( b < c - d ) \land (-f_3 = - a - b + c - d > 0);\\
(\alpha,\beta,\gamma) &= (\pi,0,0) \quad if \quad  (b > c+d) \land (-f_6 = - a + b - c - d > 0 ).\\
\label{condmin2}\end{split}
\ee

%
\subsection{Domains of the solutions}
\label{AppendixA1}
%
In this part we describe  the domains of the solutions given in eq.
(\ref{sol3+1}). We will give the numerical intervals of values of
$m_{min}$ in which the minimization solutions are well defined
for $\Dmq_{41(43)}= 0.93~{\rm eV^2}$ and
$1.78~{\rm eV^2}$ and using the best fit values reported in Table
\ref{tabNudata}. In  Tables \ref{DomainNH} and \ref{DomainIH} we
present the results of this numerical analysis in the cases of NO
and IO spectra, respectively.
\begin{table}[h]
\centering
\renewcommand{\arraystretch}{1.1}
\begin{tabular}{lc}
\toprule
Solution  &  Domain of existence in terms of $m_{min}$  \\ \midrule
\multirow{2}{*}{ $(u^{\pm},v^{\pm},t^{\pm})$ }  &   $2.363 \times 10^{-2} \, {\rm eV} < m_1 < 6.473 \times 10^{-2}\, {\rm eV}$ \\
 \vspace{5pt}
&   ($3.337 \times 10^{-2} \,{\rm eV}  < m_1 < 9.061 \times 10^{-2} \, {\rm eV} $) \\
\multirow{2}{*}{$(u^{\pm}_2,v^{\pm}_2,t^{\pm}_2)$} & None \\
 \vspace{5pt}
 & (None) \\
\multirow{2}{*}{ $(u^{\pm}_3,v^{\pm}_3,t^{\pm}_3)$ }& $5.485 \times 10^{-2} \, {\rm eV} < m_1 <6.473 \times 10^{-2} \, {\rm eV} $ \\
 \vspace{5pt}
& ($7.811 \times 10^{-2}  \, {\rm eV} < m_1 < 9.061 \times 10^{-2} \, {\rm eV} $) \\
\multirow{2}{*}{ $(v^{\pm}_4,t^{\pm}_4)$ for $A  > 0$ } &  $2.090 \times 10^{-2} \, {\rm eV} < m_1 < 2.363 \times 10^{-2} \, {\rm eV} \,\,\, \lor$ \\  \vspace{5pt} &  $\lor \,\,\,  5.485 \times 10^{-2} \, {\rm eV} < m_1 <6.473 \times 10^{-2} \, {\rm eV} $ \\
& \Big ($3.043 \times 10^{-2} \, {\rm eV} < m_1 < 3.337 \times 10^{-2} \, {\rm eV} \,\,\, \lor$ \\   \vspace{5pt} & $ \lor \,\,\, 7.811 \times 10^{-2} \, {\rm eV} < m_1 < 9.061 \times  10^{-2} \, {\rm eV} $\Big) \\
\multirow{2}{*}{ $(v^{\pm}_4,t^{\pm}_4)$ for $A  < 0$ } &  $2.363 \times 10^{-2} \, {\rm eV} < m_1 < 5.485 \times 10^{-2} \, {\rm eV} $ \\
 \vspace{5pt}
&  ($3.337 \times 10^{-2} \, {\rm eV} < m_1 < 7.811 \times 10^{-2} \, {\rm eV} $) \\
\bottomrule
\end{tabular}
\caption{Numerical
results for the domains of existence of the solutions given in  eq.
(\ref{sol3+1}) in the 3+1 NO case for $\Dmq_{41} = 0.93 \, {\rm
eV^2} $
($\Dmq_{41} = 1.78\, {\rm eV^2} $). The expression of $A$ is given
in eq. (\ref{A}).}
 \label{DomainNH}
\end{table}

\begin{table}[h]
\centering
\renewcommand{\arraystretch}{1.1}
\begin{tabular}{lc}
\toprule
Solution  &  Domain of existence in terms of $m_{min}$  \\ \midrule
\multirow{2}{*}{ $(u^{\pm},v^{\pm},t^{\pm})$ } &   $0 \, {\rm eV} < m_3 < 3.855 \cdot 10^{-2}\, {\rm eV} $ \\
 \vspace{5pt}
&  ($0 \, {\rm eV} < m_3 < 7.437 \times 10^{-2} \, {\rm eV} $) \\
\multirow{2}{*}{ $(u^{\pm}_2,v^{\pm}_2,t^{\pm}_2)$ }  & None \\
 \vspace{5pt}
 & (None) \\
\multirow{2}{*}{ $(u^{\pm}_3,v^{\pm}_3,t^{\pm}_3)$ } & $3.084 \times 10^{-2} \, {\rm eV} < m_3 < 3.855 \times 10^{-2} \, {\rm eV} $ \\
 \vspace{5pt}
& ($6.344 \times 10^{-2} \, {\rm eV} < m_3 < 7.437 \times 10^{-2} \, {\rm eV} $) \\
\multirow{2}{*}{ $(v^{\pm}_4,t^{\pm}_4)$ for $A  > 0$ } &  $3.084 \times 10^{-2} \, {\rm eV} < m_3 < 3.855 \times  10^{-2} \, {\rm eV} $ \\
 \vspace{5pt}
& ($6.344 \times 10^{-2} \, {\rm eV} < m_3 < 7.437 \times  10^{-2} \, {\rm eV} $) \\
\multirow{2}{*}{ $(v^{\pm}_4,t^{\pm}_4)$ for $A  < 0$ } &  $0 \, {\rm eV} < m_3 < 3.084 \times 10^{-2} \, {\rm eV} $ \\
 \vspace{5pt}
&($0 \, {\rm eV} < m_3 < 6.344 \times 10^{-2} \, {\rm eV} $) \\
\bottomrule
\end{tabular}
\caption{The same as in Table \ref{DomainNH}
but for the case of the  3+1 IO scheme.
}
 \label{DomainIH}
\end{table}

\section{\texorpdfstring{Extrema of \meff in the 3+2 Scheme}{Extrema of the effective Majorana mass in the 3+2 Scheme}}
\label{AppendixB}

 As in the case of the 3+1 scheme, it proves somewhat easier
to study the extrema of  $\meff^2$ than of \meff.
The expression for  $\meff^2$ for both NO and IO spectra
has the form:
\be
\meff^2 = |a+e^{i \alpha } b+e^{i
\beta } c+e^{i \gamma } d + e^{i \eta} e |^2.
\ee
%
Equating to zero the first derivatives of
 $\meff^2$ with respect to the four phases
we get the following system of four equations:
\be
\begin{split}
& a \sin (\alpha )+c \sin (\alpha -\beta )+d \sin (\alpha -\gamma )+e \sin (\alpha -\eta )  = 0,\\
& a \sin (\beta )-b \sin (\alpha -\beta )+d \sin (\beta -\gamma )+e \sin (\beta -\eta ) = 0,\\
-& a \sin (\gamma )+b \sin (\alpha -\gamma )+c \sin (\beta -\gamma )-e \sin (\gamma -\eta ) = 0,\\
-& a \sin (\eta )+b \sin (\alpha -\eta )+c \sin (\beta -\eta )+d \sin (\gamma -\eta ) = 0.\\
\end{split}
\label{3plus2genminmax}
\ee
%

The analytical study of the minima of $\meff^2$ in this case is a
non-trivial task since four phases are involved and the
non-linearity of the system of the first derivatives of $\meff^2$
with respect to the four phases makes the analysis rather
complicated. Therefore finding all possible solutions of the
minimization procedure in analytical form is a complicated problem.
Thus, we have performed the general analysis of the minimization of
\meff numerically. This allowed to determine the intervals of values
of $m_{min}$ in which the minimal value of \meff is exactly zero. It
is possible, however, to perform analytically the analysis of the
minima of \meff, corresponding to the 16 sets of CP conserving
values (either  $0$ or $\pi$) of the  four phases $ \alpha$,
$\beta$, $\gamma$ and $\eta$. This can be done by using the
Sylvester's criterion for the Hessian, evaluated for the indicated
values of the phases $0,\pi$, which determines the physical domain
of the minimization solutions. The minima thus found, as we show,
correspond to $\meff \neq 0$.
%
%

\begin{figure}[h!]
  \begin{center}
\includegraphics[width=7.55cm]{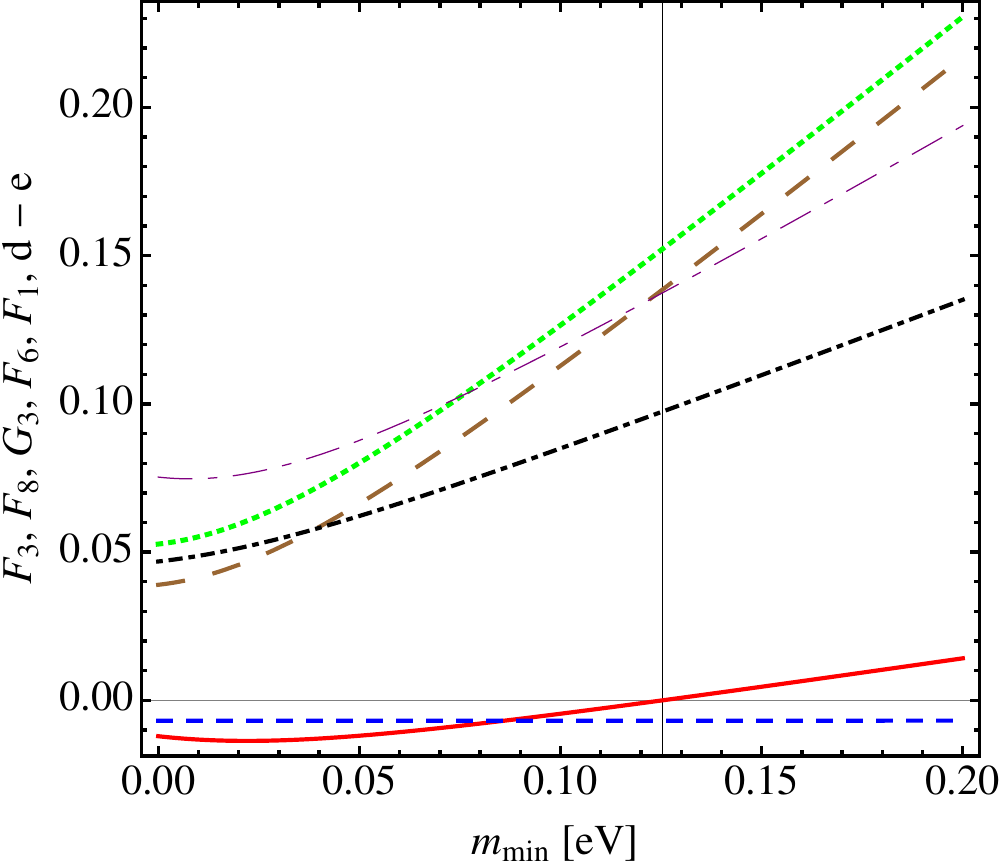}
     \end{center}
\caption{\label{fig:SylvesterIH3plus2}
The functions $F_3$ (dotted green line),
$F_8$ (long-dashed brown line), $G_3$ (short-long-dashed purple line),
$F_6$ (dot-dashed black line), $F_1$ (solid red line), $d - e$
(short-dashed blue line),
defined in eq. (\ref{eq:F1F8}), versus $m_{min}\equiv m_3$
for the oscillation parameter values reported in
Table \ref{tabNudata3plus2}.
The vertical black line corresponds to
$m_{min} = \ov m_3 \simeq 1.25 \times 10^{-1}
 \,{\rm eV}$.
}
\end{figure}

 Assuming $a,b,c,d,e > 0$ and
$a,b,c,d,e \in \mathbb{R}$ and using the Sylvester's criterion
we find that the minima of \meff take place at
\be\begin{split}
 (\alpha,\beta,\gamma,\eta) & = (\pi,\pi,\pi,\pi) \quad if \quad  F_1 = a - b - c - d - e > 0,\\
 (\alpha,\beta,\gamma,\eta) &= (0,0,0,\pi)  \quad if \quad  ( d < e) \land (c < e - d) \land (b < - c - d + e)\land\\
 & \phantom{,aaaaaaaaaaaaaaa}\land  (F_8 = a + b + c + d - e  < 0), \\
(\alpha,\beta,\gamma,\eta) &= (0,0,\pi,0)  \quad if \quad   (d > e) \land  (c < d - e)  \land  (b  < - c + d - e) \land\\
 &  \phantom{,aaaaaaaaaaaaaaa}\land  (F_3 = a + b + c - d + e < 0),\\
 (\alpha,\beta,\gamma,\eta) &= (0,\pi,0,0)  \quad if \quad  (c > d + e)  \land  (b  <  c - d - e) \land \\
 & \phantom{,aaaaaaaaaaaaaaa} \land  (G_3 = a + b - c + d + e < 0),\\
 (\alpha,\beta,\gamma,\eta) &= (\pi,0,0,0)  \quad if \quad   (b > c + d + e) \land  (F_6 = a - b + c + d + e < 0)\,.
\label{cond3+2}\end{split}
\ee
%
At the other CP conserving  values of the
phases, $(\alpha,\beta,\gamma,\eta)$ = $(0,0,0,0)$,
$(\pi,\pi,0,0)$, $(\pi,0,\pi,0)$, $(\pi,0,0,\pi)$,
$(0,\pi,\pi,0)$, $(0,\pi,0,\pi)$, $(0,0,\pi,\pi)$, $(\pi,\pi,\pi,0)$, $(\pi,\pi,0,\pi)$, $(\pi,0,\pi,\pi)$  and  $(0,\pi,\pi,\pi)$,
\meff cannot reach a minimum. In Fig. \ref{fig:SylvesterIH3plus2}
we show the dependence of the functions
$F_1$, $F_8$, $F_3$, $G_3$ and $F_6$ on
$m_{min}\equiv m_3$ for the best fit values of the neutrinos
oscillation parameters given in Tables 1 and 3.

\end{document}